%% file: ms.tex
\documentclass[]{elsarticle}
\usepackage[dvipsnames]{xcolor}
\colorlet{Reviewer1}{black}
\colorlet{Reviewer1R1}{black}
\usepackage{hyperref}

\makeatletter
\def\ps@pprintTitle{%
 \let\@oddhead\@empty
 \let\@evenhead\@empty
 \def\@oddfoot{}%
 \let\@evenfoot\@oddfoot}
\makeatother

\usepackage{numcompress}

\usepackage[english]{babel}
\usepackage[utf8]{inputenc}
\usepackage{amsmath}
\usepackage{cleveref}
\usepackage{booktabs}
\usepackage[disable]{todonotes}
\usepackage{subcaption}
\usepackage{geometry}
\usepackage{amssymb}
\usepackage{algpseudocode}
\usepackage{algorithm}
\usepackage{transparent}
\usepackage[nolist]{acronym}
\usepackage{tabularx} 
\usepackage{relsize}
\newcommand{\tableheadline}[1]{\multicolumn{1}{c}{#1}}
\newcommand{\myfloatalign}{\centering} 

\DeclareRobustCommand{\rchi}{{\mathpalette\irchi\relax}}
\newcommand{\irchi}[2]{\raisebox{\depth}{$#1\chi$}} 

\usepackage{transparent}

\newcommand{\ie}{i.e.\ }
\newcommand{\wrt}{w.r.\ to\ }

\newcommand{\GVOF}{geometrical \ac{VOF} method}

\newcommand{\UFVM}{unstructured \ac{FVM}}
\newcommand{\LSGRAD}{\ac{LS} gradient}

\newcommand{\Point}{\ensuremath{\mathbf{p}}}

\newcommand{\Dimension}{\ensuremath{D}}

\newcommand{\Cell}{\ensuremath{\Omega}}
\newcommand{\Cells}{\ensuremath{\{\Omega_k}\}_{k\in K}}
\newcommand{\Cellsl}{\ensuremath{\{\Omega_l}\}_{l\in K}}

\newcommand{\ForwardCellImage}{\ensuremath{\FlowMap{\tstart}{\tend}(\Omega_l)}}

\newcommand{\ForwardPhaseImage}{\ensuremath{\FlowMap{\tstart}{\tend}(\Omega^+_l(\tstart))}}

\newcommand{\ForwardPhaseImages}{\ensuremath{\{\ForwardPhaseImage}\}_{l\in K}}

\renewcommand{\S}{\ensuremath{\mathbf{S}}}
\newcommand{\Sf}{\ensuremath{\mathbf{S}_f}}

\newcommand{\SfMag}{\ensuremath{\| \Sf{} \|}}
\newcommand{\Such}{\ensuremath{\, : \,}}

\newcommand{\tstart}{\ensuremath{t^n}}
\newcommand{\tend}{\ensuremath{t^{n+1}}}
\newcommand{\tinterval}{\ensuremath{[\tstart,\tend]}}
\renewcommand{\vec}[1]{\ensuremath{\mathbf{#1}}}
\newcommand{\U}{\ensuremath{\mathbf{v}}}
\newcommand{\x}{\ensuremath{\mathbf{x}}}
\newcommand{\n}{\ensuremath{\mathbf{n}}}

\newcommand{\CellOther}{\ensuremath{l}}

\newcommand{\CellPointsSorted}{\ensuremath{\mathcal{P}_{k,\Sigma}}}

\newcommand{\EdgePoint}[1]{\ensuremath{\x_{pq,#1}}}

\newcommand{\EpsilonMosso}{\ensuremath{\epsilon_\textbf{k}^{MS}}}
\newcommand{\EpsilonLlsf}{\ensuremath{\epsilon_\textbf{k}^{LLSF}}}
\newcommand{\Error}{\epsilon}

\newcommand{\FlowMap}[2]{\ensuremath{\Phi_{#1}^{#2}}}

\newcommand{\GradNum}{\ensuremath{\nabla_k}}

\newcommand{\CentroidRe}{\ensuremath{\mathbf{x}_{k,R}}}
\newcommand{\CentroidAd}{\ensuremath{\mathbf{x}_{k,A}}}

\newcommand{\FluxVolume}{\ensuremath{V_{f}}}

\newcommand{\Halfspace}{\ensuremath{\mathcal{H}}}
\newcommand{\HalfspaceInterface}{\ensuremath{\mathcal{H}_{k,A}}}

\newcommand{\HalfspaceInitial}{\ensuremath{\mathcal{H}_{k,R}}}

\newcommand{\IsoPolygonCentroid}{\ensuremath{\x_{k,\lambda}}}

\newcommand{\Indicator}{\ensuremath{\rchi}}

\newcommand{\Overshoot}{\ensuremath{\PlicFraction^o}}

\newcommand{\PhaseFluxVolumeContrib}{\ensuremath{V_{f}^\VolFrac}}

\newcommand{\PlicNormal}{\ensuremath{\n_{k}}}

\newcommand{\PlicNormalCorrected}{\ensuremath{\n_{k}^m}}

\newcommand{\PlicNormalEstimatei}[1]{\ensuremath{\n_{k,#1}}}
\newcommand{\PlicNormalEstimateOther}{\PlicNormalEstimatei{\CellOther}}

\newcommand{\PlicOrigin}{\ensuremath{O_k}}
\newcommand{\PlicPositionx}{\ensuremath{x_\VolFrac}}

\newcommand{\PlicFraction}{\ensuremath{\VolFrac_k}}

\newcommand{\PlicPosition}{\ensuremath{\mathbf{p}_{k}}}
\newcommand{\PlicPolygon}{\ensuremath{\mathcal{Q}_k}}

\newcommand{\PlicPolygoni}[1]{\ensuremath{\mathcal{Q}_{#1}}}
\newcommand{\PlicPolygonCentroid}{\ensuremath{\x_{\PlicPolygon}}}
\newcommand{\PlicPolygonCentroidi}[1]{\ensuremath{\x_{\PlicPolygoni{#1}}}}
\newcommand{\PlicPolygonCentroidOther}{\ensuremath{\PlicPolygonCentroidi{\CellOther}}}
\newcommand{\PlicPolygonsNormals}{\ensuremath{\mathcal{N}_{k}}}

\newcommand{\PointVolumeFraction}{\ensuremath{\alpha_{p}}}

\newcommand{\Slab}{\ensuremath{\mathcal{S}}}
\newcommand{\Slope}{\ensuremath{\vec{s}}}

\newcommand{\Streamline}{\ensuremath{\Psi}}

\newcommand{\Undershoot}{\ensuremath{\PlicFraction^u}}

\newcommand{\VolumeOther}{\ensuremath{|\Omega_\CellOther|}}
\newcommand{\VolFrac}{\ensuremath{\alpha}}
\newcommand{\VolFracskend}{\ensuremath{\{\alpha_k(\tend)\}_{k \in K}}}
\newcommand{\VolFracskstart}{\ensuremath{\{\alpha_k(\tstart)\}_{k \in K}}}

\newcommand{\VolFracLinear}{\ensuremath{\alpha_{l}}}

\newcommand{\WeightOther}{\ensuremath{w_{\CellOther}}}

\newcommand{\CellNeighborhood}{\ensuremath{\mathcal{C}}}

\newcommand{\Nsigma}[1]{\ensuremath{\mathbf{n}_{\Sigma}}}

\renewcommand{\div}[1]{\ensuremath{\nabla\cdot{#1}}}

\graphicspath{{figures/}}

\usepackage{amsthm}
\theoremstyle{definition}

\theoremstyle{remark}

\crefname{remark}{Remark}{remark} 

\begin{document}

\begin{frontmatter}

\listoftodos

\begin{acronym}[]
  \acro{AABB}{Axis-Aligned Bounding Box}
  \acro{ALE}{Arbitrary Lagrangian-Eulerian}
  \acro{AMR}{Adaptive Mesh Refinement}
  \acro{BDS}{Backward Differencing Scheme}
  \acro{BE}{Backward Euler}
  \acro{BGL}{Boost Geometry Library}
  \acro{BT}{Barycentric Triangulation}
  \acro{BCPT}{Barycentric Convex Polygon Triangulation}
  \acro{CAD}{Computer Aided Design}
  \acro{CCI}{Cell / Cell Intersection}
  \acro{CCU}{Cell-wise Conservative Unsplit}
  \acro{CCNR}{Cell Cutting Normal Reconstruction}
  \acro{CDS}{Central Differencing Scheme}
  \acro{CG}{Computer Graphics}
  \acro{CGAL}{Computational Geometry Algorithms Library}
  \acro{CIAM}{Calcul d'Interface Affine par Morceaux}
  \acro{CLCIR}{Conservative Level Contour Interface Reconstruction} 
  \acro{CBIR}{Cubic B\'{e}zier Interface Reconstruction} 
  \acro{CVTNA}{Centroid Vertex Triangle Normal Averaging} 
  \acro{CFD}{Computational Fluid Dynamics}
  \acro{CFL}{Courant-Friedrichs-Lewy}
  \acro{CPT}{Cell-Point Taylor}
  \acro{CPU}{Central Processing Unit}
  \acro{CSG}{Computational Solid Geometry}
  \acro{CV}{Control Volume}
  \acro{DG}{Discontinuous Galerking Method}
  \acro{DR}{Donating Region}
  \acro{DRACS}{Donating Region Approximated by Cubic Splines}
  \acro{DDR}{Defined Donating Region}
  \acro{DGNR}{Distance Gradient Normal Reconstruction}
  \acro{DNS}{Direct Numerical Simulations}
  \acro{EGC}{Exact Geometric Computation}
  \acro{EI-LE}{Eulerian Implicit - Lagrangian Explicit}
  \acro{EILE-3D}{Eulerian Implicit - Lagrangian Explicit 3D}
  \acro{EILE-3DS}{Eulerian Implicit - Lagrangian Explicit 3D Decomposition Simplified}
  \acro{ELVIRA}{Efficient Least squares Volume of fluid Interface Reconstruction Algorithm}
  \acro{EMFPA}{Edge-Matched Flux Polygon Advection}
  \acro{EMFPA-SIR}{Edge-Matched Flux Polygon Advection and Spline Interface Reconstruction}
  \acro{FDM}{Finite Difference Method}
  \acro{FEM}{Finite Element Method}
  \acro{FMFPA-3D}{Face-Matched Flux Polyhedron Advection}
  \acro{FNB}{Face in Narrow Band test}
  \acro{FT}{Flux Triangulation}
  \acro{FV}{Finite Volume}
  \acro{FVM}{Finite Volume method}
  \acro{GPCA}{Geometrical Predictor-Corrector Advection}
  \acro{HTML}{HyperText Markup Language}
  \acro{HPC}{High Performance Computing}
  \acro{HyLEM}{Hybrid Lagrangian–Eulerian Method for Multiphase flow}
  \acro{IO}{Input / Output}
  \acro{IDW}{Inversed Distance Weighted}
  \acro{ISA}{iso-advector scheme}
  \acro{IDWGG}{Inversed Distance Weighted Gauss Gradient}
  \acro{LENT}{Level Set / Front Tracking}
  \acro{LE}{Lagrangian tracking / Eulerian remapping}
  \acro{LEFT}{hybrid level set / front tracking}
  \acro{LFRM}{Local Front Reconstruction Method}
  \acro{LVIRA}{Least squares Volume of fluid Interface Reconstruction Algorithm}
  \acro{LLSG}{Linear Least Squares Gradient}
  \acro{LS}{Least Squares}
  \acro{LSF}{Least Squares Fit}
  \acro{IDWLSG}{Inverse Distance Weighted Least Squares Gradient}
  \acro{LCRM}{Level Contour Reconstruction Method}
  \acro{LSG}{Least Squares Gradient}
  \acro{LSP}{Liskov Substitution Principle}
  \acro{MCE}{Mean Cosine Error}
  \acro{MoF}{Moment of Fluid}
  \acro{MS}{Mosso-Swartz}
  \acro{NIFPA}{Non-Intersecting Flux Polyhedron Advection}
  \acro{NS}{Navier-Stokes}
  \acro{NP}{Non-deterministic Polynomial}
  \acro{OOD}{Object Oriented Design}
  \acro{OD}{Owkes-Desjardins scheme}
  \acro{ODE}{ordinary differential equation}
  \acro{OD-S}{Owkes-Desjardins Sub-resolution}
  \acro{OT}{Oriented Triangulation}
  \acro{OCPT}{Oriented Convex Polygon Triangulation}
  \acro{EPT}{Edge-based Polygon Triangulation}
  \acro{PDE}{Partial Differential Equation}
  \acro{PAM}{Polygonal Area Mapping Method}
  \acro{iPAM}{improved Polygonal Area Mapping Method}
  \acro{PIR}{Patterned Interfacce Reconstruction}
  \acro{PCFSC}{Piecewise Constant Flux Surface Calculation}
  \acro{PLIC}{Piecewise Linear Interface Calculation}
  \acro{RTS}{Run-Time Selection}
  \acro{RKA}{Rider-Kothe Algorithm}
  \acro{RK}{Runge-Kutta}
  \acro{RTT}{Reynolds Transport Theorem}
  \acro{SCL}{Space Conservation Law}
  \acro{SMCI}{Surface Mesh / Cell Intersection}
  \acro{SFINAE}{Substitution Failure Is Not An Error}
  \acro{SIR}{Spline Interface Reconstruction}
  \acro{SLIC}{Simple Line Interface Calculation}
  \acro{SRP}{Single Responsibility Principle}
  \acro{STL}{Standard Template Library}
  \acro{TBDS}{Taylor-Backward Differencing Scheme}
  \acro{TBES}{Taylor-Euler Backward Scheme}
  \acro{THINC/QQ}{Tangent of Hyperbola Interface Capturing with Quadratic surface representation and Gaussian Quadrature}
  \acro{UML}{Unified Modeling Language}
  \acro{UFVFC}{Unsplit Face-Vertex Flux Calculation}
  \acro{VOF}{Volume of Fluid}
  \acro{YSR}{Youngs' / Swartz Reconstruction algorithm}
\end{acronym}


\title{Unstructured un-split geometrical Volume-of-Fluid methods\\- A review}

\author[mma]{Tomislav Mari\'{c}\corref{corr}}
\cortext[corr]{Corresponding author}
\ead{maric@mma.tu-darmstadt.de}

\author[ornl]{Douglas B. Kothe}
\ead{kothe@ornl.gov}

\author[mma]{Dieter Bothe}
\ead{bothe@mma.tu-darmstadt.de}

\address[mma]{Mathematical Modeling and Analysis Group, TU Darmstadt}
\address[ornl]{Oak Ridge National Laboratory, Exascale Computing Project}

\begin{abstract}

\emph{Note}: this is an updated preprint of the manuscript accepted for publication in \emph{Journal of Computational Physics}, DOI: \url{https://doi.org/10.1016/j.jcp.2020.109695}. Please refer to the journal version when citing this work.

\textcolor{Reviewer1}{Geometrical Volume-of-Fluid (VoF) methods mainly support structured meshes, and only a small number of contributions in the scientific literature report results with unstructured meshes and \textcolor{Reviewer1R1}{three spatial dimensions}. Unstructured meshes are traditionally used for handling geometrically complex solution domains that are prevalent when simulating problems of industrial relevance. However, three-dimensional geometrical operations are significantly more complex than their two-dimensional counterparts, \textcolor{Reviewer1R1}{which is confirmed by the ratio of publications with three-dimensional results on unstructured meshes to publications with two-dimensional results or support for structured meshes.} Additionally, unstructured meshes present challenges in serial and parallel computational efficiency, accuracy, implementation complexity, and robustness. Ongoing research is still very active, focusing on different issues: interface positioning in general polyhedra, estimation of interface normal vectors, advection accuracy, and parallel and serial computational efficiency.}

\textcolor{Reviewer1}{This survey tries to give a complete and critical overview of classical, as well as contemporary geometrical VOF methods with concise explanations of the underlying ideas and sub-algorithms, focusing primarily on unstructured meshes and three dimensional calculations. Reviewed methods are listed in historical order and compared in terms of accuracy and computational efficiency.}

\end{abstract}


\end{frontmatter}


\frenchspacing
\input{sections/introduction.tex}
\input{sections/mathematical-model-dieter.tex}
\input{sections/mathematical-model-tomislavdieter.tex}
\input{sections/interface-reconstruction.tex}
\input{sections/volume-fraction-advection.tex}
\input{sections/conclusions.tex}

\section{Acknowledgements}

Funded by the Deutsche Forschungsgemeinschaft (DFG, German Research Foundation) – Project-ID 265191195 – SFB 1194, sub-projects B01 and Z-INF. 

Calculations for this research were conducted on the Lichtenberg high performance computer of the TU Darmstadt.

The first author is grateful for the fruitful discussions with his colleague, Dr.-Ing. Dirk Gr\"{u}nding, that have helped to improve the sections \ref{subsec:fluxbased} and \ref{subsec:cellbased} of the manuscript. 

The authors extend their gratitude to Dr. Christopher B. Ivey for the tabular result data of the \ac{NIFPA}-1 method.

\clearpage

\bibliography{references}


\end{document}

%% file: sections/introduction.tex
\section{Introduction}
\label{sec:intro}

\input{sections/extended-abstract.tex}



%% file: sections/extended-abstract.tex
The \acf{VOF} method \citep{DeBar1974,Hirt1981,Rider1998} is widely used to capture interfaces in the numerical simulation of multi-phase flows, owing in part to its many potential advantages: global and local volume conservation, second-order convergence in three dimensions, numerical consistency, numerical stability, robust treatment of interface coalescence and breakup, support for unstructured domain discretization, and a straightforward parallel computation model. These characteristics, however, often remain elusive for many \ac{VOF} formulations. 

The \ac{VOF} method approximates the interface with a discrete Heaviside function represented by a volume fraction, \ie the ratio of the volume occupied by a specific phase in a multi-material computational cell, to the volume of the whole cell. Over the last two decades, different variants of the \ac{VOF} method have been developed, all of which can be categorized as taking either an \emph{algebraic} or \emph{geometric} approach to approximate interface kinematics via an algorithm for advection of volume fractions. 

Algebraic \ac{VOF} methods \citep{Zalesak1979, Ubbink1999, Muzaferija1999, Waclawczyk2008} invoke continuum-based \ac{PDE} discretization schemes for the advection of the volume fraction field. This approach is challenging and can lead to problems, owing to the volume fraction field possessing a large and abrupt change (across the interface) that causes interpolation and subsequently discretization errors, when \textcolor{Reviewer1R1}{algebraic advection algorithms} are used. The algebraic methods additionally suffer from the loss of numerical consistency caused by artificial diffusion, \ie the inability to maintain a constant (and not widening) interface width. The loss of consistency likewise leads to the loss in the convergence order. More recent developments of algebraic \ac{VOF} schemes have alleviated some aforementioned issues, \textcolor{Reviewer1R1}{but not all of them}. 

Geometric \ac{VOF} methods, instead, \textcolor{Reviewer1R1}{rely on geometrical operations to approximate the solution of the volume fraction advection equation}. \textcolor{Reviewer1R1}{All variants of the geometrical \ac{VOF} method rely on a cell-by-cell geometrical approximation of the interface, that is reconstructed in multi-material cells by the \ac{PLIC} algorithm.} \textcolor{Reviewer1R1}{Volume fraction advection is then computed from the geometrical approximation of the interface and cell-faces, cells, or phase-specific material volumes, that are traced along Lagrangian trajectories}. This requires additional and complex geometrical operations such as the triangulation and intersection of possibly non-convex self-intersecting polyhedrons with non-planar faces. These geometric approximations enable the advection of the fluid interface in a direction that is independent of the mesh geometry (\ie the direction of the face-normal vectors), which simultaneously removes \textcolor{Reviewer1R1}{mesh-anisotropy errors} and introduces support for unstructured meshes. \textcolor{Reviewer1R1}{The mesh-anisotropy errors impress the shape of the dual of the cell onto the shape of the advected interface. For example, a sphere advected in the direction of the spatial diagonal on a Cartesian mesh with the algebraic VOF method deforms into an octahedron, because the cell stencil of the algebraic method on a cubic mesh is a dual of a cube - an octahedron.} Geometrical reconstruction of the interface and the fluxed phase-specific volumes circumvents the interpolation errors of the algebraic \ac{VOF} schemes. These geometric approximations are the basis for the second-order convergence, numerical stability and consistency of the geometric \ac{VOF} methods. 

\begin{figure}[!b] 
  \centering
  \begin{subfigure}[t]{0.29\textwidth}
    \includegraphics[width=\columnwidth]{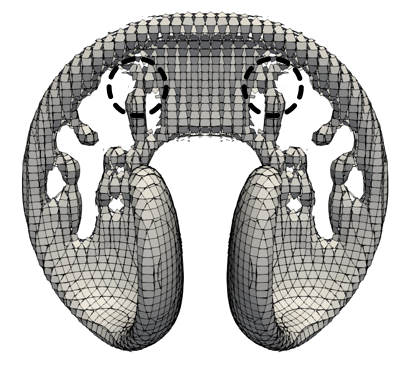}
    \caption{Youngs reconstruction and barycentric triangulation.}
    \label{fig:youngsbary}
  \end{subfigure}
  \begin{subfigure}[t]{0.29\textwidth}
    \includegraphics[width=\columnwidth]{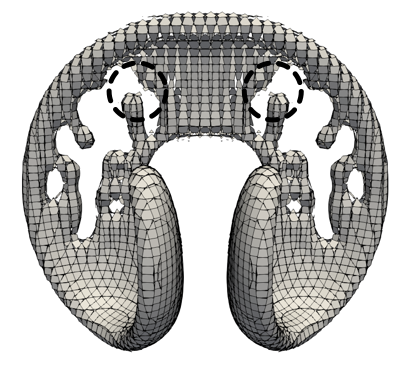}
    \caption{Youngs reconstruction and flux-based triangulation.}
    \label{fig:youngsflux}
  \end{subfigure}

  \begin{subfigure}[t]{0.29\textwidth}
    \includegraphics[width=\columnwidth]{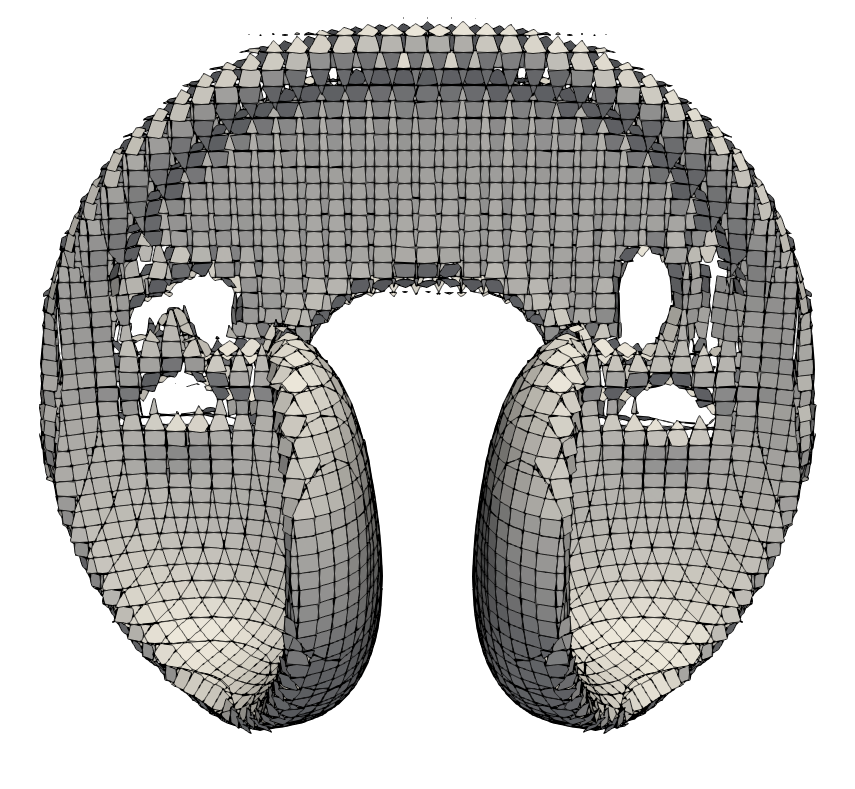}
    \caption{Swartz reconstruction with barycentric triangulation.}
    \label{fig:swartzbary}
  \end{subfigure}
  \begin{subfigure}[t]{0.29\textwidth}
    \includegraphics[width=\columnwidth]{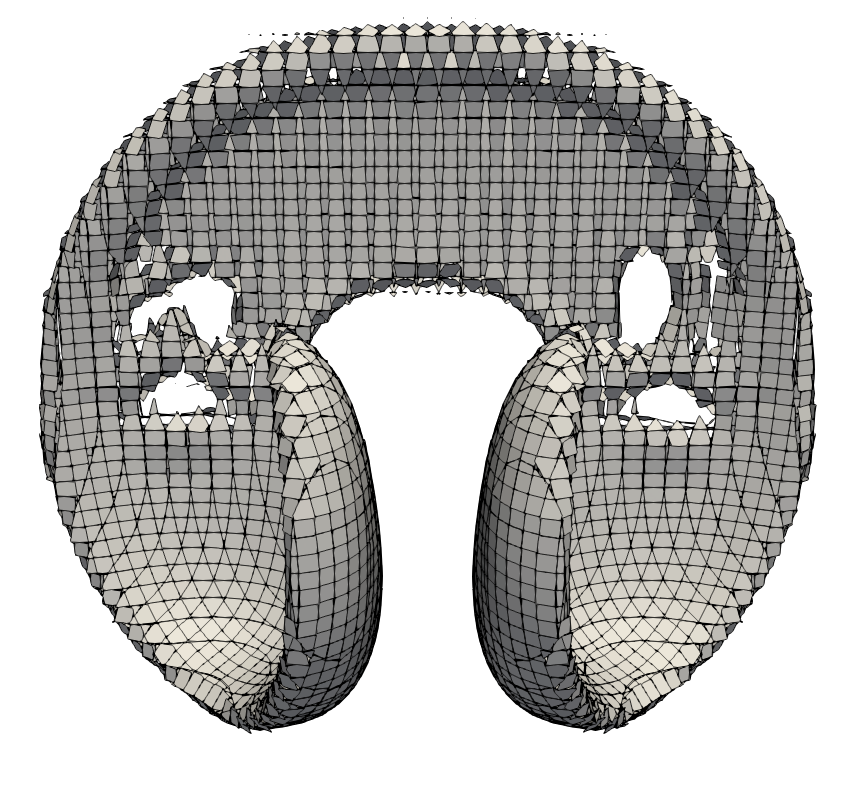}
    \caption{Swartz reconstruction with flux-based triangulation.}
    \label{fig:swartzflux}
  \end{subfigure}
  \begin{subfigure}[t]{0.29\textwidth}
  \centering
    \includegraphics[width=\columnwidth]{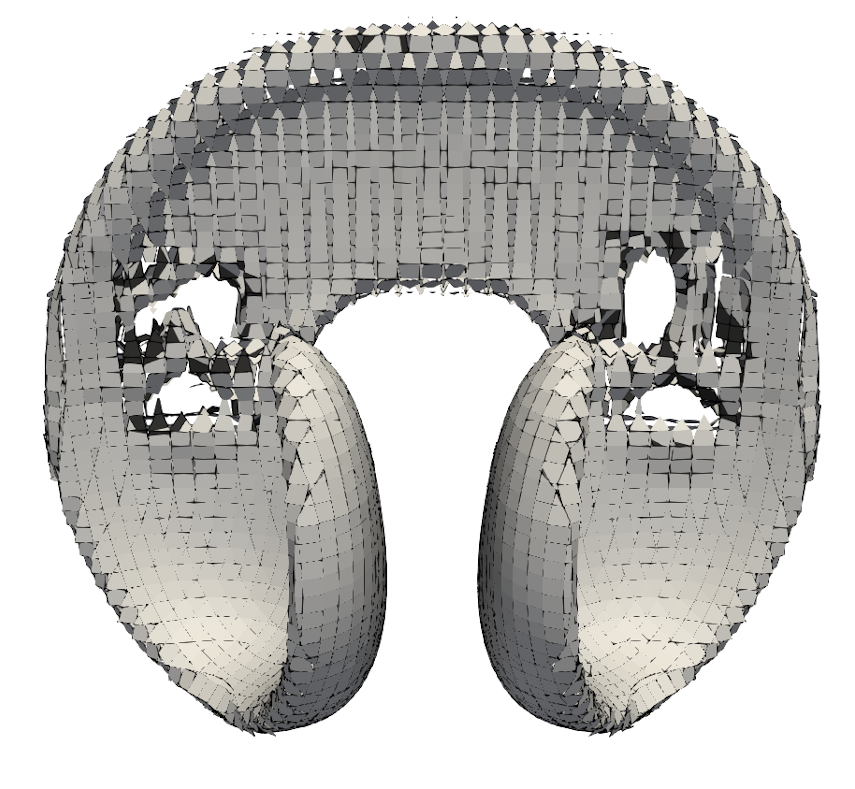}
    \caption{Swartz recontruction with barycentric and flux-based triangulation.}
    \label{fig:swartzbaryflux}
  \end{subfigure}
  \caption{3D deformation verification case \citep{Enright2002,Smolarkiewicz1982} at $t=1.5s$ with $64^3$ equidistant cubical cells combining reconstruction and flux-volume triangulation algorithms. Results are obtained using the algorithms reported in \citep{Maric2018}.}
  \label{fig:swartzyoungs}
\end{figure}

Geometric \ac{VOF} methods can be further categorized as dimensionally \emph{split} and \emph{un-split}. Dimensionally split methods best support structured meshes \citep{Scardovelli2000,Renardy2002,Popinet2003,Scardovelli2003,Aulisa2007,Popinet2009,Agbaglah2011}, because they rely on the operator-splitting approach to achieve second-order accuracy, \textcolor{Reviewer1R1}{that requires face-normal vectors to be collinear with the coordinate axes (grid lines of the structured mesh)}. The two main motivations for formulating dimensionally un-split geometric \ac{VOF} methods are (i) the ability to utilize unstructured meshes for handling geometrically complex solution domains, and (ii) the possibility of increasing the overall solution accuracy by improving the Lagrangian reconstruction of the fluxed phase-specific volume. Dimensionally un-split \ac{VOF} methods therefore have been and still are very actively investigated.

\textcolor{Reviewer1}{Algebraic \ac{VOF} methods solve a linear algebraic system to advect the interface, \textcolor{Reviewer1R1}{which is an approach that has a high level of serial and parallel computational efficiency}. Geometric \ac{VOF} methods, on the other hand, rely on different relatively complex explicit geometric sub-algorithms. Local geometrical operations increase the serial computational efficiency of the method. \textcolor{Reviewer1R1}{However, geometric data and calculations follow a moving fluid interface, which can freely leave one parallel process and enter another, easily making a parallel computation imbalanced in terms of the computational load shared by the parallel processes.}}

\textcolor{Reviewer1}{The choice of sub-algorithms of the geometric \ac{VOF} method significantly impacts the solution accuracy. An example \ac{PLIC} interface is shown in \cref{fig:swartzyoungs} for the standard 3D deformation verification case \citep{Enright2002,Smolarkiewicz1982} at $t=1.5s$. Two reconstruction algorithms are compared (Youngs \citep{Youngs1982} and simplified Swartz \citep{Maric2018}), and two triangulation algorithms (barycentric and flux-based \citep{Maric2018}). \textcolor{Reviewer1R1}{The barycentric triangulation uses the centroid of the volume and the triangles from its triangulated boundary to construct tetrahedrons that decompose the volume. The flux-based triangulation relies on the displacement vectors given by the velocity field to decompose the volume into tetrahedrons more accurately.} Solutions presented in \cref{fig:youngsbary,fig:youngsflux} are affected for the Youngs reconstruction algorithm by the chosen triangulation. \textcolor{Reviewer1R1}{Similarly, comparing \cref{fig:swartzbary,fig:swartzflux} with \cref{fig:youngsbary,fig:youngsflux}, the importance in choosing a better reconstruction algorithm is evident, because the more accurate Swartz reconstruction algorithm prevents the artificial breakup of the thin layer.}}

\textcolor{Reviewer1R1}{The effect of the triangulation is barely visible for the simplified Swartz algorithm in \cref{fig:swartzbary,fig:swartzflux}, however the effect is substantial, because it impacts \textcolor{Reviewer1R1}{convergence}. To emphasize the difference, different \textcolor{Reviewer1R1}{gray scale} is used in \cref{fig:swartzbaryflux} for the simplified Swartz algorithm, using respectively the barycentric (gray color) and flux-based (black color) triangulation. The convergence-order and absolute accuracy of the standard advection verification cases are primarily affected by the fidelity of the advection in those parts of the interface, where the topological changes occur. The impact of the sub-algorithms is large in \cref{fig:swartzyoungs}, even with a prescribed velocity. Therefore, one can safely assume that the choice of sub-algorithms will strongly impact the solution when the velocity results from \textcolor{Reviewer1R1}{solving the} two-phase Navier-Stokes system.}

Improving the sub-algorithms of the geometric \ac{VOF} method is a topic of ongoing extensive research effort. This survey article tries to give a complete and critical overview of classical as well as contemporary geometrical VOF methods with detailed self-consistent explanations of the underlying ideas and sub-algorithms. The referenced algorithms are systematically categorized and compared in terms of accuracy and computational efficiency. Links to publications used for the comparison are provided, together with brief reviews which are focused on those specific improvements reported in the literature. \textcolor{Reviewer1R1}{The aim of this survey article is to provide a solid starting-point for formulation and implementation of dimensionally un-split geometric \ac{VOF} methods.}

%% file: sections/mathematical-model-dieter.tex
\section{Geometrical Volume-of-Fluid method}
\label{sec:mathmodel}

\begin{figure}[htb] 
    \centering
    \def\svgwidth{0.5\columnwidth}
    { 
        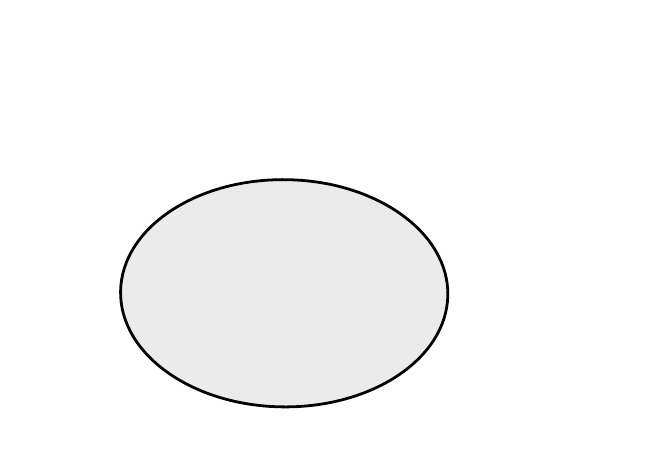
    }
    \caption{Multi-material domain.}
    \label{fig:solutiondomain}
\end{figure} 

The core idea of the \acf{VOF} method is to capture an evolving interface $\Sigma(t)$ that separates two phases $\Omega^+(t)$ and $\Omega^-(t)$, shown schematically in \cref{fig:solutiondomain}. More precisely, $\Sigma(t)$ is thus defined as the boundary of its adjacent phases, \ie  
\begin{equation}
    \Sigma(t) = \partial \Omega^+(t), 
\end{equation}
say, where a given domain $\Omega$ contains two phases $\Omega^+(t)$ and $\Omega^-(t)$, such that 
\begin{equation}
    \Omega = \Omega^+(t) \cup \Omega^-(t) \cup \Sigma(t).
\end{equation}
Furthermore, the phases are described by means of phase indicator functions. Thus, for instance, 
\begin{equation}
    \Omega^+(t) = \{\x \in \Omega: \rchi(t,\x) = 1 \}
\end{equation}
with $\rchi(t, \cdot)$ being the (phase) indicator function of $\Omega^+(t)$, \ie 
\begin{equation}
    \rchi(t,\x) = 
    \begin{cases}
        1 \quad \x \in \Omega^+(t), \\
        0 \quad \x \not\in \Omega^+(t).
    \end{cases}
\end{equation}
This continuum formulation has its discrete analogue, where the \ac{FVM} is an appropriate discretization approach. In fact, \textcolor{Reviewer1}{introducing the volume fraction of the phase $+$ inside a volume $V$ of magnitude $|V|$ at time $t$ as}
\begin{equation}
    \VolFrac(t, V) := \dfrac{1}{|V|}\int_V \rchi(t,\x) dV,
    \label{eq:volfracdef}
\end{equation}
it follows that 
\textcolor{Reviewer1R1}{\begin{equation}
    \VolFrac(t, V) = \dfrac{1}{|V|}\int_{V\cap\Omega^+(t)} 1\, dV 
    = \dfrac{|V \cap \Omega^+(t)|}{|V|}
    = \dfrac{|V^+(t)|}{|V|},
    \label{eq:alphadef}
\end{equation}}
where \textcolor{Reviewer1R1}{$V^+(t):=V\cap\Omega^+(t)$} is the volume occupied by $\Omega^+$ inside the volume V at time $t$, \textcolor{Reviewer1R1}{which we call the \emph{phase-specific volume}}.

\begin{figure}[htb] 
    \centering
    \def\svgwidth{0.6\columnwidth}
    { 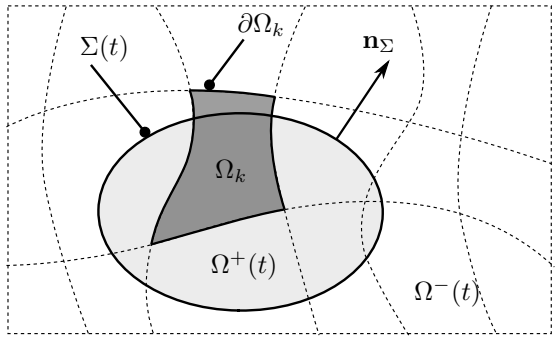
    }
    \caption{\textcolor{Reviewer1R1}{Decomposition of the multi-material solution domain into disjoint subsets $\{\Omega_k\}_{k \in K}$.}}
    \label{fig:solutiondomaindecomp}
\end{figure} 
Therefore, \cref{eq:volfracdef} defines the volume fraction of phase $\Omega^+(t)$ inside $V$, hence $\VolFrac$ denotes the \emph{volume of fluid} \textcolor{Reviewer1R1}{\emph{fraction}} inside $V$, if "fluid" refers to the phase labeled by "$+$". On the discrete level, $\Omega$ is decomposed into disjoint open sub-volumes (mesh cells, say) $\Omega_k$ for $k \in K$, as shown in \cref{fig:solutiondomaindecomp}. Given such a decomposition of $\Omega$ into $\{\Omega_k\}_{k \in K}$, we define  
\begin{equation}
    \VolFrac_k(t) := \alpha(t, \Omega_k) 
    = \dfrac{1}{|\Cell_k|} \int_{\Cell_k} \rchi(t, \x) dV 
    = \dfrac{|\Omega_k \cap \Omega^+(t)|}{|\Omega_k|}
    = \dfrac{|\Omega_k^+(t)|}{|\Omega_k|},
    \label{eq:alphakdef}
\end{equation}
\textcolor{Reviewer1R1}{where, equivalently to $V^+$ in \cref{eq:alphadef}, we call $\Cell_k^+(t)$ the \emph{phase-specific volume} inside $\Cell_k$ at time $t$.} Knowledge of $\{\VolFrac_k\}_{k \in K}$ directly allows to identify all mesh cells with nonempty intersection with the interface $\Sigma(t)$, the so-called interface or multi-material cells. Indeed, it holds that
\begin{align}
    \alpha_k(t) = 0 &\iff \Omega_k \subset \Omega^-(t), \nonumber \\
    \alpha_k(t) \in (0,1) &\iff \Omega_k \cap \Sigma(t) \ne \emptyset, \\
    \alpha_k(t) = 1 &\iff \Omega_k \subset \Omega^+(t).\nonumber 
\end{align}
This detection of all interface cells at time $t$ requires the computation of the indicator function $\rchi$ (\textcolor{Reviewer1R1}{or an approximation thereof}), for which an evolution equation for $\rchi$ is required. This equation comes from continuum physics and is usually based on the assumption of absence of phase change. In this case, fluid particles cannot cross the interface, \ie the value of $\rchi$ does not change along a trajectory, viz.  
\begin{equation}
    \rchi(t, \x(t)) \equiv \text{const}
    \label{eq:chiconst}
\end{equation}
for $\x(\cdot)$ a solution of 
\begin{equation}
    \dot{\x}(t) = \U(t, \x(t)),
    \label{eq:xoft}
\end{equation}
where $\vec{v}$ denotes the velocity field. Note that $\vec{v}$ is a two-phase velocity field, hence \cref{eq:xoft} is an \acf{ODE} with \emph{discontinuous} right-hand side. While, in general, such \ac{ODE}s can lack solvability or uniqueness, it can be shown that \cref{eq:xoft} is a well-posed \ac{ODE} if $\vec{v}$ is a physically sound two-phase velocity field even if phase change is allowed \citep{Bothe2019}.   

Consequently, the phase indicator $\rchi$ satisfies 
\begin{equation}
    \frac{D\rchi}{Dt}=0 \quad \text{(Lagrangian derivative)},
\end{equation}
in a certain sense, discussed in more detail below. In an Eulerian form, the basic and well-known transport equation  
\begin{equation}
    \partial_t \rchi + \vec{v} \cdot \nabla\rchi = 0
    \label{eq:chitransp}
\end{equation}
for the phase indicator results. Formally, this is the same as the \emph{level set equation} or, more general, the transport equation for a non-diffusive passive scalar. \textcolor{Reviewer1R1}{But in contrast to the level set equation, an interpretation of \cref{eq:chitransp} in a pointwise sense is not useful:} at points where $\rchi$ is locally constant, \cref{eq:chitransp} is trivially fulfilled, while, at points where $\rchi$ has a jump discontinuity, \cref{eq:chitransp} can only be valid in a weak sense. Even more, a classical interpretation of \cref{eq:chitransp} in the sense of distributions does not reach far enough. Instead, the theory of functions of bounded variations and related concepts from geometrical measure theory provide an appropriate mathematical framework. Besides derivatives like $\partial_t\rchi$ or $\nabla\rchi$ of the discontinuous indicator function, also nonlinear operations on such quantities are important, such as $\|\nabla\rchi\|$, where $\|\cdot\|$ denotes the Euclidean norm. In fact, it holds that 
\begin{equation}
    \int_V\phi \|\nabla\rchi\| dV =  \int_{V\cap \Sigma} \phi \, dS, 
    \label{eqn:intmagindicator}
\end{equation}
\ie $\|\nabla\rchi\|$ is the Dirac distribution \wrt $\Sigma$. \textcolor{Reviewer1R1}{Equation \ref{eqn:intmagindicator} is the theoretical basis for numerical approximations of surface quantities like, e.g., surface tension forces. For instance, choosing $\phi =1$ in (13) shows that}
\begin{equation}
    |\Sigma \cap V| = \int_V \| \nabla\rchi \| dV
\end{equation}
is the area of the surface $\Sigma \cap V$ and, mathematically, the right-hand side is the total variation of the Radon measure $\nabla \rchi$. For more information about this subject see, e.g., \citep{Giusti1984,Evans2018}. 

Within the \ac{VOF} method, the discretization of \cref{eq:chitransp} is usually based on the \ac{FVM}, which is directly related to the integral form of the phase-specific volume balance. This integral form follows by application of the Reynolds transport theorem. \textcolor{Reviewer1}{Indeed, if $\Cell_k \subset \Omega$ is a fixed control volume, then 
\begin{equation}
    \dfrac{d}{dt}\int_{\Cell_k} \rchi \, dV = 
    \dfrac{d}{dt}\int_{\Cell_k \cap \Omega^+(t)} 1 \, dV = 
    \int_{\Cell_k \cap \Sigma(t)} \U^\Sigma\cdot \n^+ \, dS,
\end{equation}}where $\U^\Sigma\cdot \n^+$ is the speed of normal displacement of $\Sigma(t)$ in the direction of the outer normal $\n^+$ to $\Omega^+$. Since the standing assumption of no phase change implies 
\begin{equation}
    \vec{v}^+\cdot \n_\Sigma =
    \vec{v}^\Sigma \cdot \n_\Sigma = 
    \vec{v}^- \cdot \n_\Sigma,
    \label{eq:nophasechange}
\end{equation}
one obtains
\begin{equation}
    \dfrac{d}{dt}\int_{\Cell_k} \rchi \, dV = 
    \int_{\Cell_k \cap \Sigma(t)} \U^+\cdot \n^+ \, dS.
    \label{eq:chirtt}
\end{equation}
Note that $\n_\Sigma$ is either $\n^+$ or $-\n^+$ $(=\n^-)$.

\begin{figure}[h] 
  \centering
  \begin{subfigure}[t]{0.4\textwidth}
    \def\svgwidth{\columnwidth}
       {\footnotesize
        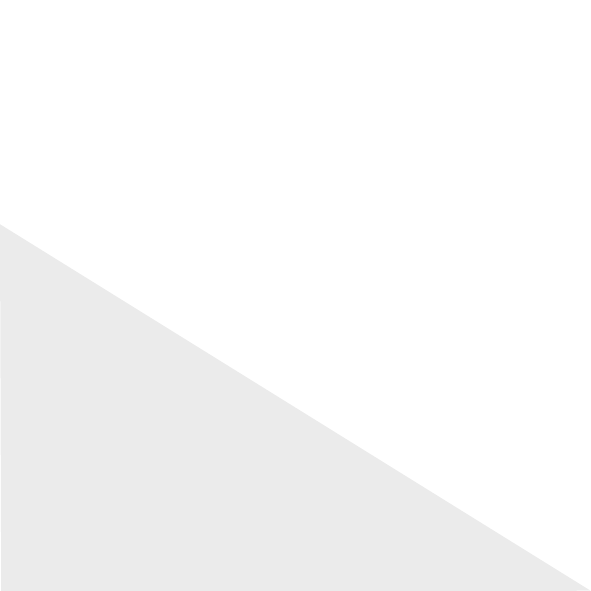
       }
      \caption{Volume $V$ as a fixed control volume, bounded by $\partial V$.}
    \label{fig:controlvolume}
  \end{subfigure}
  \quad
  \begin{subfigure}[t]{0.4\textwidth}
    \def\svgwidth{\columnwidth}
       {\footnotesize
        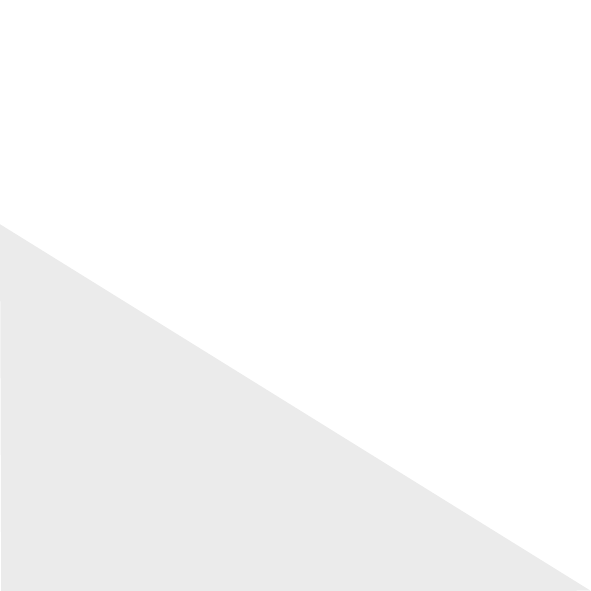
        }
      \caption{Volume $V(t)$, bounded by $\partial V(t)$, tracked as a co-moving (material) volume with the flow map $\FlowMap{t_0}{t}$, starting at time $t_0$ with $V(t_0)$.}
    \label{fig:moving-control}
  \end{subfigure}
    \caption{\textcolor{Reviewer1R1}{Fixed control volume $V$ (\cref{fig:controlvolume}) versus co-moving (material) volume $V(t)$ (\cref{fig:moving-control}).}}
\end{figure}

From here on, we also assume the flow \textcolor{Reviewer1R1}{inside $\Omega^+(t)$} to be incompressible, meaning that $\nabla\cdot\vec{v}^+=0$. Let us note in passing that the \textcolor{Reviewer1}{transport equation (\ref{eq:chitransp})} for $\rchi$ then, formally, becomes 
\begin{equation}
    \partial_t \rchi + \nabla \cdot (\rchi \vec{v}) = 0.
    \label{eq:chitranspdiv}
\end{equation}
Employing \textcolor{Reviewer1R1}{the divergence theorem with} $\nabla \cdot \vec{v}^+ = 0$, equation (\ref{eq:chirtt}) implies
\begin{equation}
    \frac{d}{dt}\int_{\Cell_k} \rchi dV = - \int_{\partial \Cell_k \cap \Omega^+(t)} \U^+ \cdot \n^+ dS.
\end{equation}
On the right-hand side, the integration runs over the "wetted part" $\partial \Omega_k \cap \Omega^+(t)$ of $\partial \Cell_k$ on which $\rchi=1$ holds, while $\rchi=0$ on the remainder of $\partial \Cell_k$. Therefore, the integral form of the transport equation for $\rchi$ \textcolor{Reviewer1R1}{in the incompressible case} finally reads  
\begin{equation}
    \frac{d}{dt}\int_{\Omega_k} \rchi \, dV = - \int_{\partial \Omega_k} \rchi \U \cdot \n \, dS.
    \label{eq:chitranspint}
\end{equation}
An important variant of \cref{eq:chitranspint} employs \emph{co-moving} (material) volumes. Since we need this concept in a precise manner below, let us recall that for a flow field $\vec{v} : J \times \overline{\Omega} \rightarrow \mathbb{R}^3, J=(a,b) \subset \mathbb{R}$, which is continuous in $(t, \x)$ and satisfies a local Lipschitz condition \wrt $\x$, the initial value problems 
\begin{equation}
    \dot{\x} = \vec{v}(t, \x(t)), \quad t \in J, \, \x(t_0) = \x_0
    \label{eq:xodej}
\end{equation}
have unique local solutions $\x(\cdot; t_0, \x_0)$ for every $t_0 \in J$, $x_0 \in \Omega$. These solutions exist for all $t\in J$ if $\vec{v}$ is linearly bounded and satisfies $\vec{v}\cdot\n = 0$ at $\partial \Omega$. Under the latter impermeability condition, initial values from $\overline{\Omega}$ are allowed. While the same result holds for two-phase flows under physically sound assumptions on the jump of $\vec{v}^+$ at $\Sigma(t)$ (see \citep{Bothe2019}), such extensions are not needed here, since we also assume no-slip at $\Sigma$, \ie
\begin{equation}
    \vec{v}^+_{||} =
    \vec{v}^-_{||} =
    \vec{v}^\Sigma_{||}, 
\end{equation}
which, together with \cref{eq:nophasechange}, implies that $\vec{v}$ is continuous at $\Sigma$ and this, together with the assumed local Lipschitz continuity of $\vec{v}^{\pm}$ on $\rm{gr}(\overline{\Omega}^{\pm})$, is sufficient. 

Now, existence of unique solutions to \cref{eq:xodej} yields the associated \emph{flow map}, \ie the map $\Phi_{t_0}^{t}$ defined as 
\begin{equation}
    \Phi_{t_0}^{t} (\x_0) := \x(t;t_0, \x_0), 
    \label{eq:flowmap}
\end{equation}
which maps the initial point $\x_0$ to the point $\x(t)$, where $\x(\cdot)=\x(\cdot; t_0, \x_0)$ is the solution for the initial condition $\x(t_0) = \x_0$. With this notation, a two-phase \emph{co-moving} (material) volume $V(t)$ is given as 
\begin{equation}
    V(t) = \Phi_{t_0}^{t} (V(t_0)) := \{ \Phi_{t_0}^{t} (\x_0) : \x_0 \in V(t_0) \}
\end{equation}
\textcolor{Reviewer1R1}{for some initial volume $V(t_0)$ illustrated in \cref{fig:moving-control}. Under the assumption of no phase change, the phase-specific volume inside $V(t)$} is also a material volume which, moreover, has constant volume if the velocity field is solenoidal for the respective phase. 

At this point it is useful to note that the flow map $\Phi_s^t$ also exists for $s>t$, since the initial value problems \ref{eq:xodej} have solutions going forward and backward in time. Forward and backward solutions are related by  means of time reversal, therefore the backward solution of \cref{eq:xodej} is the forward solution of the same \ac{ODE}, but with $-\U$ instead of $\U$. Thus, the inverted flow map $\big( \Phi_t^s \big)^{-1}$ is nothing but the flow map to the reversed velocity field $-\U$, \textcolor{Reviewer1}{and 
\begin{equation}
    \FlowMap{s}{t}\FlowMap{t}{s}(V) = \FlowMap{t}{s}\FlowMap{s}{t}(V) = V. 
    \label{eq:flowmapinv}
\end{equation}}\textcolor{Reviewer1R1}{The analogue of \cref{eq:chitranspint} for a co-moving volume starting as the cell $\Omega_k$ at time $t_0$ reads as
\begin{equation}
    \dfrac{d}{dt}\int_{\FlowMap{t_0}{t}{(\Cell_k})} \rchi \, dV = 0. 
    \label{eq:chitranspmat}
\end{equation}}\textcolor{Reviewer1R1}{Evidently, \cref{eq:chitranspmat} is equivalent to}
%
\textcolor{Reviewer1R1}{\begin{equation}
    |\FlowMap{t_0}{t}(\Cell_k)\cap \Omega^+(t)| \equiv const. \quad \text{for } t \in J.
    \label{eq:spacecons}
\end{equation}}\Cref{eq:spacecons} is the so-called space (geometric) conservation law, applied to the co-moving volume starting as the cell $\Cell_k$ at time $t_0$, which holds for \textcolor{Reviewer1R1}{solenoidal velocity fields $\nabla \cdot \U = 0$ inside the phases, and in the absence of phase change. The space conservation law further implies 
\begin{equation}
    \alpha(t,\FlowMap{t_0}{t}(V)) = \dfrac{|\FlowMap{t_0}{t}(V)\cap\Omega^+(t)|}{|\FlowMap{t_0}{t}(V)|} = \dfrac{|V\cap\Omega^+(t_0)|}{|V|} = \alpha(t_0, V),
    \label{eq:spaceconvplus}
\end{equation}}which is especially relevant for a version of the geometrical VOF method described in \cref{subsec:cellbased}. Either \cref{eq:chitranspint} or \cref{eq:chitranspmat} \textcolor{Reviewer1R1}{represents the core equation} of any \emph{geometrical} \ac{VOF} method, \ie a \ac{VOF} method which employs geometrical calculations for the approximate computation of integrals appearing in the time-integrated form of these relations. Methods covered by this review rely on the geometric approach, where a common challenge is to approximate complicated, non-convex volumes in $\mathbb{R}^3$, with non-planar boundaries that arise when integrating \cref{eq:chitranspint,eq:chitranspmat} in time. Of course, one may also try to avoid such complications by replacing the full volume integrals by temporal integrals of volumetric fluxes on the boundary of $\Omega_k$, or dimensionally (directionally) splitting the evaluation of integrals in \cref{eq:chitranspint,eq:chitranspmat}; however, this leads to larger approximation errors. 

A completely different approach, leading to the \emph{algebraic} \ac{VOF} method relies on the direct discretization of the \cref{eq:chitranspdiv} for $\rchi$, without aiming at a fully sharp interface representation on the discrete level. However, as outlined in the introduction, this assumption leads to problems with consistency and, therefore, the convergence of the algebraic method. One of the main advantages of the geometric \ac{VOF} methods is the reduction of numerical diffusion, which significantly reduces the number of interface cells in the interface normal direction. 

Discretization of \cref{eq:chitranspint,eq:chitranspmat} leads to two different categories of dimensionally un-split geometrical \ac{VOF} methods: the \emph{flux-based} versus the \emph{cell-based} method. For both methods, $\Omega$ is decomposed into disjoint $\Omega_k, k \in K$, such that the definition of $\alpha_k (t)$ given by \cref{eq:alphakdef} applies. The task of both un-split geometrical \ac{VOF} methods is the following: given a time discretization $t^0 < t^1 < t^2 < \dots < t^N$, at each point $t^n$ in the discretization, given $\VolFracskstart$, compute $\VolFracskend$ by integrating either \cref{eq:chitranspint} or \cref{eq:chitranspmat} in time over $[t^n, t^{n+1}]$. Details of the temporal integration of \cref{eq:chitranspint} and \cref{eq:chitranspmat} are explained in the following sections.

%% file: figures/mathmodel-domain.pdf_tex
\begingroup%
  \makeatletter%
  \providecommand\color[2][]{%
    \errmessage{(Inkscape) Color is used for the text in Inkscape, but the package 'color.sty' is not loaded}%
    \renewcommand\color[2][]{}%
  }%
  \providecommand\transparent[1]{%
    \errmessage{(Inkscape) Transparency is used (non-zero) for the text in Inkscape, but the package 'transparent.sty' is not loaded}%
    \renewcommand\transparent[1]{}%
  }%
  \providecommand\rotatebox[2]{#2}%
  \newcommand*\fsize{\dimexpr\f@size pt\relax}%
  \newcommand*\lineheight[1]{\fontsize{\fsize}{#1\fsize}\selectfont}%
  \ifx\svgwidth\undefined%
    \setlength{\unitlength}{189bp}%
    \ifx\svgscale\undefined%
      \relax%
    \else%
      \setlength{\unitlength}{\unitlength * \real{\svgscale}}%
    \fi%
  \else%
    \setlength{\unitlength}{\svgwidth}%
  \fi%
  \global\let\svgwidth\undefined%
  \global\let\svgscale\undefined%
  \makeatother%
  \begin{picture}(1,0.68564763)%
    \lineheight{1}%
    \setlength\tabcolsep{0pt}%
    \put(0,0){\includegraphics[width=\unitlength,page=1]{mathmodel-domain.pdf}}%
    \put(0.05992822,0.49357669){\color[rgb]{0,0,0}\makebox(0,0)[lt]{\lineheight{0}\smash{\begin{tabular}[t]{l}$\Sigma(t)$\end{tabular}}}}%
    \put(0,0){\includegraphics[width=\unitlength,page=2]{mathmodel-domain.pdf}}%
    \put(0.64673413,0.52687048){\color[rgb]{0,0,0}\makebox(0,0)[lt]{\lineheight{0}\smash{\begin{tabular}[t]{l}$\n_\Sigma$\end{tabular}}}}%
    \put(0.39857885,0.23024357){\color[rgb]{0,0,0}\makebox(0,0)[lt]{\lineheight{0}\smash{\begin{tabular}[t]{l}$\Omega^+(t)$\end{tabular}}}}%
    \put(0.66511731,0.08341812){\color[rgb]{0,0,0}\makebox(0,0)[lt]{\lineheight{0}\smash{\begin{tabular}[t]{l}$\Omega^-(t)$\end{tabular}}}}%
    \put(0,0){\includegraphics[width=\unitlength,page=3]{mathmodel-domain.pdf}}%
    \put(0.47098518,0.6249519){\color[rgb]{0,0,0}\makebox(0,0)[lt]{\lineheight{0}\smash{\begin{tabular}[t]{l}$\Omega$\end{tabular}}}}%
  \end{picture}%
\endgroup%

%% file: figures/mathmodel-domain-decomposition.pdf_tex
\begingroup%
  \makeatletter%
  \providecommand\color[2][]{%
    \errmessage{(Inkscape) Color is used for the text in Inkscape, but the package 'color.sty' is not loaded}%
    \renewcommand\color[2][]{}%
  }%
  \providecommand\transparent[1]{%
    \errmessage{(Inkscape) Transparency is used (non-zero) for the text in Inkscape, but the package 'transparent.sty' is not loaded}%
    \renewcommand\transparent[1]{}%
  }%
  \providecommand\rotatebox[2]{#2}%
  \newcommand*\fsize{\dimexpr\f@size pt\relax}%
  \newcommand*\lineheight[1]{\fontsize{\fsize}{#1\fsize}\selectfont}%
  \ifx\svgwidth\undefined%
    \setlength{\unitlength}{226.12062836bp}%
    \ifx\svgscale\undefined%
      \relax%
    \else%
      \setlength{\unitlength}{\unitlength * \real{\svgscale}}%
    \fi%
  \else%
    \setlength{\unitlength}{\svgwidth}%
  \fi%
  \global\let\svgwidth\undefined%
  \global\let\svgscale\undefined%
  \makeatother%
  \begin{picture}(1,0.55063161)%
    \lineheight{1}%
    \setlength\tabcolsep{0pt}%
    \put(0,0){\includegraphics[width=\unitlength,page=1]{mathmodel-domain-decomposition.pdf}}%
    \put(0.1412752,0.44424924){\color[rgb]{0,0,0}\makebox(0,0)[lt]{\lineheight{0}\smash{\begin{tabular}[t]{l}$\Sigma(t)$\end{tabular}}}}%
    \put(0,0){\includegraphics[width=\unitlength,page=2]{mathmodel-domain-decomposition.pdf}}%
    \put(0.55536693,0.45522837){\color[rgb]{0,0,0}\makebox(0,0)[lt]{\lineheight{0}\smash{\begin{tabular}[t]{l}$\n_\Sigma$\end{tabular}}}}%
    \put(0.33315433,0.12496006){\color[rgb]{0,0,0}\makebox(0,0)[lt]{\lineheight{0}\smash{\begin{tabular}[t]{l}$\Omega^+(t)$\end{tabular}}}}%
    \put(0.63138888,0.08569769){\color[rgb]{0,0,0}\makebox(0,0)[lt]{\lineheight{0}\smash{\begin{tabular}[t]{l}$\Omega^-(t)$\end{tabular}}}}%
    \put(0,0){\includegraphics[width=\unitlength,page=3]{mathmodel-domain-decomposition.pdf}}%
    \put(0.37240723,0.47793413){\color[rgb]{0,0,0}\makebox(0,0)[lt]{\lineheight{0}\smash{\begin{tabular}[t]{l}$\partial\Omega_k$\end{tabular}}}}%
    \put(0,0){\includegraphics[width=\unitlength,page=4]{mathmodel-domain-decomposition.pdf}}%
    \put(0.33975899,0.26404739){\color[rgb]{0,0,0}\makebox(0,0)[lt]{\lineheight{0}\smash{\begin{tabular}[t]{l}$\Omega_k$\end{tabular}}}}%
    \put(0,0){\includegraphics[width=\unitlength,page=5]{mathmodel-domain-decomposition.pdf}}%
  \end{picture}%
\endgroup%

%% file: figures/mathmodel-0.pdf_tex
\begingroup%
  \makeatletter%
  \providecommand\color[2][]{%
    \errmessage{(Inkscape) Color is used for the text in Inkscape, but the package 'color.sty' is not loaded}%
    \renewcommand\color[2][]{}%
  }%
  \providecommand\transparent[1]{%
    \errmessage{(Inkscape) Transparency is used (non-zero) for the text in Inkscape, but the package 'transparent.sty' is not loaded}%
    \renewcommand\transparent[1]{}%
  }%
  \providecommand\rotatebox[2]{#2}%
  \newcommand*\fsize{\dimexpr\f@size pt\relax}%
  \newcommand*\lineheight[1]{\fontsize{\fsize}{#1\fsize}\selectfont}%
  \ifx\svgwidth\undefined%
    \setlength{\unitlength}{170.07874016bp}%
    \ifx\svgscale\undefined%
      \relax%
    \else%
      \setlength{\unitlength}{\unitlength * \real{\svgscale}}%
    \fi%
  \else%
    \setlength{\unitlength}{\svgwidth}%
  \fi%
  \global\let\svgwidth\undefined%
  \global\let\svgscale\undefined%
  \makeatother%
  \begin{picture}(1,1)%
    \lineheight{1}%
    \setlength\tabcolsep{0pt}%
    \put(0,0){\includegraphics[width=\unitlength,page=1]{mathmodel-0.pdf}}%
    \put(0.0474086,0.18339353){\color[rgb]{0,0,0}\makebox(0,0)[lt]{\lineheight{0}\smash{\begin{tabular}[t]{l}$\Indicator(t, \cdot) = 1$\end{tabular}}}}%
    \put(0.0474086,0.93113348){\color[rgb]{0,0,0}\makebox(0,0)[lt]{\lineheight{0}\smash{\begin{tabular}[t]{l}$\Indicator(t,\cdot) = 0$\end{tabular}}}}%
    \put(0,0){\includegraphics[width=\unitlength,page=2]{mathmodel-0.pdf}}%
    \put(0.08859894,0.37591951){\color[rgb]{0,0,0}\makebox(0,0)[lt]{\lineheight{0}\smash{\begin{tabular}[t]{l}$\partial V$\end{tabular}}}}%
    \put(0,0){\includegraphics[width=\unitlength,page=3]{mathmodel-0.pdf}}%
    \put(0.09967101,0.82174227){\color[rgb]{0,0,0}\makebox(0,0)[lt]{\lineheight{0}\smash{\begin{tabular}[t]{l}$\Sigma(t)$\end{tabular}}}}%
    \put(0,0){\includegraphics[width=\unitlength,page=4]{mathmodel-0.pdf}}%
    \put(0.86751425,0.6250247){\color[rgb]{0,0,0}\makebox(0,0)[lt]{\lineheight{0}\smash{\begin{tabular}[t]{l}$\n_\Sigma$\end{tabular}}}}%
    \put(0.79203487,0.8064933){\color[rgb]{0,0,0}\makebox(0,0)[lt]{\lineheight{0}\smash{\begin{tabular}[t]{l}$\n$\end{tabular}}}}%
    \put(0,0){\includegraphics[width=\unitlength,page=5]{mathmodel-0.pdf}}%
    \put(0.47599659,0.29539801){\color[rgb]{0,0,0}\makebox(0,0)[lt]{\lineheight{0}\smash{\begin{tabular}[t]{l}$\Omega^+(t)$\end{tabular}}}}%
    \put(0.65202244,0.9326028){\color[rgb]{0,0,0}\makebox(0,0)[lt]{\lineheight{0}\smash{\begin{tabular}[t]{l}$\Omega^-(t)$\end{tabular}}}}%
    \put(0.50429419,0.72270857){\color[rgb]{0,0,0}\makebox(0,0)[lt]{\lineheight{0}\smash{\begin{tabular}[t]{l}$V$\end{tabular}}}}%
  \end{picture}%
\endgroup%

%% file: figures/mathmodel-0-cell.pdf_tex
\begingroup%
  \makeatletter%
  \providecommand\color[2][]{%
    \errmessage{(Inkscape) Color is used for the text in Inkscape, but the package 'color.sty' is not loaded}%
    \renewcommand\color[2][]{}%
  }%
  \providecommand\transparent[1]{%
    \errmessage{(Inkscape) Transparency is used (non-zero) for the text in Inkscape, but the package 'transparent.sty' is not loaded}%
    \renewcommand\transparent[1]{}%
  }%
  \providecommand\rotatebox[2]{#2}%
  \newcommand*\fsize{\dimexpr\f@size pt\relax}%
  \newcommand*\lineheight[1]{\fontsize{\fsize}{#1\fsize}\selectfont}%
  \ifx\svgwidth\undefined%
    \setlength{\unitlength}{170.07874016bp}%
    \ifx\svgscale\undefined%
      \relax%
    \else%
      \setlength{\unitlength}{\unitlength * \real{\svgscale}}%
    \fi%
  \else%
    \setlength{\unitlength}{\svgwidth}%
  \fi%
  \global\let\svgwidth\undefined%
  \global\let\svgscale\undefined%
  \makeatother%
  \begin{picture}(1,1)%
    \lineheight{1}%
    \setlength\tabcolsep{0pt}%
    \put(0,0){\includegraphics[width=\unitlength,page=1]{mathmodel-0-cell.pdf}}%
    \put(0.0474086,0.18339353){\color[rgb]{0,0,0}\makebox(0,0)[lt]{\lineheight{0}\smash{\begin{tabular}[t]{l}$\Indicator(t, \cdot) = 1$\end{tabular}}}}%
    \put(0.0474086,0.93113348){\color[rgb]{0,0,0}\makebox(0,0)[lt]{\lineheight{0}\smash{\begin{tabular}[t]{l}$\Indicator(t,\cdot) = 0$\end{tabular}}}}%
    \put(0,0){\includegraphics[width=\unitlength,page=2]{mathmodel-0-cell.pdf}}%
    \put(0.08859894,0.37591951){\color[rgb]{0,0,0}\makebox(0,0)[lt]{\lineheight{0}\smash{\begin{tabular}[t]{l}$\partial V(t)$\end{tabular}}}}%
    \put(0,0){\includegraphics[width=\unitlength,page=3]{mathmodel-0-cell.pdf}}%
    \put(0.09967101,0.82174227){\color[rgb]{0,0,0}\makebox(0,0)[lt]{\lineheight{0}\smash{\begin{tabular}[t]{l}$\Sigma(t)$\end{tabular}}}}%
    \put(0,0){\includegraphics[width=\unitlength,page=4]{mathmodel-0-cell.pdf}}%
    \put(0.86751425,0.6250247){\color[rgb]{0,0,0}\makebox(0,0)[lt]{\lineheight{0}\smash{\begin{tabular}[t]{l}$\n_\Sigma$\end{tabular}}}}%
    \put(0.79203487,0.8064933){\color[rgb]{0,0,0}\makebox(0,0)[lt]{\lineheight{0}\smash{\begin{tabular}[t]{l}$\n$\end{tabular}}}}%
    \put(0,0){\includegraphics[width=\unitlength,page=5]{mathmodel-0-cell.pdf}}%
    \put(0.47599659,0.29539801){\color[rgb]{0,0,0}\makebox(0,0)[lt]{\lineheight{0}\smash{\begin{tabular}[t]{l}$\Omega^+(t)$\end{tabular}}}}%
    \put(0.65202244,0.9326028){\color[rgb]{0,0,0}\makebox(0,0)[lt]{\lineheight{0}\smash{\begin{tabular}[t]{l}$\Omega^-(t)$\end{tabular}}}}%
    \put(0,0){\includegraphics[width=\unitlength,page=6]{mathmodel-0-cell.pdf}}%
    \put(0.16804885,0.25266253){\color[rgb]{0,0,0}\makebox(0,0)[lt]{\lineheight{0}\smash{\begin{tabular}[t]{l}$\partial V(t_0)$\end{tabular}}}}%
    \put(0,0){\includegraphics[width=\unitlength,page=7]{mathmodel-0-cell.pdf}}%
  \end{picture}%
\endgroup%

%% file: sections/mathematical-model-tomislavdieter.tex
\subsection{Flux-based un-split geometrical VOF method}
\label{subsec:fluxbased}

\begin{figure}[H] 
  \centering
    \def\svgwidth{0.5\textwidth}
       {\footnotesize
        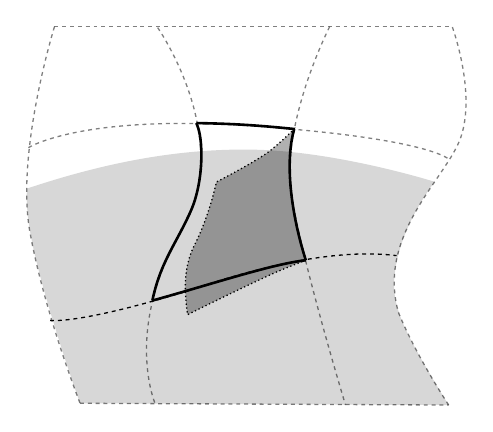
       }
      \caption{Calculation of the phase-specific volume $\PhaseFluxVolumeContrib$ fluxed through $S_f$ over time interval $[\tstart, \tend]$ using the flux-based un-split VOF method.} 
      \label{fig:fluxbased}
\end{figure}

By equations (\ref{eq:alphakdef}) and (\ref{eq:chitranspint}),
\begin{equation}
    \dfrac{d}{dt} \VolFrac_k(t) = 
    \dfrac{d}{dt} \dfrac{1}{|\Omega_k|} \int_{\Omega_k} \rchi(t, \x) \, dV = 
    - \dfrac{1}{|\Omega_k|} \int_{\partial \Omega_k} \rchi \U \cdot \n \, dS. 
   \label{eqn:volfrac-eqn}
\end{equation}
Integrating this equation in time over \tinterval{} yields 
\begin{align}
    \VolFrac_k(\tend) = \VolFrac_k(\tstart) - \dfrac{1}{|\Omega_k|} \int_{\tstart}^{\tend} \int_{\partial \Omega_k} \Indicator \U \cdot  \n \, dS \, dt.  
  \label{eqn:volfrac-ex}
\end{align}
\Cref{eqn:volfrac-ex} is still an exact equation, as no approximations have been applied so far. Notice that \cref{eqn:volfrac-ex} has two unknowns: $\VolFrac_k$ and $\Indicator$. A numerical method based on \cref{eqn:volfrac-ex} is termed a \emph{dimensionally un-split flux-based geometrical Volume of Fluid method}. The un-split flux-based \GVOF{} utilizes three-dimensional geometrical operations that are not dimensionally split to approximate the integral on the right-hand side of \cref{eqn:volfrac-ex}. The term $\int_{t^n}^{t^{n+1}} \int_{\partial \Omega_k} \Indicator \U \cdot  \n \, dS \, dt$ \textcolor{Reviewer1R1}{is, in fact, the volume of the phase "+" that is fluxed over the boundary $\partial \Omega_k$ over the time step from $t^n$ to $t^{n+1}$, the so-called \emph{fluxed phase-specific volume}, shown in \cref{fig:fluxbased}. The boundary $\partial \Omega_k$ is assumed as piecewise-smooth, composed of smooth surfaces (so-called \emph{faces}), i.e.\ 
\begin{equation}
    \partial \Omega_k = \cup_{f \in F_k} S_f.
\end{equation}
\Cref{eqn:volfrac-ex} can therefore be reformulated as
\begin{align}
    \VolFrac_k(\tend) = \VolFrac_k(\tstart) - \dfrac{1}{|\Omega_k|}  \sum_{f \in F_k} \int_{\tstart}^{\tend} \int_{S_f} \Indicator \U \cdot  \n \, dS \, dt,  
    \label{eq:alphasfint}
\end{align}
\textcolor{Reviewer1R1}{where the double integral on the r.h.s.\ gives the amount of the phase-specific volume, fluxed over the face $S_f$ during the interval $[\tstart, \tend]$. Introducing the set
\begin{equation}
    \PhaseFluxVolumeContrib := \bigcup_{t \in [\tstart, \tend]} 
    \{\x : \FlowMap{\tstart}{t}(\x) \in S_f \cap \Omega^+(t) \}
    = \bigcup_{t \in [\tstart, \tend]} \FlowMap{t}{\tstart}(S_f \cap \Omega^+(t)),
\end{equation}
i.e.\ the part of $\Omega^+(t)$ which is fluxed over $S_f$ in $[t^n, t^{n+1}]$, we can rewrite \cref{eq:alphasfint} as}
\begin{align}
    \VolFrac_k(\tend) = \VolFrac_k(\tstart) - \dfrac{1}{|\Omega_k|}  \sum_{f \in F_k} |\PhaseFluxVolumeContrib|.
    \label{eqn:phase-specific}
\end{align}
\Cref{eqn:phase-specific} is still an exact equation. To compute the sets $\PhaseFluxVolumeContrib$, we exploit the flow invariance of $\Omega^+(\cdot)$ to obtain 
\begin{equation}
    \begin{aligned}
        \PhaseFluxVolumeContrib & = \bigcup_{t \in [\tstart,\tend]} \FlowMap{t}{\tstart}(S_f) \cap \FlowMap{t}{\tstart}(\Omega^+(t))
        & = \bigcup_{t \in [\tstart,\tend]} \FlowMap{t}{\tstart}(S_f) \cap \Omega^+(t^n).
    \end{aligned}
\end{equation}
Therefore, if the \emph{flux volume} across the face $S_f$ is defined as
\begin{equation}
    V_f = \bigcup_{t \in [\tstart,\tend]}\FlowMap{t}{\tstart}(S_f), 
    \label{eq:fluxvol}
\end{equation}
$\PhaseFluxVolumeContrib$ is expressed using the phase $\Omega^+$, according to
\begin{equation}
    \PhaseFluxVolumeContrib = V_f \cap \Omega^+(\tstart), 
    \label{eq:phase-fluxvol}
\end{equation}
and inserted back into \cref{eqn:phase-specific} to solve for $\VolFracskend$. In other words, the fluxed phase-specific volume $\PhaseFluxVolumeContrib$, as shown in \cref{fig:fluxbased}, is computed as an intersection of the volume $V_f$, constructed by tracking $S_f$ backward in time, with the flow map $\FlowMap{}{}$, and the phase $\Omega^+(\tstart)$.}

\textcolor{Reviewer1R1}{It is important to shed some light on the way $\PhaseFluxVolumeContrib$ is generally calculated in a discrete setting. The magnitude of the flux volume is given by 
\begin{equation}
    |V_f| = \int_{t^n}^{t^{n+1}} \int_{S_f} \U \cdot \n \, dS \, dt. 
    \label{eq:fluxvolmag}
\end{equation}
The approximate computation of \cref{eq:fluxvolmag} depends on the chosen equation discretization method, and here we utilize the unstructured FVM. The flux volume $V_f$, given by \cref{eq:fluxvol}, is also approximated, because of the velocity interpolation and temporal integration used in the approximation of the flow map $\FlowMap{}{}$ in a discrete setting. Approximations used in \cref{eq:fluxvol} and \cref{eq:fluxvolmag} are in general very different from each other. Therefore, the flux volume $V_f$ is modified in the geometric VOF method, such that \cref{eq:fluxvolmag} is satisfied. Only if this is achieved, can \cref{eq:phase-fluxvol} be used to compute $\PhaseFluxVolumeContrib$ and inserted into \cref{eqn:phase-specific} to solve for $\VolFracskend$, while maintaining volume conservation. Specific approximations, utilized for this purpose by different flux-based geometrical VOF methods, are described in \cref{sec:advect}.}
\subsection{Cell-based un-split geometrical \ac{VOF} method}
\label{subsec:cellbased}

\textcolor{Reviewer1R1}{The direction in time in which the co-moving volume is tracked distinguishes \emph{forward tracking} from \emph{backward tracking} cell-based un-split \GVOF{}s. Both forward and backward tracking methods utilize geometrical intersections between the images of co-moving volumes and the underlying Eulerian mesh $\{\Omega_l\}_{l \in K}$, in the so-called Eulerian remap step. To help reduce the number of necessary intersections, an \emph{intersection stencil} of a volume $V$ is defined as the set of all indices of those cells that have a non-empty intersection with $V$, namely}
\textcolor{Reviewer1R1}{\begin{equation}
    \CellNeighborhood(V) := \{ l \in K : V \cap \Cell_l \ne \emptyset\}.
    \label{eq:cellstencilv}
\end{equation}}For example, the intersection stencil of the forward image of the cell $\Cell_k$ in $\Cellsl$ is  
\begin{equation}
    \CellNeighborhood(\Phi_{\tstart}^{\tend}(\Omega_k)) := \{ l \in K : \Phi_{\tstart}^{\tend}(\Omega_k) \cap \Cell_l \ne \emptyset\}. 
    \label{eq:cellstencilomegak}
\end{equation}
\textcolor{Reviewer1R1}{Note that additional indices $l \in K$ must be introduced via \cref{eq:cellstencilomegak} because the forward cell image $\Phi_{\tstart}^{\tend}(\Omega_k)$ generally overlaps with multiple cells from $\Cellsl$, and not just its pre-image (the cell $\Cell_k$). Equivalently, forward or backward images of each phase-specific volume $\Cell^+_k(t^n)$ will generally overlap with multiple cells from the mesh $\Cellsl$.} 

\subsubsection{Forward tracking}

\begin{figure}[h] 
  \centering
  \begin{subfigure}[t]{0.45\textwidth}
    \def\svgwidth{\columnwidth}
       {\footnotesize
        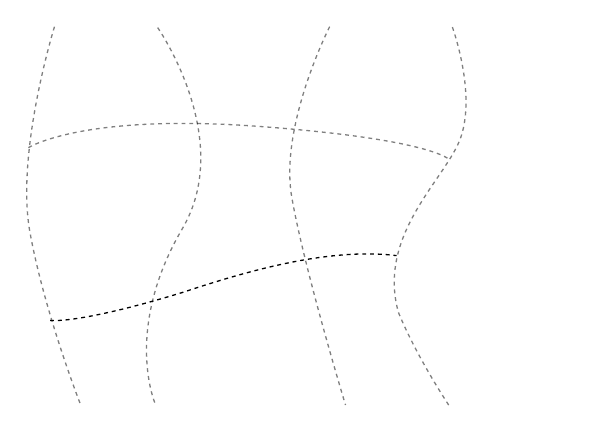
       }
      \caption{\textcolor{Reviewer1R1}{Lagrangian forward tracking of the phase-specific volume $\Cell_l^+(\tstart)$ as $\FlowMap{\tstart}{\tend}(\Cell_l^+(\tstart))$ and Eulerian remapping of volume fractions using \cref{eq:eulerremap}, by intersecting $\ForwardPhaseImage$ with $\Cell_k$, based on the intersection stencil $\CellNeighborhood^+_{n+1}(\Cell_k)$.}} 
      \label{fig:forwarda}
  \end{subfigure}
  \quad
  \begin{subfigure}[t]{0.45\textwidth}
    \def\svgwidth{\columnwidth}
       {\footnotesize
        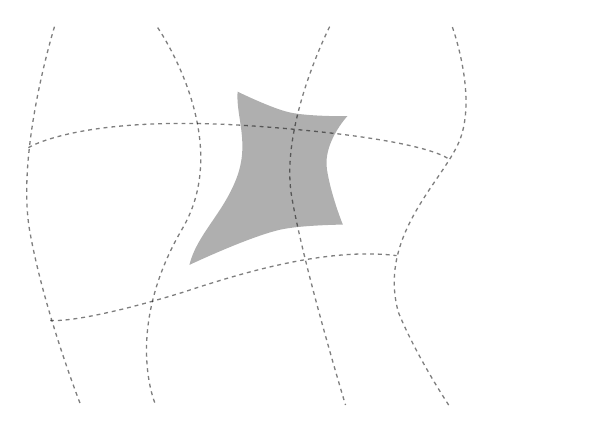
       }
      \caption{\textcolor{Reviewer1R1}{Lagrangian forward tracking of the cell $\Omega_l$ as a material volume, reconstruction of $\ForwardPhaseImage$ from $\alpha(\tend, \FlowMap{\tstart}{\tend}(\Omega_l))$ by means of \cref{eq:alphamat}, and Eulerian remapping of volume fractions by \cref{eq:eulerremap}, by intersecting $\ForwardPhaseImage$ with $\Cell_k$, based on the intersection stencil $\CellNeighborhood^+_{n+1}(\Cell_k)$.}}
    \label{fig:forwardb}
  \end{subfigure}
  \caption{Computing $\VolFracskend$ by two different forward-tracking methods.}
  \label{fig:forward-tracking}
\end{figure}
There are \textcolor{Reviewer1R1}{two types of forward tracking \GVOF{}s (\citep{Mosso1997,Dyadechko2005,Zhang2008,Chenadec2013}) and their main differences} are illustrated by \cref{fig:forward-tracking} . \textcolor{Reviewer1R1}{The first class of cell-based methods tracks only the phase-specific volume $\Cell_l^+(t^n)$ (\cref{fig:forwarda}), while the second class of methods also tracks the entire cell $\Cell_l$ (\cref{fig:forwardb}).} 

\textcolor{Reviewer1R1}{For the first class of methods, phase-specific volumes are tracked forward in time as $\FlowMap{\tstart}{\tend}(\Omega^+_l(\tstart))$, for $l \in K$, and distributed over the underlying Eulerian mesh $\Cells$ in the Eulerian remapping step to compute $\VolFracskend$.}\textcolor{Reviewer1}{The volume-conserving flow map conserves the phase-specific volume \textcolor{Reviewer1R1}{$|\Omega_l^+(\tstart)|$}}, so we have $|\FlowMap{\tstart}{\tend}\Cell_l^+(\tstart)| = |\Cell_l^+(\tstart)|$. In the general case, however, the exact flow map $\FlowMap{\tstart}{\tend}$ has to be approximated, which introduces spatial interpolation and temporal integration errors. These errors make it impossible to exactly satisfy the relation $|\FlowMap{\tstart}{\tend}\Cell_l^+(\tstart)| = |\Cell_l^+(\tstart)|$, so additional geometrical corrections must be applied to $\ForwardPhaseImage$ in order to enforce volume conservation. 

\textcolor{Reviewer1R1}{The \emph{Eulerian re-mapping} step is used to compute $\VolFracskend$ from the forward images of phase-specific volumes $\ForwardPhaseImages$ (cf. \cref{fig:forwarda}) as
\textcolor{Reviewer1}{
\begin{equation}
    \alpha_k(\tend) = \dfrac{1}{|\Omega_k|} \sum_{l \in \CellNeighborhood^+_{n+1}(\Cell_k)} | \Omega_k \cap \ForwardPhaseImage|, 
    \label{eq:eulerremap}
\end{equation}}\noindent where 
\begin{equation}
    \CellNeighborhood^+_{n+1}(\Cell_k) = \{ l \in K : \Cell_k \cap \ForwardPhaseImage \ne \emptyset \}
    \label{eq:stencilcplusn}
\end{equation}
is the intersection stencil of $\Cell_k$ in $\ForwardPhaseImages$. The part of the boundary $\partial\Cell_l^+(\tstart)$ that belongs to the interface, $\Cell_l \cap \Sigma(\tstart)$, is approximated, mostly linearly, by the geometrical VOF method (cf. \cref{sec:recon}).} A computationally efficient calculation of the sum in \cref{eq:eulerremap} in a discrete setting is described in \cref{sec:advect}.

The second class of forward tracking methods \textcolor{Reviewer1R1}{applies the space conservation law given by \cref{eq:spacecons} to the phase-specific volume $\Cell_l^+(\tstart)$ and, simultaneously, to the whole cell $\Cell_l$, given by \cref{eq:spaceconvplus} and shown in \cref{fig:forwardb}, which leads to
\begin{equation}
     |\ForwardPhaseImage| = |\Cell_l^+(\tstart)|\label{eq:phaseforwardvol} \\
\end{equation}
\textcolor{Reviewer1R1}{and }
\begin{equation}
    |\ForwardCellImage| = |\Cell_l|, \label{eq:cellforwardvol} 
\end{equation}
respectively, if the map $\FlowMap{}{}$ is volume-conserving.} Equations (\ref{eq:phaseforwardvol}) and (\ref{eq:cellforwardvol}) yield
\begin{equation}
    \alpha(\tend, \FlowMap{\tstart}{\tend}(\Omega_l))
    = \dfrac{|\ForwardPhaseImage|}{|\ForwardCellImage|} 
    = \dfrac{|\Cell_l^+(\tstart)|}{|\Cell_l|} 
    = \alpha(\tstart, \Cell_l)
    = \alpha_l(\tstart).
    \label{eq:alphamat}
\end{equation}
\Cref{eq:alphamat} states that the volume fraction \textcolor{Reviewer1R1}{in the forward cell image $\ForwardCellImage$} is the same as the volume fraction of $\Cell_l$. From \cref{eq:alphakdef}, just like for the previous forward tracking method, we know that 
\begin{equation}
    \alpha_l(\tend) = \dfrac{|\Cell_l \cap \Omega^+(\tend)|}{|\Cell_l|}.
    \label{eq:forwardalphaupdate}
\end{equation}
\textcolor{Reviewer1R1}{The phase $\Omega^+(\tend)$ is a disjoint decomposition of forward phase-specific volume images, i.e.\ 
\begin{equation}
    \Omega^+(\tend) = \bigcup_{l \in K} \ForwardPhaseImage.
   \label{eq:omegaplusunion}
\end{equation}}\textcolor{Reviewer1R1}{The final step in evaluating \cref{eq:forwardalphaupdate} is therefore the computation of $\ForwardPhaseImages$, as in the previous method. However, contrary to the first method, $\ForwardPhaseImages$ are not calculated using the flow map, as shown in \cref{fig:forwarda}. Instead, $\ForwardPhaseImages$ are \emph{approximated} on the forward image of the whole mesh, $\FlowMap{\tstart}{\tend}(\Cellsl)$, using volume fractions on the background Eulerian mesh $\Cellsl$, based on \cref{eq:alphamat}. Each reconstructed phase-specific volume $\ForwardPhaseImage$ inside the forward image of a cell $\FlowMap{\tstart}{\tend}(\Cell_l)$ generally overlaps with multiple cells from the background Eulerian mesh $\Cells$ (cf. \cref{fig:forwardb}), so the $\VolFracskend$ are computed using the Eulerian remapping step given by \cref{eq:eulerremap}. For example, $\Omega^+_l(\tend)$ is computed on the forward image of the mesh $\FlowMap{\tstart}{\tend}(\Cellsl)$ by the \ac{PLIC} interface reconstruction using $\VolFracskstart$ (cf. \cref{sec:recon}), and then used to compute $\VolFracskend$ by employing \cref{eq:eulerremap}. As before, indices $l$ and $k$ are necessary because the forward images of each phase-specific volume overlaps with more than one cell from the background Eulerian mesh.}

\subsubsection{Backward tracking}

\begin{figure}[h] 
    \centering
    \def\svgwidth{0.5\columnwidth}
    {\footnotesize
        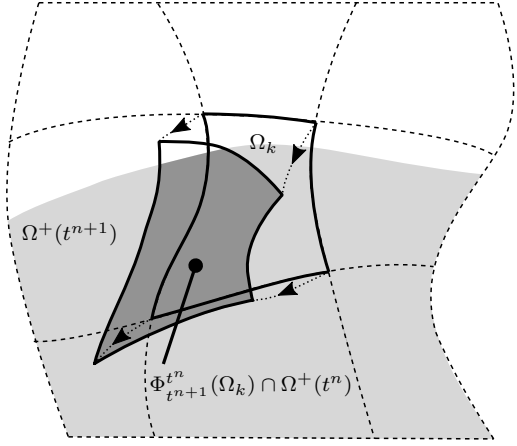
    }
    \caption{Lagragian backward tracking and Eulerian re-mapping.}
    \label{fig:backward}
\end{figure}
\textcolor{Reviewer1R1}{\Cref{eq:flowmapinv} is used to express the main idea of backward tracking based on \cref{eq:alphakdef}, namely 
\begin{equation}
    \VolFrac_k(\tend) = 
    \dfrac{|\Cell_k \cap \Omega^+(\tend)|}{|\Cell_k|}  
    = \dfrac{|\FlowMap{\tstart}{\tend}[\FlowMap{\tend}{\tstart}(\Cell_k) \cap \Omega^+(\tstart)]|}{|\Cell_k|},
    \label{eq:backidea}
\end{equation}
which states that the phase specific volume $\Cell_k^+(\tend)$ can be computed as an intersection of the \emph{pre-image} of the cell $\FlowMap{\tend}{\tstart}(\Cell_k)$ (shown schematically in \cref{fig:backward}) and $\Omega^+(\tstart)$. \textcolor{Reviewer1R1}{The backward tracking method} therefore considers the cell in the Eulerian mesh $\Cell_k$ as a forward image of some pre-image of the cell $\Cell_k$. Because an exact $\Omega^+(t)$ is not available, this still leaves the question of \textcolor{Reviewer1R1}{approximating} $\FlowMap{\tend}{\tstart}(\Cell_k) \cap \Omega^+(\tstart)$ . The sub-domain $\Omega^+(\tstart)$ is given in each interface cell $\Cell_k$ as $\Cell_k^+(\tstart)$, which leads to 
\begin{equation}
    \FlowMap{\tend}{\tstart}(\Cell_k) \cap \Omega^+(\tstart) 
    = \FlowMap{\tend}{\tstart}(\Cell_k) \cap \bigcup_{l \in \CellNeighborhood^+_n(\FlowMap{\tend}{\tstart}(\Cell_k))} \Cell_l^+(\tstart),
    \label{eq:backunion}
\end{equation}
where 
\begin{equation}
    \CellNeighborhood^+_n(\FlowMap{\tend}{\tstart}(\Cell_k)) = \{\ l \in K : \FlowMap{\tend}{\tstart}(\Cell_k) \cap \Cell_l^+(\tstart) \ne \emptyset \}
\end{equation}
is the intersection stencil of the pre-image $\FlowMap{\tend}{\tstart}(\Cell_k)$ in $\{ \Cell_l^+(\tstart) \}_{l \in K}$. \Cref{eq:backunion} inserted into \cref{eq:backidea} leads to
\textcolor{Reviewer1R1}{\begin{equation}
    \begin{aligned}
        \VolFrac_k(\tend) & = \dfrac{|\FlowMap{\tstart}{\tend}[\FlowMap{\tend}{\tstart}(\Cell_k) \cap \bigcup_{l \in \CellNeighborhood^+_n(\FlowMap{\tend}{\tstart}(\Cell_k))} \Cell_l^+(\tstart)]|}{|\Cell_k|} \\
         & = \dfrac{|\bigcup_{l \in \CellNeighborhood^+_n(\FlowMap{\tend}{\tstart}(\Cell_k))}\FlowMap{\tstart}{\tend}[\FlowMap{\tend}{\tstart}(\Cell_k) \cap \Cell_l^+(\tstart))]|}{|\Cell_k|} \\
    \end{aligned}
    \label{eq:backprep}
\end{equation}
The intersections $\FlowMap{\tend}{\tstart}(\Cell_k) \cap \Cell_l^+(\tstart)$ are denoted with the darker gray shade in \cref{fig:backward}, and their union is the phase-specific volume of the phase $\Omega^+(t^n)$ in the pre-image of the cell $\Cell_k$. Since the volumes $\FlowMap{\tend}{\tstart}(\Cell_k) \cap \Cell_l^+(\tstart)$ form a disjoint decomposition of the phase specific volume in the pre-image of $\Cell_k$, the union of their forward images is also a union of disjoint sets, which leads to  
\begin{equation}
    \VolFrac_k(\tend)  = \dfrac{1}{|\Cell_k|}\sum_{l \in \CellNeighborhood^+_n(\FlowMap{\tend}{\tstart}(\Cell_k))} |\FlowMap{\tend}{\tstart}[\FlowMap{\tend}{\tstart}(\Cell_k) \cap \Cell_l^+(\tstart)]|.
    \label{eq:backsum}
\end{equation}
The flow map $\FlowMap{\tstart}{\tend}$ is volume-conserving, so $|\FlowMap{\tend}{\tstart}[\FlowMap{\tend}{\tstart}(\Cell_k) \cap \Cell_l^+(\tstart)]| = |\FlowMap{\tend}{\tstart}(\Cell_k) \cap \Cell_l^+(\tstart)|$, which leads to the final equation for $\VolFrac_k(\tend)$, namely
\begin{equation}
    \VolFrac_k(\tend)  = \dfrac{1}{|\Cell_k|}\sum_{l \in \CellNeighborhood^+_n(\FlowMap{\tend}{\tstart}(\Cell_k))} |\FlowMap{\tend}{\tstart}(\Cell_k) \cap \Cell_l^+(\tstart)|.
    \label{eq:backfinal}
\end{equation}}Interestingly, once the phase specific volume in the pre-image $\FlowMap{\tend}{\tstart}(\Cell_k)$ is computed, there is no need to map this volume forward into $\Cell_k$ at $\tend$: its magnitude is sufficient to compute $\VolFrac_k(\tend)$. Since the phase specific volume in the pre-image of $\Cell_k$ is a union of intersections, the magnitude of the phase-specific volume is the sum of the magnitudes of these intersections. The backward tracking method therefore performs Eulerian remapping of $\Omega^+(\tstart)$ on the pre-image $\FlowMap{\tstart}{\tend}\Cell_k$ and uses the magnitude of the phase-specific volume in the pre-image to compute $\VolFrac_k(\tend)$. In the PLIC geometrical VOF method, however, velocity interpolation and temporal integration errors are sources of volume conservation errors in $\FlowMap{}{}$, while the piecewise-linear approximation of $\Cell_l^+(\tstart)$ limits the spatial convergence to second order.}

%% file: figures/flux-based.pdf_tex
\begingroup%
  \makeatletter%
  \providecommand\color[2][]{%
    \errmessage{(Inkscape) Color is used for the text in Inkscape, but the package 'color.sty' is not loaded}%
    \renewcommand\color[2][]{}%
  }%
  \providecommand\transparent[1]{%
    \errmessage{(Inkscape) Transparency is used (non-zero) for the text in Inkscape, but the package 'transparent.sty' is not loaded}%
    \renewcommand\transparent[1]{}%
  }%
  \providecommand\rotatebox[2]{#2}%
  \newcommand*\fsize{\dimexpr\f@size pt\relax}%
  \newcommand*\lineheight[1]{\fontsize{\fsize}{#1\fsize}\selectfont}%
  \ifx\svgwidth\undefined%
    \setlength{\unitlength}{141.8952713bp}%
    \ifx\svgscale\undefined%
      \relax%
    \else%
      \setlength{\unitlength}{\unitlength * \real{\svgscale}}%
    \fi%
  \else%
    \setlength{\unitlength}{\svgwidth}%
  \fi%
  \global\let\svgwidth\undefined%
  \global\let\svgscale\undefined%
  \makeatother%
  \begin{picture}(1,0.87627065)%
    \lineheight{1}%
    \setlength\tabcolsep{0pt}%
    \put(0,0){\includegraphics[width=\unitlength,page=1]{flux-based.pdf}}%
    \put(0.69633658,0.4991632){\makebox(0,0)[lt]{\lineheight{1.25}\smash{\begin{tabular}[t]{l}$S_f$\end{tabular}}}}%
    \put(0,0){\includegraphics[width=\unitlength,page=2]{flux-based.pdf}}%
    \put(0.46599365,0.20065734){\rotatebox{0.16620669}{\makebox(0,0)[lt]{\lineheight{1.25}\smash{\begin{tabular}[t]{l}$\PhaseFluxVolumeContrib = \FluxVolume  \cap \Omega^+(t^n)$\end{tabular}}}}}%
    \put(0.41941172,0.56142208){\makebox(0,0)[lt]{\lineheight{1.25}\smash{\begin{tabular}[t]{l}$\Omega_k$\end{tabular}}}}%
    \put(0,0){\includegraphics[width=\unitlength,page=3]{flux-based.pdf}}%
    \put(0.53010195,0.72783605){\makebox(0,0)[lt]{\lineheight{1.25}\smash{\begin{tabular}[t]{l}$\partial\Omega_k$\end{tabular}}}}%
    \put(0.14912677,0.44777325){\makebox(0,0)[lt]{\lineheight{1.25}\smash{\begin{tabular}[t]{l}$\Omega^+(\tstart)$\end{tabular}}}}%
    \put(0,0){\includegraphics[width=\unitlength,page=4]{flux-based.pdf}}%
  \end{picture}%
\endgroup%

%% file: figures/mathmodel-forward-tracking-volume.pdf_tex
\begingroup%
  \makeatletter%
  \providecommand\color[2][]{%
    \errmessage{(Inkscape) Color is used for the text in Inkscape, but the package 'color.sty' is not loaded}%
    \renewcommand\color[2][]{}%
  }%
  \providecommand\transparent[1]{%
    \errmessage{(Inkscape) Transparency is used (non-zero) for the text in Inkscape, but the package 'transparent.sty' is not loaded}%
    \renewcommand\transparent[1]{}%
  }%
  \providecommand\rotatebox[2]{#2}%
  \newcommand*\fsize{\dimexpr\f@size pt\relax}%
  \newcommand*\lineheight[1]{\fontsize{\fsize}{#1\fsize}\selectfont}%
  \ifx\svgwidth\undefined%
    \setlength{\unitlength}{176.00115967bp}%
    \ifx\svgscale\undefined%
      \relax%
    \else%
      \setlength{\unitlength}{\unitlength * \real{\svgscale}}%
    \fi%
  \else%
    \setlength{\unitlength}{\svgwidth}%
  \fi%
  \global\let\svgwidth\undefined%
  \global\let\svgscale\undefined%
  \makeatother%
  \begin{picture}(1,0.70646501)%
    \lineheight{1}%
    \setlength\tabcolsep{0pt}%
    \put(0,0){\includegraphics[width=\unitlength,page=1]{mathmodel-forward-tracking-volume.pdf}}%
    \put(0.29297806,0.14061775){\color[rgb]{0,0,0}\makebox(0,0)[lt]{\lineheight{0}\smash{\begin{tabular}[t]{l}$\Omega_l^+(t^n)$\end{tabular}}}}%
    \put(0,0){\includegraphics[width=\unitlength,page=2]{mathmodel-forward-tracking-volume.pdf}}%
    \put(0.51148717,0.55286553){\color[rgb]{0,0,0}\makebox(0,0)[lt]{\lineheight{0}\smash{\begin{tabular}[t]{l}$\ForwardPhaseImage$\end{tabular}}}}%
    \put(0,0){\includegraphics[width=\unitlength,page=3]{mathmodel-forward-tracking-volume.pdf}}%
    \put(0.55515931,0.19169677){\color[rgb]{0,0,0}\makebox(0,0)[lt]{\lineheight{0}\smash{\begin{tabular}[t]{l}$\Omega_k \cap \ForwardPhaseImage$\end{tabular}}}}%
    \put(0.63796485,0.41359856){\color[rgb]{0,0,0}\makebox(0,0)[lt]{\lineheight{0}\smash{\begin{tabular}[t]{l}$\Omega_k$\end{tabular}}}}%
    \put(0,0){\includegraphics[width=\unitlength,page=4]{mathmodel-forward-tracking-volume.pdf}}%
    \put(0.34669384,0.4423288){\color[rgb]{0,0,0}\makebox(0,0)[lt]{\lineheight{0}\smash{\begin{tabular}[t]{l}$\Omega_l$\end{tabular}}}}%
    \put(0,0){\includegraphics[width=\unitlength,page=5]{mathmodel-forward-tracking-volume.pdf}}%
  \end{picture}%
\endgroup%

%% file: figures/mathmodel-forward-tracking-recon.pdf_tex
\begingroup%
  \makeatletter%
  \providecommand\color[2][]{%
    \errmessage{(Inkscape) Color is used for the text in Inkscape, but the package 'color.sty' is not loaded}%
    \renewcommand\color[2][]{}%
  }%
  \providecommand\transparent[1]{%
    \errmessage{(Inkscape) Transparency is used (non-zero) for the text in Inkscape, but the package 'transparent.sty' is not loaded}%
    \renewcommand\transparent[1]{}%
  }%
  \providecommand\rotatebox[2]{#2}%
  \newcommand*\fsize{\dimexpr\f@size pt\relax}%
  \newcommand*\lineheight[1]{\fontsize{\fsize}{#1\fsize}\selectfont}%
  \ifx\svgwidth\undefined%
    \setlength{\unitlength}{176.00115967bp}%
    \ifx\svgscale\undefined%
      \relax%
    \else%
      \setlength{\unitlength}{\unitlength * \real{\svgscale}}%
    \fi%
  \else%
    \setlength{\unitlength}{\svgwidth}%
  \fi%
  \global\let\svgwidth\undefined%
  \global\let\svgscale\undefined%
  \makeatother%
  \begin{picture}(1,0.70646501)%
    \lineheight{1}%
    \setlength\tabcolsep{0pt}%
    \put(0,0){\includegraphics[width=\unitlength,page=1]{mathmodel-forward-tracking-recon.pdf}}%
    \put(0.63779576,0.41513903){\color[rgb]{0,0,0}\makebox(0,0)[lt]{\lineheight{0}\smash{\begin{tabular}[t]{l}$\Omega_k$\end{tabular}}}}%
    \put(0,0){\includegraphics[width=\unitlength,page=2]{mathmodel-forward-tracking-recon.pdf}}%
    \put(0.55034909,0.59041685){\color[rgb]{0,0,0}\makebox(0,0)[lt]{\lineheight{0}\smash{\begin{tabular}[t]{l}$\ForwardCellImage$\end{tabular}}}}%
    \put(0,0){\includegraphics[width=\unitlength,page=3]{mathmodel-forward-tracking-recon.pdf}}%
    \put(1.25393511,0.44489772){\color[rgb]{0,0,0}\makebox(0,0)[lt]{\begin{minipage}{0.01741571\unitlength}\raggedright \end{minipage}}}%
    \put(1.25175811,0.45687106){\color[rgb]{0,0,0}\makebox(0,0)[lt]{\begin{minipage}{0.14259161\unitlength}\raggedright \end{minipage}}}%
    \put(1.25005732,0.52961614){\color[rgb]{0,0,0}\makebox(0,0)[lt]{\lineheight{1.25}\smash{\begin{tabular}[t]{l} \end{tabular}}}}%
    \put(0,0){\includegraphics[width=\unitlength,page=4]{mathmodel-forward-tracking-recon.pdf}}%
    \put(0.34348866,0.44525613){\color[rgb]{0,0,0}\makebox(0,0)[lt]{\lineheight{0}\smash{\begin{tabular}[t]{l}$\Omega_l$\end{tabular}}}}%
    \put(0.1130893,0.37964427){\makebox(0,0)[lt]{\lineheight{1.25}\smash{\begin{tabular}[t]{l}$\Omega^+_l(t^n)$\end{tabular}}}}%
    \put(0.08103762,0.6109727){\makebox(0,0)[lt]{\lineheight{1.25}\smash{\begin{tabular}[t]{l}$\ForwardPhaseImage$\end{tabular}}}}%
    \put(0,0){\includegraphics[width=\unitlength,page=5]{mathmodel-forward-tracking-recon.pdf}}%
  \end{picture}%
\endgroup%

%% file: figures/mathmodel-backward-trackingb.pdf_tex
\begingroup%
  \makeatletter%
  \providecommand\color[2][]{%
    \errmessage{(Inkscape) Color is used for the text in Inkscape, but the package 'color.sty' is not loaded}%
    \renewcommand\color[2][]{}%
  }%
  \providecommand\transparent[1]{%
    \errmessage{(Inkscape) Transparency is used (non-zero) for the text in Inkscape, but the package 'transparent.sty' is not loaded}%
    \renewcommand\transparent[1]{}%
  }%
  \providecommand\rotatebox[2]{#2}%
  \newcommand*\fsize{\dimexpr\f@size pt\relax}%
  \newcommand*\lineheight[1]{\fontsize{\fsize}{#1\fsize}\selectfont}%
  \ifx\svgwidth\undefined%
    \setlength{\unitlength}{141.90124512bp}%
    \ifx\svgscale\undefined%
      \relax%
    \else%
      \setlength{\unitlength}{\unitlength * \real{\svgscale}}%
    \fi%
  \else%
    \setlength{\unitlength}{\svgwidth}%
  \fi%
  \global\let\svgwidth\undefined%
  \global\let\svgscale\undefined%
  \makeatother%
  \begin{picture}(1,0.87623376)%
    \lineheight{1}%
    \setlength\tabcolsep{0pt}%
    \put(0,0){\includegraphics[width=\unitlength,page=1]{mathmodel-backward-trackingb.pdf}}%
    \put(1.55526494,0.55180993){\color[rgb]{0,0,0}\makebox(0,0)[lt]{\begin{minipage}{0.02160084\unitlength}\raggedright \end{minipage}}}%
    \put(1.55256479,0.56666054){\color[rgb]{0,0,0}\makebox(0,0)[lt]{\begin{minipage}{0.17685742\unitlength}\raggedright \end{minipage}}}%
    \put(0.48137235,0.56452761){\color[rgb]{0,0,0}\makebox(0,0)[lt]{\lineheight{0}\smash{\begin{tabular}[t]{l}$\Omega_k$\end{tabular}}}}%
    \put(0,0){\includegraphics[width=\unitlength,page=2]{mathmodel-backward-trackingb.pdf}}%
    \put(0.08049108,0.40382992){\color[rgb]{0,0,0}\makebox(0,0)[lt]{\lineheight{0}\smash{\begin{tabular}[t]{l}$\Omega^+(t^{n+1})$\end{tabular}}}}%
    \put(-0.08226906,0.85180118){\color[rgb]{0,0,0}\makebox(0,0)[lt]{\begin{minipage}{1.24795825\unitlength}\raggedright \end{minipage}}}%
    \put(1.2606151,0.77459484){\color[rgb]{0,0,0}\makebox(0,0)[lt]{\begin{minipage}{0.84041012\unitlength}\raggedright \end{minipage}}}%
    \put(0.30515321,0.13881431){\color[rgb]{0,0,0}\makebox(0,0)[lt]{\lineheight{0}\smash{\begin{tabular}[t]{l}$\FlowMap{\tend}{\tstart}(\Omega_k)\cap\Omega^+(\tstart)$\end{tabular}}}}%
    \put(0,0){\includegraphics[width=\unitlength,page=3]{mathmodel-backward-trackingb.pdf}}%
  \end{picture}%
\endgroup%

%% file: sections/interface-reconstruction.tex
\section{Interface reconstruction}
\label{sec:recon}

The first implementation of the \GVOF{} \citep{DeBar1974} employed \textcolor{Reviewer1R1}{the \acf{PLIC}, that approximates the interface linearly in each multi-material cell}. Later developments such as \citep{Hirt1981} have simplified the interface approximation to \emph{piecewise constant} \ac{SLIC}. A detailed overview of the earliest publications on this topic can be found in \citep[page 6, table 1]{Rider1998}, together with a table summarizing most important contributions. A review of reconstruction algorithms is also available in \citep{Pilliod2004,Aulisa2007}. A comparison between the order of convergence and relative computational costs for more recent developments is shown in \cref{tab:plic-algorithms}.

\ac{PLIC} algorithms have prevailed over \ac{SLIC} algorithms because of their many advantages. A more accurate interface approximation is provided by \ac{PLIC} algorithms than by \ac{SLIC} algorithms, resulting in \textcolor{Reviewer1}{second-order convergent interface advection} on \textcolor{Reviewer1R1}{structured meshes} \citep{Renardy2002,Francois2006,Popinet2009,Owkes2015}. The \ac{PLIC} algorithms enable efficient computations and increased accuracy over a wide span of spatial scales when they support local dynamic \ac{AMR} \citep{Popinet2003,Ahn2009,Agbaglah2011}. The piecewise planar interface approximation supports the numerical simulation of the transport of insoluble surfactants on the fluid interface \citep{James2004}, which is not possible with the piecewise constant interface approximation given by the \ac{SLIC} algorithm. The \ac{SLIC} algorithms generate a substantial amount of \emph{jetsam} (\emph{flotsam}) \citep{Aulisa2007}. Jetsam (flotsam) are elements of the interface that are \emph{artificially separated} and transported with the flow velocity. \citet{Noh1976} have introduced "jetsam" (\emph{jettisoned goods}) and "flotsam" (\emph{floating wreckage}) for artificially separated interface elements, according to \citet{Kothe1996}. \ac{PLIC} algorithms based on error minimization can be directly applied to unstructured meshes \citep{Mosso1997, Mosso1996, Dyadechko2005, Liovic2006, Mosso2008}, since they rely on linear traversal of the surrounding cells, \textcolor{Reviewer1R1}{without accessing cells in any specific direction}. Reconstructing a piecewise linear interface while strictly satisfying conservation of volume \textcolor{Reviewer1R1}{requires accurate volume truncation and interface positioning algorithms}. The \ac{PLIC} algorithms do have one \textcolor{Reviewer1}{general disadvantage}: so-called artificial numerical surface tension. It is introduced by the interface reconstruction algorithm in the form of rounding of sharp corners, that occurs during repeated reconstructions of the interface \citep{Zhang2008,Ahn2009}. The advantages of \ac{PLIC} algorithms make them still the prevailing choice in $3D$, compared to still more complex and computationally expensive higher-order interface approximations. 

\begin{table}[h]
  \myfloatalign
  \begin{tabularx}{\textwidth}{Xlc} 
    \toprule
    \tableheadline{Algorithm} & \tableheadline{Convergence} & \tableheadline{Cost} \\ \midrule
    \citet{Youngs1982} & 1.0-1.8,\citep{Aulisa2007} &  1 \\
    Mosso-Swartz, \citet{Mosso1996} & 2.0 & 3-4 \\
    \ac{LVIRA}, \citet{Pilliod2004} & 2.0 & 9,\citep{Ahn2007}\\
    \ac{ELVIRA}, \citet{Pilliod2004} & 1.9-2.2 & 900,\citep{Lopez2008}\\
    \ac{CVTNA}, \citet{Liovic2006} & 2.0 & 50 \\
    \ac{CLCIR},\ac{CBIR}, \citet{Lopez2008} & 2.0-2.11 & 3 \\
    \ac{CIAM}, \citet{Scardovelli2003} & 1.0-2.28 & 1 \\
    \ac{LSF}, \citet{Scardovelli2003,Aulisa2007} & 2.0 & 1.5 \\
    \ac{MoF}, \citet{Dyadechko2005},\citep{Dyadechko2008} & 2.0 & 7,\citep{Ahn2007}\\
    \ac{PIR}, \citet{Mosso2008} & 2.0 & 10 \\
    \midrule
    \bottomrule
  \end{tabularx}
  \caption[PLIC reconstruction algorithms]{PLIC reconstruction algorithms. The convergence order of the reconstruction errors are reported for circular and spherical interfaces. For \ac{CIAM} and \ac{LSF} the error convergence order is reported for an ellipse by the authors. Additional citations are listed for those algorithms whose relative costs are not reported in the original publications.}
  \label{tab:plic-algorithms}
\end{table}

The Youngs' gradient-based algorithm is taken as the reference for the relative computational cost in all the referenced publications. It is important to note that the algorithm cost does depend on the implementation. \textcolor{Reviewer1R1}{However, the costs in \citep{Ahn2007} are reported on the same software platform, which makes them more objective.} The relative costs reported in \citep{Pilliod2004,Scardovelli2003,Liovic2006,Lopez2008} have been reported on structured Cartesian meshes. 

From \cref{tab:plic-algorithms}, \textcolor{Reviewer1R1}{it follows that only a sub-set of reconstruction algorithms can be used on unstructured meshes.} The main constraint enforced by the algorithms on unstructured meshes is the inability to exercise access to mesh elements in a specific direction, e.g., accessing different face centers by changing their $y$ coordinate. Algorithms that rely on more than the first level of addressing experience a substantial increase in computational complexity on unstructured meshes also in terms of algorithm parallelization using the domain decomposition and message passing parallel programming model. Such computational complexity restrictions should be considered when choosing a \ac{PLIC} reconstruction algorithm for unstructured meshes. Consequently, if the algorithm's relative cost reported in \cref{tab:plic-algorithms} is already high on structured meshes, it can be disregarded as a candidate for unstructured meshes.  

To reconstruct the \textcolor{Reviewer1}{interface in each multi-material cell $\Cell_k$, the interface normal and position vectors ($\PlicNormal,\PlicPosition$) are computed}. The aim of each reconstruction algorithm is to accurately compute those two parameters. At first, $\PlicNormal$ is approximated by the \emph{interface orientation algorithm}. Then, the interface plane is positioned by the \emph{interface positioning algorithm} that calculates $\PlicPosition$. \textcolor{Reviewer1}{In order to achieve second-order convergence in the $L_1$ error norm of the volume fraction field for the interface advection, second-order convergence of the interface reconstruction must be ensured. Error convergence of the reconstruction is verified either by some error norm of the difference between \textcolor{Reviewer1R1}{the reconstructed and the exact interface normal}, or by some error norm of the volume of symmetric difference between the volume bounded by the reconstructed interface, and the volume bounded by the exact interface.} In the following sub-sections, a sub-set of the \ac{PLIC} reconstruction algorithms from \cref{tab:plic-algorithms} are outlined and categorized into contributions to the interface orientation and positioning.

\subsection{Interface orientation}

\textcolor{Reviewer1R1}{All interface orientation algorithms rely on the volume fraction field $\VolFrac$ to approximate $\PlicNormal$.} Some second-order convergent algorithms rely \textcolor{Reviewer1R1}{exclusively} on the $\VolFrac$ field, while others employ additional geometrical calculations to \textcolor{Reviewer1R1}{increase convergence}. 

\subsubsection{Youngs' algorithm} 
\label{sec:gvof:youngs}

\textcolor{Reviewer1R1}{This algorithm, originally developed by \citet{Youngs1982}, defines $\PlicNormal$ as}
\textcolor{Reviewer1}{
\begin{equation}
\PlicNormal = -\dfrac{\GradNum \VolFrac}{\|\GradNum \VolFrac\|}.
  \label{eqn:youngs}
\end{equation}
The discrete gradient $\GradNum$ is crucial for maintaining accuracy \citep{Cerne2002}.} On unstructured meshes, \textcolor{Reviewer1R1}{the gradient} $\GradNum$ is usually approximated using the unstructured Finite Volume discretization. However, using the \ac{FVM} for gradient operator discretization is by no means a requirement for the \GVOF{} - other discretization methods can be used as well. More details on the gradient operator discretization practice on unstructured meshes using \UFVM{} can be found in \citep[ch. 2]{Moukalled2016}. 

\citet{Mavriplis2003} has concluded that a wider stencil \ac{IDWLSG} approximation delivers accurate results on equidistant hexahedral unstructured meshes. A wider stencil gradient results in a more accurate aerodynamic drag estimation on unstructured meshes and the accuracy and convergence of the \LSGRAD{} deteriorates strongly on unstructured tetrahedral meshes.

\citet{Aulisa2007} have proposed a gradient calculation that uses finite differences to compute the components of the volume fraction gradient at cell corners from \textcolor{Reviewer1R1}{cell-centered} values obtained by averaging finite difference operations. This gradient approximation cannot be applied without modification on unstructured meshes as it relies on \textcolor{Reviewer1R1}{accessing cells in a specific direction}. 

Similar to \citet{Mavriplis2003}, \citet{Ahn2007} have proposed a \acf{LS} minimization to estimate $\GradNum\VolFrac$ on unstructured meshes. However, this minimization differs from the \ac{IDWLSG} proposed by \citet{Mavriplis2003} in the fact that their \ac{LLSG} does not rely on inverse distance weighting. Instead, they have applied a second-order linear approximation of the volume fraction field $\VolFrac$ using the Taylor series expansion \textcolor{Reviewer1}{from the centroid $\x_k$ of the cell $\Cell_k$: 
\begin{equation}
    \VolFrac(\x) \dot{=} \VolFrac_k + \GradNum\VolFrac \cdot(\x - \x_k). 
  \label{eqn:volfrac-linear}
\end{equation}}\textcolor{Reviewer1}{A volume fraction error for the cells in the intersection stencil $\CellNeighborhood(\Omega_k)$ given by \cref{eq:cellstencilv} is then defined as
\begin{equation}
    \Error_{LLSG}(\Cell_k) = \sum_{l \in \CellNeighborhood(\Omega_k)} \left(\VolFrac_{\Cell_l} - \dfrac{\int_{\Cell_l} \VolFrac(\x) dV }{|\Cell_l|} \right)^2, 
  \label{eqn:volfrac-error-linear}
\end{equation}
using \cref{eqn:volfrac-linear} for $\VolFrac(\x)$.} The \ac{LS} minimization of $\Error_{LLSG}$ results in a $3\times3$ linear algebraic system that is solved for the $3$ components of $\GradNum\VolFrac$ in each cell $k$. The \ac{LLSG} gradient estimation was used for the $\VolFrac$ field by \citet{Garimella2007} for their \ac{ALE} method. On polyhedral unstructured meshes, the explicit construction of a wider gradient stencil as proposed by \citet{Mavriplis2003} is redundant \textcolor{Reviewer1}{because a polyhedral cell is connected to all adjacent cells by its faces.}

The only difference between the \ac{IDWLSG} and \ac{LLSG} algorithm is the introduction of inverse distance weights in the minimized error functional
\begin{equation}
    \Error_{IDWLSG}(\Cell_k) = \sum_{l \in \CellNeighborhood(\Omega_k)} \left[ w_l \left(\VolFrac_{\Cell_l} - \dfrac{\int_{\Cell_l} \VolFracLinear(\x) dV }{|\Cell_l|} \right)\right]^2,
\end{equation}
where $w_l$ is the inversed distance weight \textcolor{Reviewer1R1}{given by}
\begin{equation}
    w_l = \dfrac{\frac{1}{\|\x_k - \x_l|^p}}{\sum_{\tilde{l} \in \CellNeighborhood(\Cell_k)} \frac{1}{\|\x_k - \x_{\tilde{l}}|^p}}.
\end{equation}
The weight exponent is set to $p=1$, \textcolor{Reviewer1R1}{so adjacent cells have} the same influence on the gradient approximation. 

\citet{Correa2011} compare different gradient operator approximations on unstructured meshes in detail. Their research is aimed at accurate volume rendering in the field of \ac{CG}. Nevertheless, their findings can be directly used for the gradient approximation on unstructured meshes in order to obtain a reasonably accurate initial \ac{PLIC} interface orientation. The following implications made by \citet{Correa2011} should be taken into consideration: 

\begin{enumerate}
  \item Inversed distance based gradient approximations are generally more accurate, especially on unstructured meshes with non-equidistant cells.
  \item Inversed distance based methods are more cost effective compared to regression based methods. 
  \item An increase in the discretization stencil size is important for improving the absolute accuracy of the approximated gradient on hexahedral meshes. 
\end{enumerate}


\subsubsection{Mosso-Swartz algorithm} 

The derivation and numerical analysis of the \ac{MS} algorithm was done by \citet{Swartz1989} and the algorithmic formulation for unstructured meshes was done by \citet{Mosso1996}. For a cell $\Cell_k$, \textcolor{Reviewer1}{the initial $\PlicNormal$ is computed using \cref{eqn:youngs}}. An interface polygon is defined as the intersection between the cell $\Omega_k$ and the \ac{PLIC} plane $(\PlicPosition,\PlicNormal)$, i.e.
\begin{equation}
  \PlicPolygon = \{ \x \in \Omega_k : (\x - \PlicPosition) \cdot \PlicNormal = 0 \}.
  \label{eqn:plic-polygon}
\end{equation}
\textcolor{Reviewer1R1}{Generally, $\{\Omega_k\}_{k \in K}$ are volumes bounded by polygons, so the PLIC polygon $\PlicPolygon$ is a set of intersection points between the polygonal boundary of $\Omega_k$ and the plane $(\PlicPosition,\PlicNormal)$. The centroid of the interface polygon is, therefore, given as}
\begin{equation}
  \PlicPolygonCentroid = \dfrac{1}{|\PlicPolygon|}\sum_{q=1}^{|\PlicPolygon|} \x_q.
  \label{eqn:plic-polygon-centroid}
\end{equation}
\textcolor{Reviewer1}{We define at this point the \emph{interface-cell stencil} of a volume $V$ in the mesh $\{\Omega_l\}_{l \in K}$ as 
\textcolor{Reviewer1R1}{\begin{equation}
    \CellNeighborhood^\alpha(V) = \{l \in K : V \cap \Omega_l \ne \emptyset \text{ and } 0 < \alpha_l < 1 \}, 
    \label{eq:cellstencilint}
\end{equation}}i.e.\ the set of labels of all cells $\Omega_l$ that have a non-empty intersection with the volume $V$ and that also have a non-empty intersection with the interface. Using the interface-cell stencil given by \cref{eq:cellstencilint}, estimated interface normal vectors are computed for each cell $\Omega_k$ from the centroids of interface polygons in the interface-cell stencil $\CellNeighborhood^\alpha(\Omega_k)$, as  
\begin{equation}
    \PlicPolygonsNormals = \{ \PlicNormalEstimateOther \Such \PlicNormalEstimateOther \cdot (\PlicPolygonCentroid - \PlicPolygonCentroidOther) = 0, \quad l \in C^\alpha(\Omega_k), l \ne k \}, 
  \label{eqn:plic-polygons-normals}
\end{equation}
To satisfy the orthogonality condition in \cref{eqn:plic-polygons-normals}, each estimated normal $\PlicNormalEstimateOther$ is computed as the vector $\PlicPolygonCentroidOther - \PlicPolygonCentroid$, rotated $90^\circ$ in the positive direction around the axis $(\PlicPolygonCentroidOther - \PlicPolygonCentroid) \times \PlicNormal$. The modified interface normal $\PlicNormalCorrected$ is obtained iteratively by a least-squares minimization 
\begin{equation}
    \EpsilonMosso = \sum_{l \in \CellNeighborhood(\Omega_k)} (\PlicNormalCorrected - \PlicNormalEstimateOther)^2 \rightarrow min.
  \label{eqn:plic-normal-diff-error}
\end{equation}
The initial iteration starts with $\PlicNormalCorrected = \PlicNormal$ as defined for the Youngs algorithm by \cref{eqn:youngs}. Once the new interface normal vector $\PlicNormalCorrected$ is obtained, the interface is reconstructed, resulting in a new set of interface polygons $\PlicPolygon$. To achieve second-order convergence, the steps given by equations \ref{eqn:plic-polygon} to \ref{eqn:plic-normal-diff-error} are repeated four times ($m=1,2,3,4$). \citet{Mosso1996} do not provide the motivation \textcolor{Reviewer1R1}{for} choosing 4 iterations. This has also been discussed by \citet[Table 3.]{Hernandez2008} for the simplified CLCIR method (LLCIR): a single iteration of the Mosso algorithm increases the convergence order on coarser meshes, but is not sufficient to maintain second-order convergence on fine meshes.} \textcolor{Reviewer1R1}{\citet{Dyadechko2005} propose the arithmetic average}
\textcolor{Reviewer1}{
\begin{equation}
    \PlicNormalCorrected = \dfrac{\sum_{l \in \CellNeighborhood^\alpha(\Omega_k)} \PlicNormalEstimateOther}{|\CellNeighborhood^\alpha(\Omega_k)|},
\end{equation}}as an alternative way to compute the modified normal, which reduces the computational effort introduced by the four outer iterations, the artificial smoothing of the interface as well as the required mesh resolution. \citet{Dyadechko2005} name this modification of the Mosso-Swartz algorithm as the \emph{Swartz} algorithm. 

\subsubsection{Conservative level contour interface reconstruction} 

The \ac{CLCIR} family of algorithms was originally developed by \citet{Lopez2008}. Like the modification of the Swartz-Mosso algorithm developed by \citet{Dyadechko2005}, the \ac{CLCIR} algorithm relies on estimating the interface normal by performing a local average of the normal vectors from the surrounding interface polygons. The difference between the \ac{CLCIR} and Swartz-Mosso algorithm variants lies in the way the estimated normal vectors $\PlicNormalEstimateOther$ are computed. \ac{CLCIR} algorithms rely on an iso-contour reconstruction to compute the estimated normal vectors. \textcolor{Reviewer1}{To triangulate the iso-contour, a volume-weighted average of volume fractions from surrounding cells is computed at each cell corner-point $\x_p$ as  
\begin{equation} 
    \VolFrac(\x_p) = \PointVolumeFraction = \dfrac{\sum_{l \in \CellNeighborhood(\x_p)}\VolFrac_{\CellOther}\VolumeOther}{\sum_{\tilde{l} \in \CellNeighborhood(\x_p)} |\Omega_{\tilde{l}}|},
    \label{eqn:clcir:pointvolfraction}
\end{equation}
where 
\textcolor{Reviewer1R1}{\begin{equation}
    \CellNeighborhood(\x_p) = \{k \in K : \x_p \in \Cell_k \}
\end{equation}}is the point-cell stencil: a set of labels of all cells in $\Cells$ that contain the cell-corner point $\x_p$. From the cell corner-point values defined by \ref{eqn:clcir:pointvolfraction}, an iso-contour (iso-surface) point $\EdgePoint{\lambda}$ is defined on each cell edge spanned by two cell corner-points $(\x_p, \x_q)$ according to
\begin{equation}
  \EdgePoint{\lambda} = \x_{p} + s(\x_{q} - \x_{p}) \text{, where } s \in [0,1], \, \VolFrac(\EdgePoint{\lambda}) = \VolFrac_\lambda
\end{equation}
where $\VolFrac_\lambda$ is a global iso-value.} The parameter $s$ is found using root finding methods if the field used for the iso-contour reconstruction is interpolated \textcolor{Reviewer1}{with higher-order interpolation methods}. \citet{Lopez2008} have used a linear approximation of $\VolFrac$ along the edge, which results in an explicit expression 
\begin{equation}
  s = \dfrac{\VolFrac_\lambda - \VolFrac_{p}}{\VolFrac_q - \VolFrac_p},
\end{equation}
for the $s$ parameter, where $\VolFrac_{p} = \VolFrac(\x_{p}),\VolFrac_{q} = \VolFrac(\x_{q})$ are given by \ref{eqn:clcir:pointvolfraction} \textcolor{Reviewer1}{with $\VolFrac_\lambda = 0.5$}. Once all the edge points with $\VolFrac_\lambda = 0.5$ have been calculated, the polygonization of the iso-contour is computed by triangulating edge points $\EdgePoint{\lambda}$ in each cell $c$, while \textcolor{Reviewer1}{enforcing outward orientation of triangle normal area vectors. The triangulation starts with the calculation of the centroid of the iso-contour in each cell,
\begin{equation}
    \IsoPolygonCentroid = \dfrac{1}{|E^\lambda_k|} \sum_{(p,q) \in E^\lambda_k} \x_{pq}, 
\end{equation}
where $E^\lambda_k$ is the set of edges of the cell $\Cell_k$, \textcolor{Reviewer1R1}{represented by the pairs} $(\x_p, \x_q)$, that contain iso-contour points, because their volume fraction values given by \ref{eqn:clcir:pointvolfraction} satisfy $\alpha_p < \alpha_\lambda < \alpha_q$, i.e.\ 
\begin{equation}
    E^\lambda_k = \{ (p,q) : \alpha_p < \alpha_\lambda < \alpha_q \}.
\end{equation}
The iso-contour centroid $\IsoPolygonCentroid$ in the cell $\Cell_k$ is subsequently used to compute a triangulation by connecting this centroid with the centroids of neighboring cells from the interface-cell stencil $\CellNeighborhood^\alpha(\Omega_k)$, into a set of triangles 
\begin{equation}
    T_k = \{ \tau_n = (\x_{k,\lambda}, \x_{l,\lambda}, \x_{m,\lambda}) \quad l,m \in \CellNeighborhood^\alpha(\Omega_k), 
    ((\x_{l,\lambda} - \x_{k,\lambda}) \times (\x_{m,\lambda} - \x_{k,\lambda}))\cdot \PlicNormal > 0 \},
    \label{eq:clcir:triangulation}
\end{equation}
where the interface-cell stencil $\CellNeighborhood^\alpha$ is given by \cref{eq:cellstencilint}. The condition in \cref{eq:clcir:triangulation} ensures that the normal of each triangle $\tau_n$, namely
\begin{equation}
    \n_{\tau_n} := (\x_{l,\lambda} - \x_{k,\lambda}) \times (\x_{m,\lambda} - \x_{k,\lambda}),
\end{equation}
in the triangulation $T_k$ remains oriented in the same direction as the initial PLIC normal $\PlicNormal$ given by the Youngs' algorithm by \cref{eqn:youngs}, i.e.\
\begin{equation}
    \n_{\tau_n} \cdot \PlicNormal > 0.
\end{equation}
The normals of the triangles from the triangulation $T_k$ are weighted by the angles at the centroid $\x_{k,\lambda}$ to compute the modified interface normal vector
\begin{equation}
    \PlicNormalCorrected = \dfrac{\sum_{n = 1\dots |T_k|} \n_{\tau_n} \beta_n }{\sum_{\tilde{n} = 1\dots |T_k|} \beta_{\tilde{n}}}.
  \label{eqn:clcir-corrected}
\end{equation}}This approximation of the modified interface normal vector is similar to the arithmetic average proposed by \citet{Dyadechko2005} in their modification of the Swartz-Mosso algorithm. The corrected normal is used for the interface positioning sub-step of the interface reconstruction algorithm that results in the new interface polygons. \citet{Lopez2008} have proposed a repetition of the iso-contour polygonization step as well as a B\'{e}zier spline interface approximation to increase the interface orientation accuracy and convergence. Their results \citep[tables 1 and 2]{Lopez2008} do not show significant improvements in convergence and absolute accuracy by performing a repeated polygonization step, nor by adding the B\'{e}zier triangle patch interpolation (\ac{CBIR}). The details of the B\'{e}zier interface approximation are omitted here and can be found in \citep{Lopez2008}. The \ac{CLCIR} method shows a promising and stable second-order of convergence across different mesh densities and is to be considered a good candidate for unstructured meshes, as well as the \ac{MS} modification proposed by \citet{Dyadechko2005}, since both methods disregard computationally expensive outer iteration steps. 

\subsubsection{Linear least squares fit algorithm} 

The \acf{LSF} algorithm was proposed by \citet{Scardovelli2003} and extended to $3D$ by \citet{Aulisa2007}. The algorithm starts by an initial estimate of the normal orientation using the Youngs' algorithm given by \cref{eqn:youngs}. The initial $\PlicNormal$ is then used to position the interface while upholding the prescribed volume fraction value $\VolFrac$. \textcolor{Reviewer1}{A positioned interface plane intersected with the cell $\Cell_k$ results in the interface polygon $\PlicPolygon$ given by \cref{eqn:plic-polygon} with the polygon centroid $\PlicPolygonCentroid$ given by \cref{eqn:plic-polygon-centroid}.} \textcolor{Reviewer1R1}{A second-order convergence is obtained by solving a minimization problem, constructed from the information available in the interface-cell stencil $\CellNeighborhood(\Cell_k)$, as follows.}
%
%
A distance between the interface polygon centroid in the current cell $c$ and the centroid in the neighbor cell $n$ is defined as
\textcolor{Reviewer1R1}{\begin{equation}
    d_{k, \CellOther} = \| \PlicPolygonCentroidOther - \PlicPolygonCentroid \|_2,
\end{equation}}
and it is used to compute the average distance to neighboring centroid in the cell $\Omega_k$ as 
\textcolor{Reviewer1R1}{\begin{equation}
    \tilde{d}_{k} = \dfrac{1}{|\CellNeighborhood^\alpha(\Omega_k)|}\sum_{l \in \CellNeighborhood^\alpha(\Omega_k)} d_{k, \CellOther}, 
\end{equation}}\textcolor{Reviewer1R1}{where $\CellNeighborhood^\alpha(\Omega_k)$ is given by \cref{eq:cellstencilint}. The individual and the average distance define the variance of the distance as 
\begin{equation}
    \sigma_k^2 = \dfrac{1}{|\CellNeighborhood^\alpha(\Omega_k)|(|\CellNeighborhood^\alpha(\Omega_k)| - 1)}\sum_{l \in \CellNeighborhood^\alpha(\Omega_k)} (d_{k,\CellOther} - \tilde{d}_k)^2.
\end{equation}The variance is used to compute the individual weight of each neighboring centroid $\PlicPolygonCentroidOther$ as  
\begin{equation}
    \WeightOther = \exp\left({\frac{-d_\CellOther^2}{a\sigma_k^2}}\right),
\end{equation}
with a free parameter $a$ that \citet{Scardovelli2003,Aulisa2007} set to $0.75$.} The individual weight is then normalized according to
\textcolor{Reviewer1R1}{\begin{equation}
    w_{l,n} = \dfrac{\WeightOther}{\sum_{\tilde{l}} w_{\tilde{l}}}.
\end{equation}}\textcolor{Reviewer1R1}{Finally, the weighted distance error 
\begin{equation}
    \EpsilonLlsf = \sum_{l \in \CellNeighborhood^\alpha(\Omega_k)} w_{l,n} \left[ \PlicNormalCorrected \cdot (\PlicPolygonCentroidOther - \PlicPolygonCentroid) \right]^2 
    \label{eq:lsf}
\end{equation}} is minimized. \textcolor{Reviewer1R1}{This minimization makes the LSF algorithm similar to the Mosso-Swartz algorithm, in the sense that the $\n_k^m$ modified normal is calculated as a result of a least-squares minimization problem. The \ac{MS} algorithm minimizes the difference between the normal vectors and the \ac{LSF} algorithm minimizes the distance to a plane. The error $\EpsilonLlsf$ in \cref{eq:lsf} is minimized with respect to the three components of the corrected interface orientation vector $\PlicNormalCorrected$, resulting in a $3\times3$ linear algebraic equation system that is then solved for the components of $\PlicNormalCorrected$.} 

The \ac{LSF} method ensures a stable second-order convergence for a reconstructed sphere and its absolute accuracy is comparable to \ac{CLCIR} \citep[Table 1]{Lopez2008}. \Cref{tab:plic-algorithms} places the \ac{LSF} algorithm into the class of efficient algorithms with the computational cost that is reported to be only $1.5$ times larger than the cost of the Youngs' algorithm, \textcolor{Reviewer1R1}{making LSF an interesting candidate for unstructured meshes.} 

\subsubsection{Moment of fluid algorithm} 

\begin{figure}
  \centering
  \def\svgwidth{0.3\textwidth}
     {
      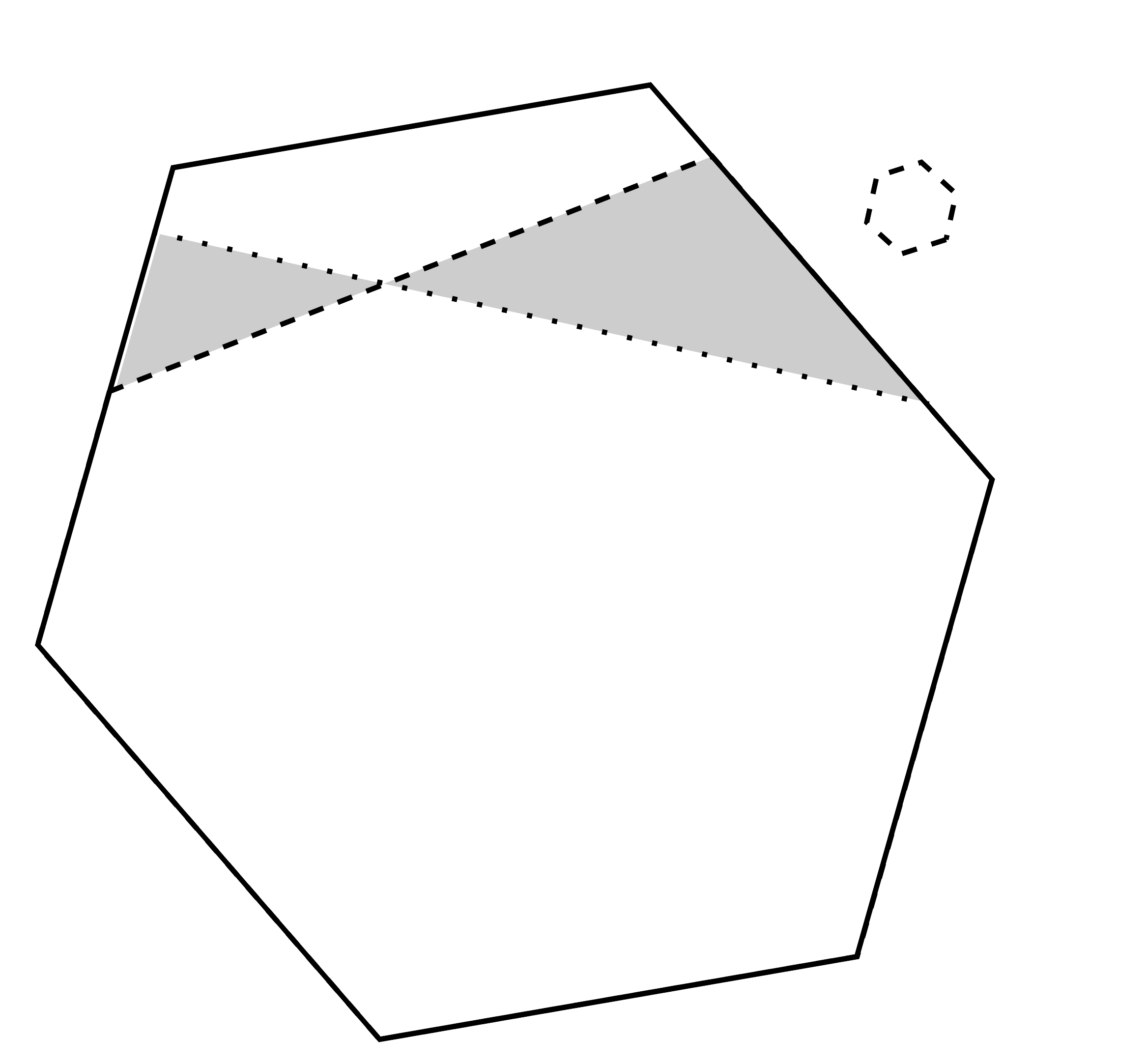
     }
     \caption{The reconstructed centroid $\CentroidRe$ and advected phase centroid $\CentroidAd$ in an interface cell used by the \ac{MoF} method.}
     \label{fig:mof-centroid}
\end{figure}

The \acf{MoF} orientation algorithm \textcolor{Reviewer1}{proposed by} \citet{Dyadechko2005} relies on the Youngs algorithm for the initial estimate of the interface normal. The second-order convergent improvement of the initial estimate is obtained by minimizing the distance between the \textcolor{Reviewer1R1}{\emph{reconstructed centroid of the phase-specific volume} $\CentroidRe$ and the \emph{advected phase centroid} $\CentroidAd$, shown schematically in \cref{fig:mof-centroid}.}

\textcolor{Reviewer1}{The reconstructed phase centroid $\CentroidRe$ is computed from the intersection between the positive half-space of the reconstructed interface, and the cell $\Omega_k$.} The positive half-space of a reconstructed \ac{PLIC} plane $\PlicNormal,\PlicPosition$ is defined as 
\textcolor{Reviewer1R1}{\begin{equation}
    \Halfspace_k (t) := \{ \x : \PlicNormal(t) \cdot (\x - \PlicPosition(t) ) \ge 0 \}.
    \label{eq:halfspace}
\end{equation}}
\noindent\textcolor{Reviewer1R1}{The centroid of the phase-specific volume is therefore the centroid of the set 
\begin{equation}
    R_k (t) = \HalfspaceInitial (t) \cap \Cell_k, 
\end{equation} i.e.\  
\begin{equation}
    \CentroidRe (t) = \dfrac{1}{|R_k(t)|} \int_{R_k(t)} \x dV.
\end{equation}}

\textcolor{Reviewer1R1}{The advected phase volume centroid is initialized as the centroid of the phase-specific volume $\Omega^+_k$, defined by the initial indicator function $\rchi(\cdot, t_0)$, i.e.\
\begin{equation}
    \CentroidAd(t_0) = \dfrac{1}{|\Cell_k|}\int_{\Cell_k} \rchi(\cdot, t_0) dV.
    \label{eq:advectioncentroidex}
\end{equation}
As the interface evolves, phase-specific volumes are contributed from cells in the interface-cell stencil $\CellNeighborhood^\alpha(\Cell_k)$ into the cell $\Cell_k$ (cf. \cref{subsec:cellbased,subsec:fluxbased}). Each phase-specific volume has a centroid attributed to it that is additionally tracked along its Lagrangian trajectory. The final advected centroid $\CentroidAd(t)$ is computed as an average of all centroids of phase-specific volumes that were advected into $\Cell_k$ from the candidate cells in $\CellNeighborhood^\alpha(\Omega_k)$ given by \ref{eq:cellstencilv}. The \ac{MoF} method is therefore not exclusively a reconstruction method, because it extends the advection of the interface by introducing and tracking centroids of phase-specific volumes. The advection aspect of the \ac{MoF} advection is addressed in \cref{sec:advect}.}

\textcolor{Reviewer1}{In \cref{fig:mof-centroid}, $A_k$ represents the intersection between the input domain and the interface-cell in the initial time step. The error of the initial interface orientation is schematically shown as the shaded region in \cref{fig:mof-centroid}. Assuming the advected phase centroid $\CentroidAd$ is available, the goal of the \ac{MoF} orientation algorithm is to minimize this shaded region by modifying the direction of the normal vector $\PlicNormal$. The difference between the advected and the reconstructed centroid, 
\begin{equation}
  \epsilon^{MOF}_k = \| \CentroidRe - \CentroidAd \|^2,
\end{equation}
is minimized to compute the corrected interface normal $\PlicNormal^A$, resulting in a new advected phase volume centroid $\CentroidAd$. The new volume
\begin{equation}
    A_k = \Halfspace^A(\PlicPosition, \PlicNormal^A) \cap \Cell_k, 
\end{equation}
is computed using $\Halfspace^A$, the positive halfspace of the PLIC plane passing through $\PlicPosition$ with the normal $\PlicNormal^A$. Then, the new centroid $\CentroidAd$ can be computed as  
\begin{equation}
  \CentroidAd = \dfrac{1}{|A_k|} \int_{A_k} \x dV.
\end{equation}}\noindent
The \ac{MoF} orientation algorithm \citep{Dyadechko2005,Ahn2008,Dyadechko2008} improves the reconstruction in two important ways. The reconstruction procedure is local to the interface cell, making the \emph{reconstruction} algorithm of the \ac{MoF} method massively parallel. This does not incur perfect linear scaling however, because the amount of time required by the reconstruction will be linearly proportional to the number of interface cells handled by the parallel process. Not requiring parallel communication for the interface reconstruction is unlike all the aforementioned normal orientation algorithms. Note, however, that the reconstruction does require the centroid to be available and the parallel computation therefore has an additional overhead of both tracing and communicating centroids of phase-specific volumes. Absolute accuracy of the method is much higher than for other orientation methods, making it a better choice for problems where local dynamic refinement is required. Additionally, the centroid-based optimization results in a more accurate automatic nested reconstruction for situations involving more than two phases \citep{Dyadechko2008}. 

\subsection{Interface positioning}

\begin{figure}[htb]
  \centering
  \def\svgwidth{0.5\textwidth}
     {
      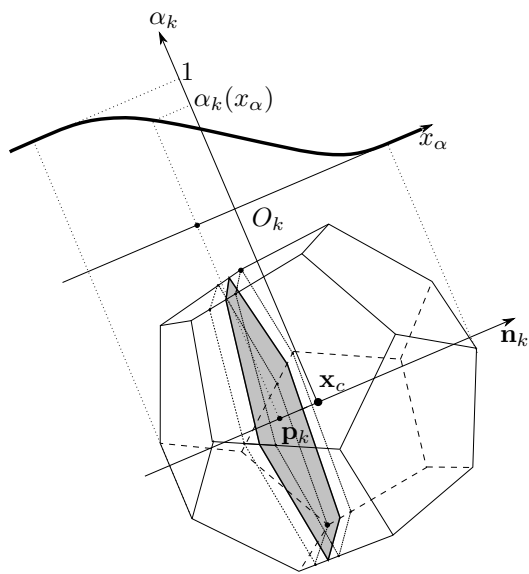
     }
  \caption{A schematic representation of the interface positioning. The volume fraction function $\PlicFraction(\PlicPosition)$ is shown for a fixed interface normal orientation $\PlicNormal$. Irrespective of $\PlicNormal$ and the shape of the cell, the $\PlicFraction$ function has diminishing gradients at the interval endpoints.}
  \label{fig:plic-positioning}
\end{figure}

Whereas reconstruction algorithms differ strongly in the choice of the interface orientation algorithm, the same interface positioning algorithms are shared by many reconstruction algorithms. The aim of a positioning algorithm is to compute the position of the interface plane $\PlicPosition$ using a known orientation $\PlicNormal$. To achieve this, a piecewise-linear approximation of the interface is used to \textcolor{Reviewer1R1}{reformulate} the volume fraction as 
\begin{equation}
  \PlicFraction = \PlicFraction(\PlicNormal, \PlicPosition). 
  \label{eqn:fraction-normal-position}
\end{equation}
The interface orientation algorithm provides $\PlicNormal$, and $\PlicFraction$ is given either by pre-processing or the advection algorithm. Hence, $\PlicPosition$ remains as the only unknown variable in \cref{eqn:fraction-normal-position}. \Cref{fig:plic-positioning} shows the volume fraction as a function of the interface position $\PlicPosition$ along a given orientation vector $\PlicNormal$. It becomes obvious by inspecting \cref{fig:plic-positioning} that \cref{eqn:fraction-normal-position} can be further simplified to a scalar equation
\begin{equation}
  \PlicFraction = \PlicFraction(\PlicPositionx), 
  \label{eqn:fraction-x}
\end{equation}
where $\PlicPositionx$ is the coordinate on the $\PlicNormal$ axis with respect to an \emph{arbitrarily chosen} origin $\PlicOrigin$. \textcolor{Reviewer1R1}{Note that floating-point operations used by the geometrical operations for the positioning are more accurate if a point inside the cell $\Omega_k$ (e.g. the cell centroid $\x_k$) is chosen as the origin of the coordinate system, as this reduces the difference between the coordinates of cell corner-points and thus increases the accuracy of floating-point operations \citep{Shewchuk2013}.} \textcolor{Reviewer1R1}{To compute the interface position \cref{eqn:fraction-x} is reformulated as}
\begin{equation}
  \PlicPositionx = \PlicPositionx(\PlicFraction). 
  \label{eqn:x-fraction}
\end{equation}
\citet{Dyadechko2005} have proven that $\PlicNormal$ and $\PlicFraction$ are sufficient to compute $\PlicPositionx$ (therefore also $\PlicPosition$) because the function given by \cref{eqn:fraction-x} is a strictly monotone function. \textcolor{Reviewer1} {Currently, there is no function of the form given by \cref{eqn:x-fraction} that can, given a volume fraction, return an interface position in \textcolor{Reviewer1R1}{an arbitrarily shaped cell $\Omega_k$ without performing some kind of a search}: either iterative root finding, or a search for a closed interval that contains $\PlicFraction$ (\emph{bracketing interval}).} 

\begin{figure}[h]
   \centering
   \begin{subfigure}[b]{0.3\textwidth}
       \centering
       \includegraphics[width=0.8\textwidth]{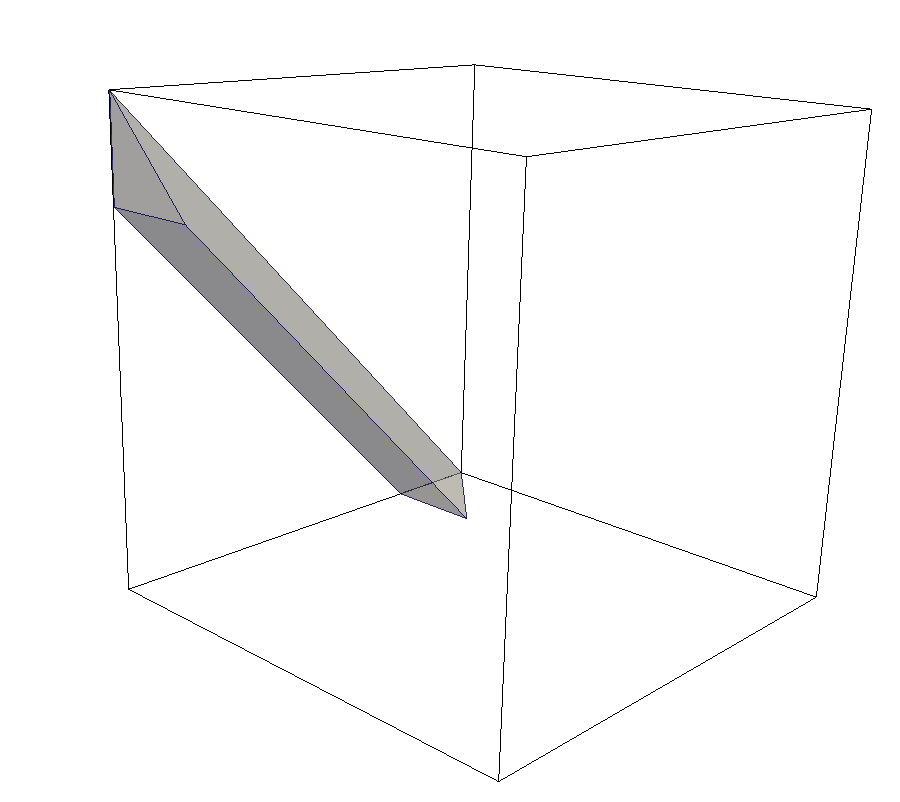}
       \caption{Intersected tetrahedron.}
       \label{fig:fill-level-tet}
   \end{subfigure}
   \begin{subfigure}[b]{0.3\textwidth}
       \centering
       \includegraphics[width=0.8\textwidth]{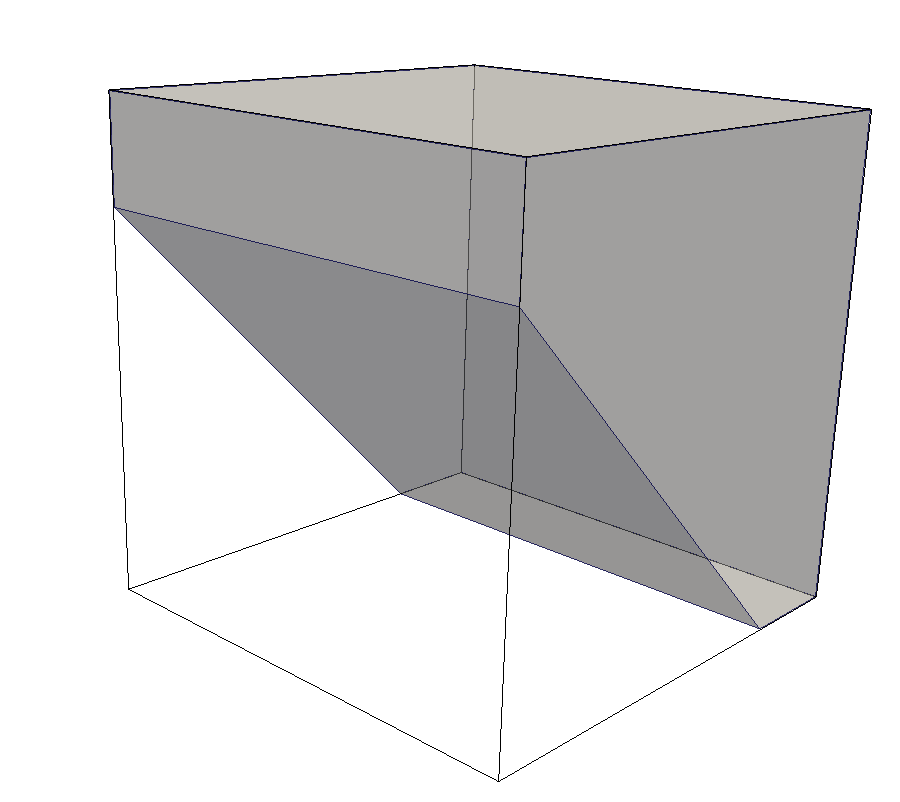}
       \caption{Intersected cube.} 
       \label{fig:fill-level-cube}
   \end{subfigure}
   \begin{subfigure}[b]{0.3\textwidth}
       \centering
       \includegraphics[width=0.8\textwidth]{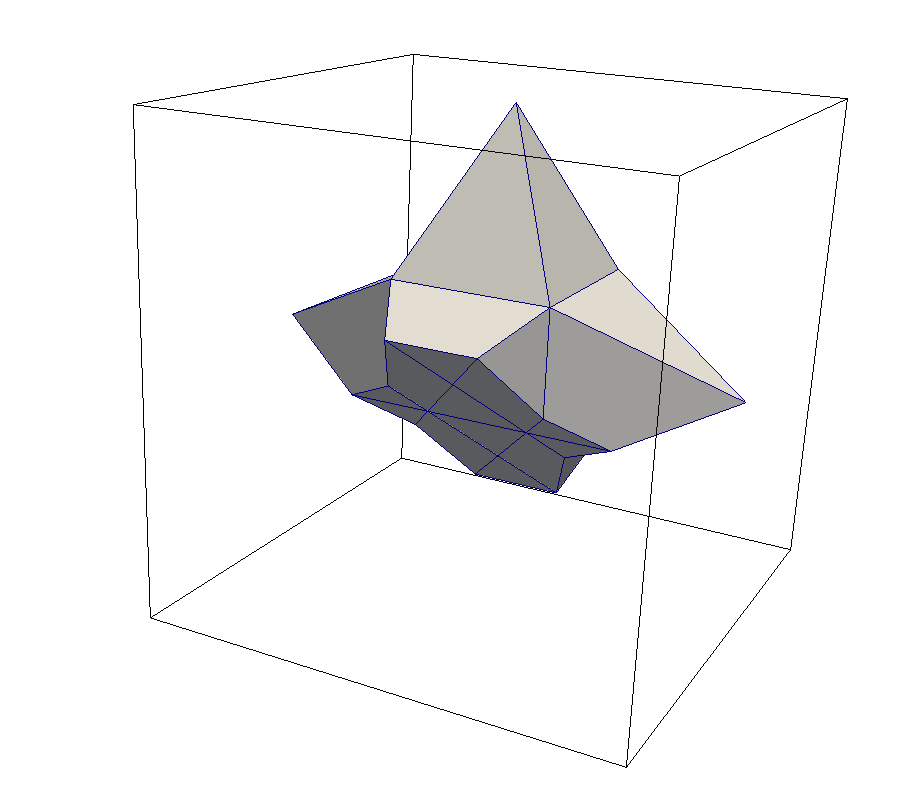}
       \caption{Intersected star shape.}
       \label{fig:fill-level-star}
   \end{subfigure}

   \begin{subfigure}[b]{\textwidth}
      \centering
      \footnotesize
      \def\svgwidth{0.8\textwidth}
         {
          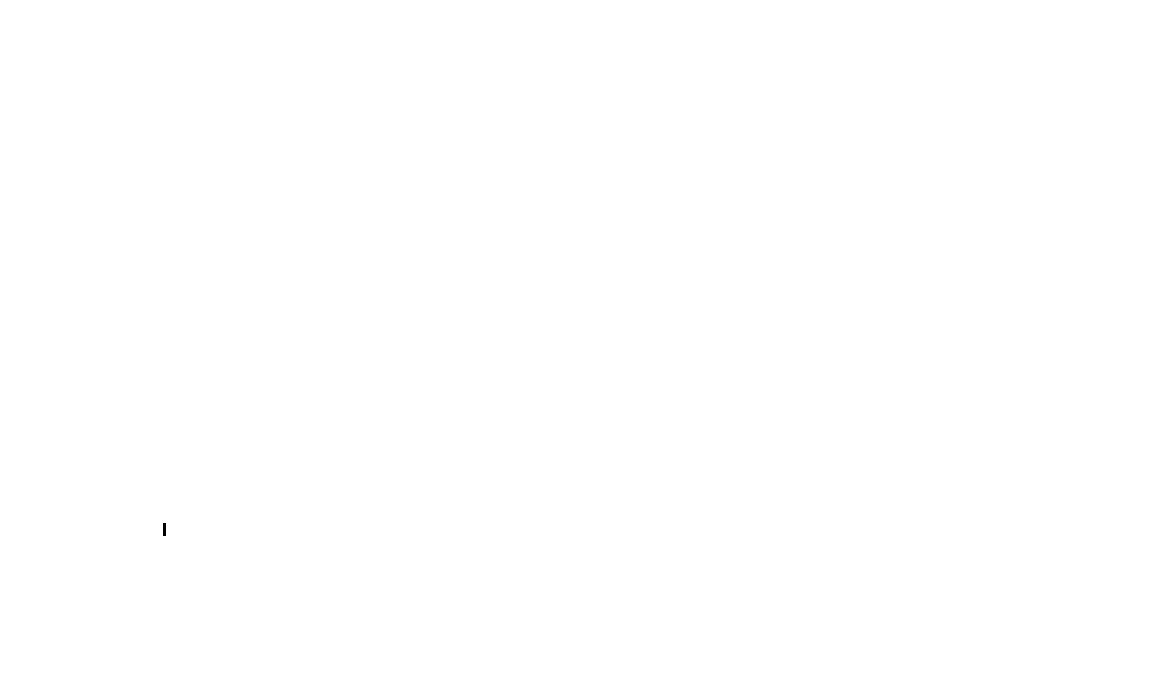
         }
   \end{subfigure}
    \caption{Volume fraction $\PlicFraction$ as a function of the interface plane position $\PlicPosition$ computed using polyhedrons (\ref{fig:fill-level-tet},\ref{fig:fill-level-cube}, \ref{fig:fill-level-star}) and $\PlicNormal = (1,1,0)$.}
   \label{fig:fill-levels}
\end{figure}

\textcolor{Reviewer1R1}{The graphs of $\PlicFraction(\PlicNormal, \PlicPosition)$ are shown in \cref{fig:fill-level-tet,fig:fill-level-cube,fig:fill-level-star} for different primitive cell shapes using the same $\PlicNormal$.} In every case, the function has a diminishing derivative at two positions $\PlicPosition$: the point where the intersection between the half-space defined by the \ac{PLIC} interface plane and the cell is the complete cell, and another point where the intersection result is an empty set. The diminishing derivative can cause divergence of slope-based numerical methods used to solve \cref{eqn:x-fraction} for $\PlicPosition$. \textcolor{Reviewer1R1}{Note that there is one special case where the derivative does not diminish: when $\PlicNormal$ is collinear with a planar face of $\Omega_k$, which might happen if a planar interface is initialized on a hexahedral mesh.} 

Several interface positioning algorithms have been developed so far to compute $\PlicPositionx$. An iterative approach based on the Brent's method \citep{Brent2013} has been used by \citet{Rider1998}. \Cref{eqn:x-fraction} is evaluated iteratively, starting with an initial guess $\PlicPositionx^0$, until $\PlicPositionx$ is computed, such that the prescribed volume fraction $\PlicFraction$ is obtained up to a prescribed tolerance. The Brent method \citep{Brent1971} is applied because it provides a stable solution to the root finding problem even when subjected to the \textcolor{Reviewer1R1}{diminishing function derivative} at the endpoints of the interval as shown in \cref{fig:plic-positioning} schematically and in \cref{fig:fill-levels} for actual volume fractions. \citet{Rider1998} already state that Newton's iterative method \textcolor{Reviewer1R1}{lowers the number of iterations provided an accurate initial starting position} for the Brent's method. In order to better estimate $\PlicPositionx^0$ in the first step, \citet{Rider1998} sort the cell points with respect to the projection on $\PlicNormal$, defining
\textcolor{Reviewer1R1}{\begin{equation}
    \CellPointsSorted = (\Point_1, \Point_2, \dots \Point_n) \quad \text{such that } (\Point_i \cdot \PlicNormal) \le (\Point_{i+i} \cdot \PlicNormal), \Point_i \text{ is a cell corner-point.}
  \label{eqn:cellpoints-sorted}
\end{equation}}
To each $\Point_i \in \CellPointsSorted$, a volume fraction is assigned as  
\begin{equation}
  \VolFrac_{c,i} = \dfrac{|\Cell_k \cap \Halfspace(\PlicNormal, \Point_i)|}{|V_k|},
  \label{eqn:pointfraction}
\end{equation}
where $\Halfspace$ is a positive half-space at point $\Point_i$ whose orientation is defined by $\PlicNormal$. The \emph{cut-out slab} is defined as 
\begin{equation}
  \Slab_{c,i} = \{ \Cell_k \cap \Halfspace(\PlicNormal, \Point_i) \cap \Halfspace(-\PlicNormal, \Point_{i+1}) \Such i = 1,\dots, |\CellPointsSorted|\}.
\end{equation}
From \cref{eqn:cellpoints-sorted,eqn:pointfraction}, we have 
\begin{equation}
  \VolFrac_{c,i} > \VolFrac_{c,i+1}
\end{equation}
and, as a consequence,
\begin{equation}
  \exists ! \, \Slab_{c,i} \Such \VolFrac_{c,i+1} \le \VolFrac_k < \VolFrac_{c,i} \iff \PlicPosition \in \Slab_{c,i}.
    \label{eq:slabs}
\end{equation}
A cut-out slab $\Slab_{c,i}$ is then chosen which contains the interface position. At this point, \citet{Rider1998} apply Brent's method \citep{Brent2013} to locate the interface within the slab.

A semi-analytical approach was extended to arbitrary cell shapes by \citet{Lopez2008an} \textcolor{Reviewer1R1}{that sorts the slabs according \cref{eq:slabs} (similar to \citet{Rider1998}) and positions the interface within the slab that contains it.} Contrary to the algorithm proposed by \citet{Rider1998}, once $\Slab_{c,i}$ is found, an analytical expression is used to compute $\PlicPosition$ explicitly. This semi-analytical (bracketing) approach is faster on cubic cells compared to the Brent method used in \citep{Rider1998}. Furthermore, \citet{Lopez2008an} state that the sorting and slab calculation step is still computationally the most expensive part of the positioning algorithm. 

A simplified iterative approach was proposed by \citet{Ahn2008} that also  works well with cells of arbitrary shape. Additionally, the average number of iterations for cubic cells is reported in \citep{Ahn2008} to be smaller than $8$ - which is the number of vertices of the cube. This makes the stabilized iterative algorithm comparable to the semi-analytical, even for $\Omega_h$ with hexahedral cells, \textcolor{Reviewer1R1}{since the total number of iterations} is smaller than for the semi-analytical algorithm where it is necessary to create the cut-out slabs.  

Another semi-iterative interface positioning algorithm was proposed by \citet{Diot2014} for planar and axis-symmetric convex cells. Their approach is faster than the standard Brent's iterative method and the algorithm proposed by \citet{Dyadechko2005} that relies on the calculation of the interface position $\PlicPositionx$ within a slab $\Slab_{c,i}$ using interpolation. Table 2 contains results for the axis-symmetric geometry and shows exactly what is to be expected: the average global iteration number for both the method of \citet{Dyadechko2005} and \citet{Diot2014} are the same, since they both rely on the same sorting and slab calculation steps. However, no reference is made to the stabilized secant/bisection method of \citet{Ahn2008}, or the method proposed by \citet{Lopez2008an}. 

\citet{Diot2016} have extended their 2D and axis-symmetric method \citep{Diot2014} to 3D for convex cells of arbitrary shape. Following their work in \citep{Diot2014}, explicit analytic expressions for computing a volume of 3D slabs $\Slab_{c,i}$ are proposed. Only a comparison with an iterative method based on Brent's root finding method is provided in the results section of the article. Accuracy and efficiency comparisons with \citet{Lopez2008an,Ahn2008} are not provided.  

\textcolor{Reviewer1}{More recently, \citet{Lopez2016} have optimized the \textcolor{Reviewer1R1}{bracketing} procedure used to calculate the cut-out slabs using linear interpolation of the fill volume, based on signed distances between the cell corner-points and the \ac{PLIC} interface. Additionally, \citet{Lopez2016} have improved the analytical expression for the volume calculation in general convex polyhedrons, that additionally increases the efficiency of interface positioning. Their Coupled Interpolation-Bracketed Analytical Volume Enforcement (CIBRAVE) is compared \citep[fig. 8]{Lopez2016} with the Brent's method, \citet{Diot2016}  and \citet{Ahn2008}, and shows a significant improvement in terms of the relative CPU time.}

\textcolor{Reviewer1}{
A new iterative positioning algorithm has recently been proposed by \cite{Chen2019}. An exact derivative $\frac{d\VolFrac}{dx}(x)$ is computed using the area of the so-called \emph{cap polygon}, namely
\begin{equation}
    \dfrac{d\VolFrac}{dx} = \dfrac{|\Pi(\n_k, \mathbf{p}(x)) \cap \Omega_k|}{|\Omega_k|}.
    \label{eqn:alphaderiv}
\end{equation}
The exact derivative in \cref{eqn:alphaderiv} is used with the Newton method to achieve significantly faster convergence compared to the Brent's method or the stabilized secant method of \cite{Ahn2008}. A \emph{cap-polygon} is defined as the intersection between the interface plane $\Pi(\n_k, \mathbf{p}(x))$ and the cell $\Omega_k$. Its area equals the derivative of $\VolFrac$ with respect to the $x$-coordinate of the positioning axis in \cref{fig:plic-positioning}.}

\textcolor{Reviewer1}{Recently, \citet{Lopez2019} have proposed an analytical expression for interface positioning in non-convex polyhedrons. The polyhedron is modeled using a connectivity table that defines each face as an ordered list of indices from a global set of points. The advantage of the connectivity table is twofold: the algorithm can be applied to non-convex polyhedrons because the connectivity table supports disjoint sets and the divergence theorem can be used for the volume calculation. The divergence theorem increases computational efficiency, as it requires less floating point and memory operations than tetrahedral decomposition. Connectivity does come with a cost: it significantly increases the implementation complexity of the algorithm compared to tetrahedral decomposition. The authors report an increase in computational efficiency in one order of magnitude, compared to Brent's method with tetrahedral decomposition. The non-convexity of the polyhedron caused by the non-planarity of its faces is adressed by triangulation \citep[fig. 18]{Lopez2019}.}

\subsection{Reconstruction comparison}
\label{sec:res:reconstruction}

The accuracy of the reconstruction can be influenced by the accuracy of the initialization algorithm \textcolor{Reviewer1}{if the latter is not accurate enough. When using high mesh resolutions, high accuracy of the initialization must be ensured to avoid the influence of the initialization on the reconstruction and, subsequently, advection errors.}

\textcolor{Reviewer1}{The volume fraction can be exactly initialized on structured and unstructured meshes for the spherical interface and it is done approximatively for more complex surfaces \citep{Kromer2019,Jones2019}.} \textcolor{Reviewer1R1}{Volume fractions can also be calculated by approximating $\Omega^+(t)$ with an unstructured mesh, and intersecting this mesh with the mesh used to approximate the solution domain $\Omega$ \citep{Ahn2009}.} Furthermore, different error measures and different verification tests are used to verify the reconstruction algorithm, which makes it difficult to summarize the results and compare methods. Reconstruction of a circle or a sphere is a widely used verification case so it is presented in this section as a basis for comparing reconstruction algorithms. \citet{Aulisa2007} have used a sphere of radius $r=0.325$ randomly placed around the center point $(0.5, 0.5, 0.5)$ in the unit box domain. \citet{Lopez2008} have used the same radius, and have placed the sphere at $(0.525,0.464, 0.516)$.

\citet{Aulisa2007} rely on the numerical quadrature approach to initialize the volume fraction field for the \ac{LSF} algorithm. The volume fraction is initialized by reformulating the sharp indicator function $\Indicator(\x)$ in a cubic cell as a \emph{height function} with respect to a face $S_f$, that is then integrated as   
\textcolor{Reviewer1}{
\begin{equation} 
  \PlicFraction = \dfrac{1}{|\Cell_k|} \int_{\Cell_k} \Indicator(\x) \, dV = \dfrac{1}{|\Cell_k|} \int_{S_f} z^*(x,y) \, dS,
  \label{eqn:aulisa:init}
\end{equation}}
such that
\begin{equation} 
  z^*(x,y) = \min(h, \max(f(x,y) - (k-1)h, 0)), 
\end{equation}
where $f(x,y) - z = 0$ is the surface equation that replaces the sharp phase indicator function $\Indicator(\x, 0)$, and $h$ is the length of the cubic cell. \Cref{eqn:aulisa:init} is numerically integrated using the Simpson's quadrature by dividing $S_f$ into $30^2$ square sub-intervals.  

\citet{Lopez2008} use a mesh refinement technique for initializing the volume fraction field, originally proposed by \citet{Francois2006} and \citet{Cummins2005}. A cubic cell is subdivided into octants (quadrants in 2D) in a single refinement level. In total, four refinement levels are used per multi-material cell. In the lowest refinement levels, the interface is approximated linearly and the cell is intersected with the linear interface approximation to compute the volume fraction, 
\textcolor{Reviewer1}{
\begin{equation} 
  \PlicFraction = \dfrac{1}{|\Cell_k|} \int_{\Cell_k} \Indicator(\x) \, dV = \dfrac{1}{|\Cell_k|} \sum_{k=1}^{N_{s}} \dfrac{|\Halfspace_k \cap \Cell_k|}{|\Cell_k|}.
  \label{eqn:approx:init}
\end{equation}}\textcolor{Reviewer1}{In \cref{eqn:approx:init},} $N_s$ is the number of sub-cells resulting from cell sub-division, $\Halfspace_k$ is the halfspace that approximates the interface in the cell $k$. \Cref{tab:recon:sphere:clcir} holds reconstruction error values computed as the volume of symmetric difference, 
\textcolor{Reviewer1}{
\begin{equation}
  E_r^1 = \int_{\Cell_k} |\Indicator(\x) - \Halfspace_k| \, dV,  
  \label{eqn:rec:error}
\end{equation}}
between the \ac{PLIC} interface and the exact interface, where $\Halfspace_k(\x)$ is the positive halfspace given by the \ac{PLIC} plane (\cref{eq:halfspace}). \Cref{eqn:rec:error} can be integrated numerically using the exact indicator function as in \cref{eqn:aulisa:init} for hexahedral cells with planar faces.  

\begin{figure}[h]
   \centering
   \begin{subfigure}[b]{0.5\textwidth}
      \centering
      \def\svgwidth{0.49\textwidth}
         {
          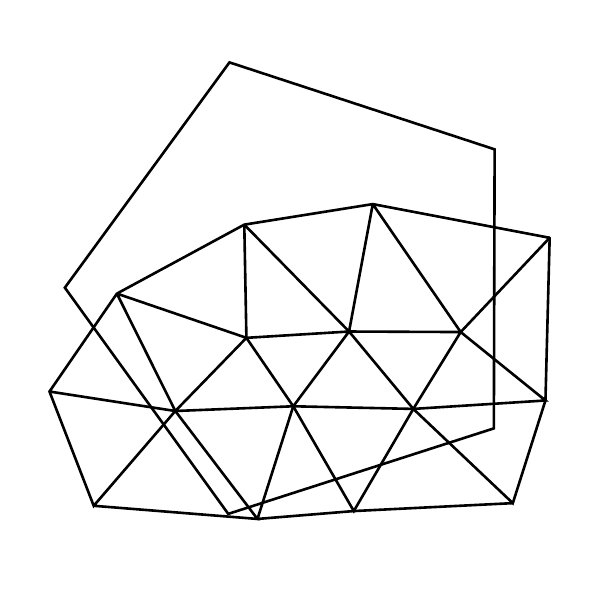
         }
       \caption{Initialization with mesh-mesh intersection.}
       \label{fig:mof:init}
   \end{subfigure}
   \begin{subfigure}[b]{0.49\textwidth}
      \centering
      \def\svgwidth{0.5\textwidth}
         {
          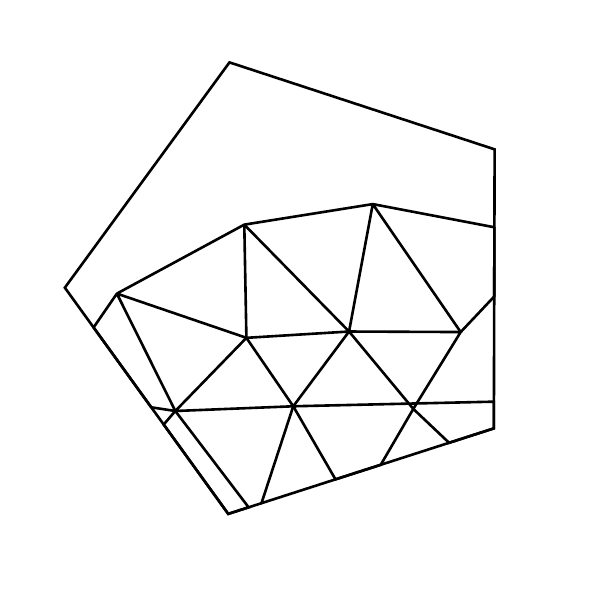
         }
       \caption{Volume of symmetric difference.}
       \label{fig:mof:symm}
   \end{subfigure}
    \caption{\textcolor{Reviewer1}{Error calculation of the \ac{MoF} method by \citet{Ahn2009}.}}
   \label{fig:mof:recerr}
\end{figure}

For polyhedral cells, \citet{Ahn2007} distribute points uniformly in multi-material cells. The points are then categorized as \emph{inside} or \emph{outside} based on the provided geometrical model of the interface, or an explicit indicator function, e.g.\ a signed distance function. The volume fraction in the multi-material cell is then given as the ratio of the number of inside points and the total number of points in a cell. Later, \citet{Ahn2009} use an intersection between two meshes to geometrically calculate the initial centroids and volume fractions for the \ac{MoF} method. \textcolor{Reviewer1R1}{The mesh of the solution domain $\Omega:=\cup_{k \in K} \Omega_k$ is intersected with an unstructured mesh that is used to approximate $\Omega^+$. The $\Omega^+$ phase is decomposed into disjoint sub-volumes, i.e.\ 
\begin{equation}
    \tilde{\Omega}^+ = \cup_{l \in L} \tilde{\Omega}^+_l.
\end{equation}
Here, $\tilde{\Omega}^+$ denotes an approximation of $\Omega^+$, because the fluid interface ($\Sigma := \partial \Omega^+$) is approximated in $\tilde{\Omega}^+$ as a set of mutually connected polygons. The volume fraction $\PlicFraction$ is then calculated using 
\begin{equation}
    \PlicFraction = \dfrac{|\tilde{\Omega}_k^+|}{|\Cell_k|} = \dfrac{1}{|\Cell_k|} \sum_{l \in \CellNeighborhood^+_n(\Omega_k)}|\Cell_k \cap \tilde{\Omega}^+_l| ,
    \label{eq:alphameshint}
\end{equation} since $\tilde{\Omega}_k^+ =\cup_ {l \in \CellNeighborhood^+_n(\Omega_k)} \Cell_k \cap \tilde{\Omega}^+_l$. Furthermore, 
\begin{equation}
    \CellNeighborhood^+_n(\Cell_k) = \{ l \in K : \Cell_k \cap \tilde{\Omega}^+_l \ne \emptyset \}
\end{equation}
is the intersection stencil of $\Cell_k$ in $\tilde{\Omega}^+$.} \textcolor{Reviewer1R1}{Calculating $\tilde{\Omega}_k^+$ is therefore important for computing initial $\PlicFraction$ by \cref{eq:alphameshint}, needed for the reconstruction, but also to express the resulting reconstruction error as the volume of symmetric difference 
\begin{equation}
    E_r^2 = \sum_{k \in K} |\tilde{\Omega}_k^+ \cap -\Halfspace_k| + |\Omega_k \cap \Halfspace_k| - |\tilde{\Omega}_k^+ \cap \Halfspace_k|,
\end{equation}where the "$-$" sign changes the orientation of the halfspace $\Halfspace_k$. Contributions to the $E_r^2$ error are illustrated as shaded areas in \cref{fig:mof:symm}.}

\begin{table}
  \centering
  \footnotesize
  \input{tables/Ahn2007.tex}
  \caption{Reconstruction error $E_r^2$ computed for a spherical interface by \citet{Ahn2007}.} 
  \label{tab:recon:sphere:mof}
\end{table}

$E_r^2$ errors for a sphere of radius $r=0.5 - \frac{1}{11}$, positioned at $(0.5 + \frac{1}{29}, 0.5 + \frac{1}{31}, 0.5 + \frac{1}{39})$ are digitized in \cref{tab:recon:sphere:mof} from the diagram reported by \citet[fig. 12]{Ahn2007}. The radius and position of the sphere, as well as the number of mesh elements $N^3$ are different from those used by \citet{Aulisa2007} and \citet{Lopez2008}, so a direct comparison with \ac{LSF}, \ac{CLCIR}, \ac{CBIR} is not possible, even though results in \cref{tab:recon:sphere:mof} were \textcolor{Reviewer1}{generated on unstructured meshes with cubic cells}. \textcolor{Reviewer1}{A comparison between the Youngs', \ac{LSF}, \ac{CLCIR} and \ac{CBIR} reconstruction algorithms performed by \citet[table 1]{Lopez2008} is presented here in \cref{tab:recon:sphere:clcir}.} 

\begin{table}
  \centering
  \footnotesize
  \input{tables/CLCIR-recon.tex}
  \caption{Reconstruction error $E_r^1$ computed for a spherical interface by \citet{Lopez2008}.} 
  \label{tab:recon:sphere:clcir}
\end{table}

Overall, \ac{MoF} maintains the best absolute accuracy, with increasing costs, lower than, but comparable to LVIRA for a serial execution on unstructured hexahedral meshes when higher resolution is used \citep[table 1]{Ahn2007}. The \ac{LVIRA} algorithm, listed in \cref{tab:plic-algorithms}, was used on unstructured meshes by \citet{Jofre2014}, however the computational efficiency of LVIRA is not reported. To the best of the authors' knowledge, \citet[table 1]{Ahn2007} is the only report containing absolute CPU times of the reconstruction algorithm for $3D$ reconstruction on unstructured meshes. Relative timings for $3D$ reconstruction can be found in \citet[table III]{Lopez2008}. Overall, the (E)LVIRA algorithms require very long execution times. \textcolor{Reviewer1}{Even though \citet{Jofre2014} rely on (E)LVIRA to ensure overall second-order convergence, their work on load balancing of the parallel execution \citep{Jofre2015} relies on the Youngs' reconstruction algorithm.}

The data summarized in this section is available at {\footnotesize \url{http://dx.doi.org/10.25534/tudatalib-162}}.

%% file: figures/MOF.pdf_tex
\begingroup%
  \makeatletter%
  \providecommand\color[2][]{%
    \errmessage{(Inkscape) Color is used for the text in Inkscape, but the package 'color.sty' is not loaded}%
    \renewcommand\color[2][]{}%
  }%
  \providecommand\transparent[1]{%
    \errmessage{(Inkscape) Transparency is used (non-zero) for the text in Inkscape, but the package 'transparent.sty' is not loaded}%
    \renewcommand\transparent[1]{}%
  }%
  \providecommand\rotatebox[2]{#2}%
  \newcommand*\fsize{\dimexpr\f@size pt\relax}%
  \newcommand*\lineheight[1]{\fontsize{\fsize}{#1\fsize}\selectfont}%
  \ifx\svgwidth\undefined%
    \setlength{\unitlength}{672bp}%
    \ifx\svgscale\undefined%
      \relax%
    \else%
      \setlength{\unitlength}{\unitlength * \real{\svgscale}}%
    \fi%
  \else%
    \setlength{\unitlength}{\svgwidth}%
  \fi%
  \global\let\svgwidth\undefined%
  \global\let\svgscale\undefined%
  \makeatother%
  \begin{picture}(1,0.92745536)%
    \lineheight{1}%
    \setlength\tabcolsep{0pt}%
    \put(0,0){\includegraphics[width=\unitlength,page=1]{MOF.pdf}}%
    \put(0.85986071,0.73233697){\color[rgb]{0,0,0}\makebox(0,0)[lt]{\lineheight{0}\smash{\begin{tabular}[t]{l}$R_k$\end{tabular}}}}%
    \put(0,0){\includegraphics[width=\unitlength,page=2]{MOF.pdf}}%
    \put(0.85986071,0.82760781){\color[rgb]{0,0,0}\makebox(0,0)[lt]{\lineheight{0}\smash{\begin{tabular}[t]{l}$A_k$\end{tabular}}}}%
    \put(0,0){\includegraphics[width=\unitlength,page=3]{MOF.pdf}}%
    \put(0.51713658,0.45037296){\color[rgb]{0,0,0}\makebox(0,0)[lt]{\lineheight{0}\smash{\begin{tabular}[t]{l}$\CentroidRe$\end{tabular}}}}%
    \put(0.30389786,0.25844013){\color[rgb]{0,0,0}\makebox(0,0)[lt]{\lineheight{0}\smash{\begin{tabular}[t]{l}$\CentroidAd$\end{tabular}}}}%
    \put(0,0){\includegraphics[width=\unitlength,page=4]{MOF.pdf}}%
    \put(0.78125063,0.02566894){\color[rgb]{0,0,0}\makebox(0,0)[lt]{\lineheight{0}\smash{\begin{tabular}[t]{l}$\Cell_k$\end{tabular}}}}%
    \put(0.52614339,0.55658762){\color[rgb]{0,0,0}\makebox(0,0)[lt]{\lineheight{0}\smash{\begin{tabular}[t]{l}$\HalfspaceInterface$\end{tabular}}}}%
    \put(0.15185844,0.53255118){\color[rgb]{0,0,0}\makebox(0,0)[lt]{\lineheight{0}\smash{\begin{tabular}[t]{l}$\HalfspaceInitial$\end{tabular}}}}%
  \end{picture}%
\endgroup%

%% file: figures/interface-positioning.pdf_tex
\begingroup%
  \makeatletter%
  \providecommand\color[2][]{%
    \errmessage{(Inkscape) Color is used for the text in Inkscape, but the package 'color.sty' is not loaded}%
    \renewcommand\color[2][]{}%
  }%
  \providecommand\transparent[1]{%
    \errmessage{(Inkscape) Transparency is used (non-zero) for the text in Inkscape, but the package 'transparent.sty' is not loaded}%
    \renewcommand\transparent[1]{}%
  }%
  \providecommand\rotatebox[2]{#2}%
  \ifx\svgwidth\undefined%
    \setlength{\unitlength}{563.77198792bp}%
    \ifx\svgscale\undefined%
      \relax%
    \else%
      \setlength{\unitlength}{\unitlength * \real{\svgscale}}%
    \fi%
  \else%
    \setlength{\unitlength}{\svgwidth}%
  \fi%
  \global\let\svgwidth\undefined%
  \global\let\svgscale\undefined%
  \makeatother%
  \begin{picture}(1,0.99036935)%
    \put(0,0){\includegraphics[width=\unitlength,page=1]{interface-positioning.pdf}}%
    \put(0.55929783,0.33196214){\color[rgb]{0,0,0}\makebox(0,0)[lb]{\smash{$\x_c$}}}%
    \put(0,0){\includegraphics[width=\unitlength,page=2]{interface-positioning.pdf}}%
    \put(0.87724893,0.41145759){\color[rgb]{0,0,0}\makebox(0,0)[lb]{\smash{$\PlicNormal$}}}%
    \put(0.2602057,0.97158891){\color[rgb]{0,0,0}\makebox(0,0)[lb]{\smash{$\PlicFraction$}}}%
    \put(0,0){\includegraphics[width=\unitlength,page=3]{interface-positioning.pdf}}%
    \put(0.31559554,0.87326156){\color[rgb]{0,0,0}\makebox(0,0)[lb]{\smash{$1$}}}%
    \put(0,0){\includegraphics[width=\unitlength,page=4]{interface-positioning.pdf}}%
    \put(0.73425028,0.75776479){\color[rgb]{0,0,0}\makebox(0,0)[lb]{\smash{$\PlicPositionx$}}}%
    \put(0.44086801,0.61448838){\color[rgb]{0,0,0}\makebox(0,0)[lb]{\smash{$\PlicOrigin$}}}%
    \put(0,0){\includegraphics[width=\unitlength,page=5]{interface-positioning.pdf}}%
    \put(0.33873231,0.82721069){\color[rgb]{0,0,0}\makebox(0,0)[lb]{\smash{$\PlicFraction(\PlicPositionx)$}}}%
    \put(0,0){\includegraphics[width=\unitlength,page=6]{interface-positioning.pdf}}%
    \put(0.4920362,0.24041769){\color[rgb]{0,0,0}\makebox(0,0)[lb]{\smash{$\PlicPosition$}}}%
    \put(0,0){\includegraphics[width=\unitlength,page=7]{interface-positioning.pdf}}%
  \end{picture}%
\endgroup%

%% file: figures/tetPolyStarHalfspaceIntersection.pdf_tex
\begingroup%
  \makeatletter%
  \providecommand\color[2][]{%
    \errmessage{(Inkscape) Color is used for the text in Inkscape, but the package 'color.sty' is not loaded}%
    \renewcommand\color[2][]{}%
  }%
  \providecommand\transparent[1]{%
    \errmessage{(Inkscape) Transparency is used (non-zero) for the text in Inkscape, but the package 'transparent.sty' is not loaded}%
    \renewcommand\transparent[1]{}%
  }%
  \providecommand\rotatebox[2]{#2}%
  \newcommand*\fsize{\dimexpr\f@size pt\relax}%
  \newcommand*\lineheight[1]{\fontsize{\fsize}{#1\fsize}\selectfont}%
  \ifx\svgwidth\undefined%
    \setlength{\unitlength}{330.67527008bp}%
    \ifx\svgscale\undefined%
      \relax%
    \else%
      \setlength{\unitlength}{\unitlength * \real{\svgscale}}%
    \fi%
  \else%
    \setlength{\unitlength}{\svgwidth}%
  \fi%
  \global\let\svgwidth\undefined%
  \global\let\svgscale\undefined%
  \makeatother%
  \begin{picture}(1,0.59834581)%
    \lineheight{1}%
    \setlength\tabcolsep{0pt}%
    \put(0,0){\includegraphics[width=\unitlength,page=1]{tetPolyStarHalfspaceIntersection.pdf}}%
    \put(0.11860601,0.10263626){\color[rgb]{0,0,0}\makebox(0,0)[lt]{\lineheight{1.25}\smash{\begin{tabular}[t]{l}0\end{tabular}}}}%
    \put(0.13254029,0.10263626){\color[rgb]{0,0,0}\makebox(0,0)[lt]{\lineheight{1.25}\smash{\begin{tabular}[t]{l}\textit{.}\end{tabular}}}}%
    \put(0.14028149,0.10263626){\color[rgb]{0,0,0}\makebox(0,0)[lt]{\lineheight{1.25}\smash{\begin{tabular}[t]{l}00\end{tabular}}}}%
    \put(0,0){\includegraphics[width=\unitlength,page=2]{tetPolyStarHalfspaceIntersection.pdf}}%
    \put(0.2144842,0.10263626){\color[rgb]{0,0,0}\makebox(0,0)[lt]{\lineheight{1.25}\smash{\begin{tabular}[t]{l}0\end{tabular}}}}%
    \put(0.22841847,0.10263626){\color[rgb]{0,0,0}\makebox(0,0)[lt]{\lineheight{1.25}\smash{\begin{tabular}[t]{l}\textit{.}\end{tabular}}}}%
    \put(0.23615969,0.10263626){\color[rgb]{0,0,0}\makebox(0,0)[lt]{\lineheight{1.25}\smash{\begin{tabular}[t]{l}25\end{tabular}}}}%
    \put(0,0){\includegraphics[width=\unitlength,page=3]{tetPolyStarHalfspaceIntersection.pdf}}%
    \put(0.31036238,0.10263626){\color[rgb]{0,0,0}\makebox(0,0)[lt]{\lineheight{1.25}\smash{\begin{tabular}[t]{l}0\end{tabular}}}}%
    \put(0.32429665,0.10263626){\color[rgb]{0,0,0}\makebox(0,0)[lt]{\lineheight{1.25}\smash{\begin{tabular}[t]{l}\textit{.}\end{tabular}}}}%
    \put(0.33203784,0.10263626){\color[rgb]{0,0,0}\makebox(0,0)[lt]{\lineheight{1.25}\smash{\begin{tabular}[t]{l}50\end{tabular}}}}%
    \put(0,0){\includegraphics[width=\unitlength,page=4]{tetPolyStarHalfspaceIntersection.pdf}}%
    \put(0.40624056,0.10263626){\color[rgb]{0,0,0}\makebox(0,0)[lt]{\lineheight{1.25}\smash{\begin{tabular}[t]{l}0\end{tabular}}}}%
    \put(0.42017481,0.10263626){\color[rgb]{0,0,0}\makebox(0,0)[lt]{\lineheight{1.25}\smash{\begin{tabular}[t]{l}\textit{.}\end{tabular}}}}%
    \put(0.42791602,0.10263626){\color[rgb]{0,0,0}\makebox(0,0)[lt]{\lineheight{1.25}\smash{\begin{tabular}[t]{l}75\end{tabular}}}}%
    \put(0,0){\includegraphics[width=\unitlength,page=5]{tetPolyStarHalfspaceIntersection.pdf}}%
    \put(0.50211874,0.10263626){\color[rgb]{0,0,0}\makebox(0,0)[lt]{\lineheight{1.25}\smash{\begin{tabular}[t]{l}1\end{tabular}}}}%
    \put(0.51605303,0.10263626){\color[rgb]{0,0,0}\makebox(0,0)[lt]{\lineheight{1.25}\smash{\begin{tabular}[t]{l}\textit{.}\end{tabular}}}}%
    \put(0.52379421,0.10263626){\color[rgb]{0,0,0}\makebox(0,0)[lt]{\lineheight{1.25}\smash{\begin{tabular}[t]{l}00\end{tabular}}}}%
    \put(0,0){\includegraphics[width=\unitlength,page=6]{tetPolyStarHalfspaceIntersection.pdf}}%
    \put(0.59799692,0.10263626){\color[rgb]{0,0,0}\makebox(0,0)[lt]{\lineheight{1.25}\smash{\begin{tabular}[t]{l}1\end{tabular}}}}%
    \put(0.61193121,0.10263626){\color[rgb]{0,0,0}\makebox(0,0)[lt]{\lineheight{1.25}\smash{\begin{tabular}[t]{l}\textit{.}\end{tabular}}}}%
    \put(0.61967243,0.10263626){\color[rgb]{0,0,0}\makebox(0,0)[lt]{\lineheight{1.25}\smash{\begin{tabular}[t]{l}25\end{tabular}}}}%
    \put(0,0){\includegraphics[width=\unitlength,page=7]{tetPolyStarHalfspaceIntersection.pdf}}%
    \put(0.6938751,0.10263626){\color[rgb]{0,0,0}\makebox(0,0)[lt]{\lineheight{1.25}\smash{\begin{tabular}[t]{l}1\end{tabular}}}}%
    \put(0.70780939,0.10263626){\color[rgb]{0,0,0}\makebox(0,0)[lt]{\lineheight{1.25}\smash{\begin{tabular}[t]{l}\textit{.}\end{tabular}}}}%
    \put(0.71555057,0.10263626){\color[rgb]{0,0,0}\makebox(0,0)[lt]{\lineheight{1.25}\smash{\begin{tabular}[t]{l}50\end{tabular}}}}%
    \put(0,0){\includegraphics[width=\unitlength,page=8]{tetPolyStarHalfspaceIntersection.pdf}}%
    \put(0.78975328,0.10263626){\color[rgb]{0,0,0}\makebox(0,0)[lt]{\lineheight{1.25}\smash{\begin{tabular}[t]{l}1\end{tabular}}}}%
    \put(0.80368753,0.10263626){\color[rgb]{0,0,0}\makebox(0,0)[lt]{\lineheight{1.25}\smash{\begin{tabular}[t]{l}\textit{.}\end{tabular}}}}%
    \put(0.81142879,0.10263626){\color[rgb]{0,0,0}\makebox(0,0)[lt]{\lineheight{1.25}\smash{\begin{tabular}[t]{l}75\end{tabular}}}}%
    \put(0,0){\includegraphics[width=\unitlength,page=9]{tetPolyStarHalfspaceIntersection.pdf}}%
    \put(0.88563146,0.10263626){\color[rgb]{0,0,0}\makebox(0,0)[lt]{\lineheight{1.25}\smash{\begin{tabular}[t]{l}2\end{tabular}}}}%
    \put(0.89956576,0.10263626){\color[rgb]{0,0,0}\makebox(0,0)[lt]{\lineheight{1.25}\smash{\begin{tabular}[t]{l}\textit{.}\end{tabular}}}}%
    \put(0.90730702,0.10263626){\color[rgb]{0,0,0}\makebox(0,0)[lt]{\lineheight{1.25}\smash{\begin{tabular}[t]{l}00\end{tabular}}}}%
    \put(0,0){\includegraphics[width=\unitlength,page=10]{tetPolyStarHalfspaceIntersection.pdf}}%
    \put(0.04824818,0.15219671){\color[rgb]{0,0,0}\makebox(0,0)[lt]{\lineheight{1.25}\smash{\begin{tabular}[t]{l}0\end{tabular}}}}%
    \put(0.06218246,0.15219671){\color[rgb]{0,0,0}\makebox(0,0)[lt]{\lineheight{1.25}\smash{\begin{tabular}[t]{l}\textit{.}\end{tabular}}}}%
    \put(0.06992367,0.15219671){\color[rgb]{0,0,0}\makebox(0,0)[lt]{\lineheight{1.25}\smash{\begin{tabular}[t]{l}0\end{tabular}}}}%
    \put(0,0){\includegraphics[width=\unitlength,page=11]{tetPolyStarHalfspaceIntersection.pdf}}%
    \put(0.04824818,0.22840441){\color[rgb]{0,0,0}\makebox(0,0)[lt]{\lineheight{1.25}\smash{\begin{tabular}[t]{l}0\end{tabular}}}}%
    \put(0.06218246,0.22840441){\color[rgb]{0,0,0}\makebox(0,0)[lt]{\lineheight{1.25}\smash{\begin{tabular}[t]{l}\textit{.}\end{tabular}}}}%
    \put(0.06992367,0.22840441){\color[rgb]{0,0,0}\makebox(0,0)[lt]{\lineheight{1.25}\smash{\begin{tabular}[t]{l}2\end{tabular}}}}%
    \put(0,0){\includegraphics[width=\unitlength,page=12]{tetPolyStarHalfspaceIntersection.pdf}}%
    \put(0.04824818,0.3046121){\color[rgb]{0,0,0}\makebox(0,0)[lt]{\lineheight{1.25}\smash{\begin{tabular}[t]{l}0\end{tabular}}}}%
    \put(0.06218246,0.3046121){\color[rgb]{0,0,0}\makebox(0,0)[lt]{\lineheight{1.25}\smash{\begin{tabular}[t]{l}\textit{.}\end{tabular}}}}%
    \put(0.06992367,0.3046121){\color[rgb]{0,0,0}\makebox(0,0)[lt]{\lineheight{1.25}\smash{\begin{tabular}[t]{l}4\end{tabular}}}}%
    \put(0,0){\includegraphics[width=\unitlength,page=13]{tetPolyStarHalfspaceIntersection.pdf}}%
    \put(0.04824818,0.38081979){\color[rgb]{0,0,0}\makebox(0,0)[lt]{\lineheight{1.25}\smash{\begin{tabular}[t]{l}0\end{tabular}}}}%
    \put(0.06218246,0.38081979){\color[rgb]{0,0,0}\makebox(0,0)[lt]{\lineheight{1.25}\smash{\begin{tabular}[t]{l}\textit{.}\end{tabular}}}}%
    \put(0.06992367,0.38081979){\color[rgb]{0,0,0}\makebox(0,0)[lt]{\lineheight{1.25}\smash{\begin{tabular}[t]{l}6\end{tabular}}}}%
    \put(0,0){\includegraphics[width=\unitlength,page=14]{tetPolyStarHalfspaceIntersection.pdf}}%
    \put(0.04824818,0.45702749){\color[rgb]{0,0,0}\makebox(0,0)[lt]{\lineheight{1.25}\smash{\begin{tabular}[t]{l}0\end{tabular}}}}%
    \put(0.06218246,0.45702749){\color[rgb]{0,0,0}\makebox(0,0)[lt]{\lineheight{1.25}\smash{\begin{tabular}[t]{l}\textit{.}\end{tabular}}}}%
    \put(0.06992367,0.45702749){\color[rgb]{0,0,0}\makebox(0,0)[lt]{\lineheight{1.25}\smash{\begin{tabular}[t]{l}8\end{tabular}}}}%
    \put(0,0){\includegraphics[width=\unitlength,page=15]{tetPolyStarHalfspaceIntersection.pdf}}%
    \put(0.04824818,0.53323518){\color[rgb]{0,0,0}\makebox(0,0)[lt]{\lineheight{1.25}\smash{\begin{tabular}[t]{l}1\end{tabular}}}}%
    \put(0.06218246,0.53323518){\color[rgb]{0,0,0}\makebox(0,0)[lt]{\lineheight{1.25}\smash{\begin{tabular}[t]{l}\textit{.}\end{tabular}}}}%
    \put(0.06992367,0.53323518){\color[rgb]{0,0,0}\makebox(0,0)[lt]{\lineheight{1.25}\smash{\begin{tabular}[t]{l}0\end{tabular}}}}%
    \put(0,0){\includegraphics[width=\unitlength,page=16]{tetPolyStarHalfspaceIntersection.pdf}}%
    \put(0.65352399,0.47831553){\color[rgb]{0,0,0}\makebox(0,0)[lt]{\lineheight{1.25}\smash{\begin{tabular}[t]{l}$\alpha_{tet}$\end{tabular}}}}%
    \put(0.65194009,0.44038263){\color[rgb]{0,0,0}\makebox(0,0)[lt]{\lineheight{1.25}\smash{\begin{tabular}[t]{l}$\alpha_{cube}$\end{tabular}}}}%
    \put(0.65271674,0.40338425){\color[rgb]{0,0,0}\makebox(0,0)[lt]{\lineheight{1.25}\smash{\begin{tabular}[t]{l}$\alpha_{star}$\end{tabular}}}}%
    \put(0.8206887,0.21378302){\color[rgb]{0,0,0}\makebox(0,0)[lt]{\begin{minipage}{0.36552492\unitlength}\raggedright \end{minipage}}}%
    \put(0.05673388,0.04264139){\color[rgb]{0,0,0}\makebox(0,0)[lt]{\lineheight{1.25}\smash{\begin{tabular}[t]{l}Distance from $\mathbf{p}_0 \in \mathcal{P}_{c,\Sigma}$ from \cref{eqn:cellpoints-sorted}.\end{tabular}}}}%
    \put(0.00768806,0.58868547){\color[rgb]{0,0,0}\makebox(0,0)[lt]{\begin{minipage}{1.03678936\unitlength}\raggedright \end{minipage}}}%
  \end{picture}%
\endgroup%

%% file: figures/MoFerror-00.pdf_tex
\begingroup%
  \makeatletter%
  \providecommand\color[2][]{%
    \errmessage{(Inkscape) Color is used for the text in Inkscape, but the package 'color.sty' is not loaded}%
    \renewcommand\color[2][]{}%
  }%
  \providecommand\transparent[1]{%
    \errmessage{(Inkscape) Transparency is used (non-zero) for the text in Inkscape, but the package 'transparent.sty' is not loaded}%
    \renewcommand\transparent[1]{}%
  }%
  \providecommand\rotatebox[2]{#2}%
  \newcommand*\fsize{\dimexpr\f@size pt\relax}%
  \newcommand*\lineheight[1]{\fontsize{\fsize}{#1\fsize}\selectfont}%
  \ifx\svgwidth\undefined%
    \setlength{\unitlength}{170.07874016bp}%
    \ifx\svgscale\undefined%
      \relax%
    \else%
      \setlength{\unitlength}{\unitlength * \real{\svgscale}}%
    \fi%
  \else%
    \setlength{\unitlength}{\svgwidth}%
  \fi%
  \global\let\svgwidth\undefined%
  \global\let\svgscale\undefined%
  \makeatother%
  \begin{picture}(1,1)%
    \lineheight{1}%
    \setlength\tabcolsep{0pt}%
    \put(0,0){\includegraphics[width=\unitlength,page=1]{MoFerror-00.pdf}}%
    \put(0.2640836,0.01316408){\color[rgb]{0,0,0}\makebox(0,0)[lt]{\lineheight{1.25}\smash{\begin{tabular}[t]{l}$\tilde{\Omega}^+$\end{tabular}}}}%
    \put(0,0){\includegraphics[width=\unitlength,page=2]{MoFerror-00.pdf}}%
    \put(0.18722813,0.87208458){\color[rgb]{0,0,0}\makebox(0,0)[lt]{\lineheight{1.25}\smash{\begin{tabular}[t]{l}$\Omega_k$\end{tabular}}}}%
  \end{picture}%
\endgroup%

%% file: figures/MoFerror-01.pdf_tex
\begingroup%
  \makeatletter%
  \providecommand\color[2][]{%
    \errmessage{(Inkscape) Color is used for the text in Inkscape, but the package 'color.sty' is not loaded}%
    \renewcommand\color[2][]{}%
  }%
  \providecommand\transparent[1]{%
    \errmessage{(Inkscape) Transparency is used (non-zero) for the text in Inkscape, but the package 'transparent.sty' is not loaded}%
    \renewcommand\transparent[1]{}%
  }%
  \providecommand\rotatebox[2]{#2}%
  \newcommand*\fsize{\dimexpr\f@size pt\relax}%
  \newcommand*\lineheight[1]{\fontsize{\fsize}{#1\fsize}\selectfont}%
  \ifx\svgwidth\undefined%
    \setlength{\unitlength}{170.07874016bp}%
    \ifx\svgscale\undefined%
      \relax%
    \else%
      \setlength{\unitlength}{\unitlength * \real{\svgscale}}%
    \fi%
  \else%
    \setlength{\unitlength}{\svgwidth}%
  \fi%
  \global\let\svgwidth\undefined%
  \global\let\svgscale\undefined%
  \makeatother%
  \begin{picture}(1,1)%
    \lineheight{1}%
    \setlength\tabcolsep{0pt}%
    \put(0,0){\includegraphics[width=\unitlength,page=1]{MoFerror-01.pdf}}%
    \put(0.01745253,0.00031653){\color[rgb]{0,0,0}\makebox(0,0)[lt]{\lineheight{1.25}\smash{\begin{tabular}[t]{l}$\tilde{\Omega}^+_k = \tilde{\Omega}^+\cap \Omega_k$\end{tabular}}}}%
    \put(0,0){\includegraphics[width=\unitlength,page=2]{MoFerror-01.pdf}}%
    \put(0.18722813,0.87208458){\color[rgb]{0,0,0}\makebox(0,0)[lt]{\lineheight{1.25}\smash{\begin{tabular}[t]{l}$\Omega_k$\end{tabular}}}}%
    \put(0,0){\includegraphics[width=\unitlength,page=3]{MoFerror-01.pdf}}%
    \put(0.12129325,0.28720908){\color[rgb]{0,0,0}\makebox(0,0)[lt]{\lineheight{1.25}\smash{\begin{tabular}[t]{l}$\Halfspace_k$\end{tabular}}}}%
    \put(0,0){\includegraphics[width=\unitlength,page=4]{MoFerror-01.pdf}}%
    \put(0.07450543,0.67694571){\color[rgb]{0,0,0}\makebox(0,0)[lt]{\lineheight{1.25}\smash{\begin{tabular}[t]{l}$\tilde{\Omega}^+_k \cap - \Halfspace_k$\end{tabular}}}}%
    \put(0,0){\includegraphics[width=\unitlength,page=5]{MoFerror-01.pdf}}%
    \put(0.58860821,0.89159674){\color[rgb]{0,0,0}\makebox(0,0)[lt]{\lineheight{1.25}\smash{\begin{tabular}[t]{l}$\Omega_k \cap \Halfspace_k \setminus \tilde{\Omega}^+_k $\end{tabular}}}}%
  \end{picture}%
\endgroup%

%% file: tables/Ahn2007.tex
\begin{tabular}{lllll}
\toprule
   &    &    Youngs &     LVIRA &       MoF \\
N & Value types &           &           &           \\
\midrule
4  & $E_r^2$ &  6.19e-02 &  1.53e-01 &  1.43e-02 \\
   & O &      - &      - &      - \\
8  & $E_r^2$ &  4.38e-03 &  5.35e-03 &  3.64e-03 \\
   & O &   3.82048 &   4.83813 &   1.97382 \\
16 & $E_r^2$ &  2.25e-03 &  1.34e-03 &  9.43e-04 \\
   & O &  0.964419 &   1.99247 &   1.94884 \\
32 & $E_r^2$ &  1.11e-03 &  3.12e-04 &  2.33e-04 \\
   & O &     1.012 &   2.10852 &   2.01529 \\
\bottomrule
\end{tabular}

%% file: tables/CLCIR-recon.tex
\begin{tabular}{llllll}
\toprule
    &    &    Youngs &       LSF &     CLCIR &  CLC-CBIR \\
N & Value type &           &           &           &           \\
\midrule
10  & $E_r^1$ &  1.89e-03 &  1.92e-03 &  2.38e-03 &  2.43e-03 \\
    & O &      1.84 &      2.01 &      2.11 &      2.11 \\
20  & $E_r^1$ &  5.28e-04 &  4.77e-04 &  5.50e-04 &  5.64e-04 \\
    & O &      1.45 &      2.00 &      2.08 &      2.12 \\
40  & $E_r^1$ &  1.93e-04 &  1.19e-04 &  1.30e-04 &  1.30e-04 \\
    & O &      1.17 &      2.00 &      2.01 &      2.03 \\
80  & $E_r^1$ &  8.60e-05 &  2.98e-05 &  3.23e-05 &  3.18e-05 \\
    & O &      1.06 &      2.00 &      2.01 &      2.02 \\
160 & $E_r^1$ &  4.12e-05 &  7.46e-06 &  8.00e-06 &  7.82e-06 \\
    & O &      1.02 &      - &      2.00 &      2.00 \\
320 & $E_r^1$ &  2.03e-05 &  - &   2.00e-06 &      1.95e-06 \\
    & O &      - &      - &       - &      - \\
\bottomrule
\end{tabular}

%% file: sections/volume-fraction-advection.tex
\section{Volume fraction advection}
\label{sec:advect}

\subsection{Geometric Volume-of-Fluid methods}

\textcolor{Reviewer1}{Early developments of volume fraction advection algorithms have reduced the complexity of geometric operations by adopting a \emph{dimensionally split} approach. The split advection algorithm solves the volume fraction equation in $\Dimension$ steps, where $\Dimension$ is the spatial dimension. Volume fraction values are updated once per splitting step, requiring an additional interface approximation (reconstruction) in \textcolor{Reviewer1R1}{each of these steps}. Computing $D$ interface reconstructions and advection steps per time step in $D$ dimensions \textcolor{Reviewer1R1}{increases} the computational cost of the dimensionally split approach. Splitting additionally requires alignment of normal area vectors with coordinate axes and is less volume conservative than the un-split approach. Unstructured meshes do not fulfill this requirement, because, in general, face-normal vectors in an unstructured mesh are not collinear with coordinate axes. A detailed review of dimensionally split algorithms on structured meshes is given by \citep{Tryggvason2011}, and more details on dimensionally-split algorithms are available in \citep{Kothe1996, Rider1998, Scardovelli1999, Scardovelli2003, Aulisa2007}. In their comprehensive review of methods for \ac{DNS} of two-phase flows, \citet{Scardovelli1999} stated that the dimensionally un-split algorithm ensures better accuracy compared to split algorithm, especially regarding asymmetries in the shape of the advected interface. Here, un-split geometric Volume-of-Fluid advection algorithms are covered in historical order.}

%
\begin{figure}[htb] 
    \centering
    \def\svgwidth{0.5\textwidth}
       {\footnotesize
        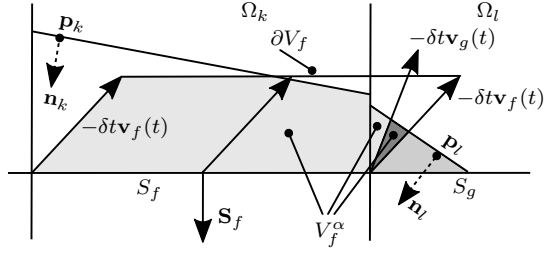
       }
   \caption{A schematic representation of the $2D$ Rider-Kothe algorithm \citep{Rider1998}.}
   \label{fig:riderkothe}
\end{figure}

\textcolor{Reviewer1R1}{In addition to the \ac{PLIC} interface reconstruction covered in \cref{sec:recon}, all flux-based geometric Volume-of-Fluid methods share the same approach for approximatively solving \cref{eqn:phase-specific}. As described in \cref{sec:recon}, PLIC reconstruction approximates $\Sigma = \partial \Omega^+(\tstart)$ as a piecewise-planar surface, i.e.\ a halfspace $\Halfspace_k:=(\PlicPosition(\tstart),\PlicNormal(\tstart))$ in each cell $\Omega_k$ (cf. \cref{fig:riderkothe}). The phase-specific volume $\PhaseFluxVolumeContrib$, fluxed through the face $S_f$ of the cell $\Omega_k$, is approximated using \cref{eq:phase-fluxvol}, 
\begin{equation}
    \PhaseFluxVolumeContrib = V_f \cap \Omega^+(\tstart) \approx \bigcup_{l \in \CellNeighborhood(S_f)} \tilde{\Omega}^+_l(\tstart) \cap V_f. 
    \label{eqn:phasevol-union}
\end{equation}
The volume $V_f$ is computed by differently by each geometrical VOF method by approximating the r.h.s. of \cref{eq:fluxvol} and  
\begin{equation}
    \CellNeighborhood(S_f) = \{ l \in L : S_f \cap \Omega_l \ne \emptyset \},
\end{equation}
is the face-cell stencil of the face $S_f$ in $\Cells$. Furthermore, the volume $\tilde{\Omega}^+_l(\tstart)$ in \cref{eqn:phasevol-union} is approximated using the PLIC halfspace in each cell $\Omega_l$ from $\CellNeighborhood(S_f)$ as 
\begin{equation}
    \tilde{\Omega}_l^+(\tstart) = 
        \begin{cases} 
            \Halfspace_l(\tstart) \cap \Omega_l & \quad \text {if } 0 < \VolFrac_l(\tstart) < 1, \\
            \Omega_l & \quad \text{if } \VolFrac_l(\tstart) = 1, \\
            \emptyset & \quad \text{if } \VolFrac_l(\tstart) = 0.
        \end{cases}
    \label{eqn:phasevol-approx}
\end{equation}
Inserting \ref{eqn:phasevol-approx} in \ref{eqn:phasevol-union}, and the result into \cref{eqn:phase-specific}, approximatively solves \cref{eqn:volfrac-ex} for $\{\VolFrac_k(t^{n+1})\}_{k \in K}$. Available un-split VOF advection algorithms differ in the way they approach the geometrical approximations of $\FluxVolume$ and $\PhaseFluxVolumeContrib$.}

\citet{Kothe1996} and later \citet{Rider1998} proposed a two-dimensional Eulerian flux-based un-split algorithm that uses a constant velocity distribution across an edge (face in \textcolor{Reviewer1}{$3D$}) $\U_f$ to construct the flux volume $\FluxVolume$ and consequently the fluxed phase-specific volume $\PhaseFluxVolumeContrib$. 

However, they noted that \citet{Pilliod2004} were the first to develop a directionally un-split multidimensional algorithm (\citep[page 3, table 1]{Kothe1996},\citep[page 6, table 1]{Rider1998}). The computation of the fluxed phase-specific volumes by the \ac{RKA} is shown schematically in \cref{fig:riderkothe}. Compared to using point velocities, the use of constant velocities by the \ac{RKA} causes an overlap between the flux volumes and subsequently the fluxed phase-specific volumes for two point-adjacent edges, i.e.\  faces. The overlap is \emph{fluxed twice} and shown schematically as the shaded triangle in \cref{fig:riderkothe}. Fluxing the same volume multiple times this way causes \emph{overshoots} and \emph{undershoots}. An overshoot is defined for each cell $c$ as
\begin{equation}
    \Overshoot = \max(\PlicFraction - 1, 0)
\end{equation}
and an undershoot is defined as 
\begin{equation}
    \Undershoot = \max(-\PlicFraction, 0).
\end{equation}
If \textcolor{Reviewer1R1}{$\Overshoot = \Undershoot = 0, \forall k \in K$, volume fractions $\{\VolFrac_k\}_{k \in K}$ are said to be "numerically bounded" and the method is stable.} Undershoots and overshoots have been handled in the \ac{RKA} using explicit conservative redistribution of the volume fraction $\VolFrac$. The second-order convergent reconstruction algorithm \ac{ELVIRA} \citep{Pilliod2004} has been used to reconstruct the \ac{PLIC} interface. Since \citet{Rider1998} have used edge (face) centered velocities to construct the flux volumes, there is no need to correct the geometrical flux volumes for volume conservation. However, this is only true if the edge-centered velocity field upholds the discrete divergence-free condition $\sum_f F_f = 0$. \citet{Rider1998} do propose a correction for volume conservation, but for different reasons: they expect the face-centered velocity field to uphold the discrete divergence-free condition only \emph{up to a specified tolerance of a linear solver used to obtain it}. They were aiming at applying their algorithm with velocity fields that result from an approximated solution of the incompressible single-field \ac{NS} equation system. In that case, the following correction is necessary:
\begin{equation}
  \partial_t \VolFrac + \nabla \cdot{(\U\VolFrac)} = \VolFrac\div{\U}. 
  \label{eqn:rk:corr}
\end{equation}
However, if $\U_f$ is exactly divergence-free, no correction for volume conservation is required for the geometrical flux volumes used by the \ac{RKA}. Using an edge (face) centered velocity simplifies the geometrical form of the geometrical flux volume because it remains convex. Still, the velocity field varies over the edge (face), and the \ac{RKA} assumes \emph{a constant} flux velocity over the edge (face). With this assumption, the flux polyhedron can only be non-convex if the face of the control volume is non-convex. The assumption of the constant edge velocity simplifies geometric operations for computing phase volume contributions, especially in three dimensions. The authors proposed two variants of their algorithm: a fully dimensionally un-split algorithm, and the un-split variant constructed from the dimensionally split algorithm. In the latter algorithm, overlaps of phase-specific volumes that appear at cell corners are corrected by additional intersections. In both cases the cell corner velocities are determined from the edge center velocities based on their signs. \citet{Rider1998} have shown that their un-split algorithm is more accurate than the operator split variant.  

\citet{Mosso1997} were the first to propose a cell-based (re-mapping) \ac{LE} method for the dimensionally un-split volume fraction advection that uses a forward projection. They reported a test case involving a translation of a circle on an unstructured irregular hexahedral mesh using a time-periodic and spatially constant velocity field that moves the circle back to the original position. The algorithm shows promising results \citep[figure 5]{Mosso1997} in the fact that the shape and the area of the circle are maintained, even on a non-orthogonal unstructured hexahedral mesh. 

\citet{Mosso1996} described the application of their \ac{LE} method to the problem of a rotating planar and circular interface and the numerical errors that are related to exact, forward Euler, backward Euler and trapezoidal integration of mesh point displacements. They concluded that the use of the trapezoidal integration for the forward projection step of the \ac{LE} method removes artificial expansion and contraction of the interface. In other words, the trapezoidal integration of point displacements conserves volume - a conclusion that is also drawn later by \citet{Chenadec2013}. This conclusion is a direct consequence of the fact that the first-order accurate Euler quadrature exhibits \emph{artificial error canceling when applied to harmonic functions} \citep{Wiedeman2002}. 

\begin{figure}[htb] 
  \centering
  \def\svgwidth{0.5\textwidth}
   {\footnotesize
     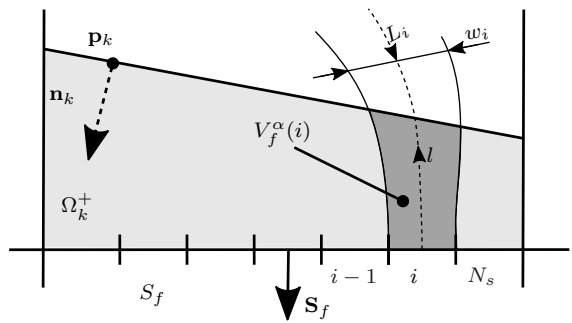
   }
  \caption{A schematic representation of the $2D$ stream scheme \citep{Harvie2000}.}
  \label{fig:streamscheme}
\end{figure}

\citet{Harvie2000} have proposed the \emph{stream scheme}: a two-dimensional Eulerian flux-based \GVOF{} that uses a continuous velocity field approximation \emph{within a cell}, and a geometrically reconstructed interface to compute the fluxed phase-specific volumes by approximating the flux volume $\FluxVolume$ as a set of \emph{stream tubes}. The velocity field is given by a streamline function formulated as 
\textcolor{Reviewer1R1}{\begin{equation}
  \Streamline(x,y) = \chi_b yx + \chi_xx + \chi_yy,
  \label{eqn:streamscheme}
\end{equation}leading to
\begin{equation}
  \U = 
  \begin{pmatrix}
    \partial_y \Streamline \\ 
    -\partial_x \Streamline
    \end{pmatrix}
  = \begin{pmatrix}
  \chi_bx + \chi_y  \\
  -\chi_by - \chi_x 
  \end{pmatrix}.
\end{equation}}\textcolor{Reviewer1R1}{Given by a stream function, this velocity field is automatically divergence-free.} This velocity field is continuous in the face-normal direction, while it is discontinuous in the tangential direction and at cell corner points. Fluxed phase-specific volumes are calculated based on stream tubes given by the velocity field and a discretization of the face in $N_s$ segments, as shown in \cref{fig:streamscheme}. Streamlines define the stream tube of the fluid particle that crosses a face with coordinate $l$ and width $w$. The stream tube width is determined by $N_s$ and the velocity field $\U$ from the volume conservation law \cite{Harvie2000}. Stream tube geometry is not explicitly approximated. Instead, fluxed phase-specific volumes are integrals of the phase indicator function along the streamline. \textcolor{Reviewer1R1}{The \ac{PLIC} interface approximates the phase indicator function by approximating $\Omega_k^+(\tstart)$ from \cref{eq:alphakdef}.}

\noindent\textcolor{Reviewer1R1}{The magnitude of the fluxed phase-specific volume is approximated by the stream scheme as}
\textcolor{Reviewer1R1}{\begin{equation}
    |\PhaseFluxVolumeContrib| = \sum_{i = 1}^{N_s} |\PhaseFluxVolumeContrib(i)| = \sum_{i = 1}^{N_s} \int_0^{L_i} w_i(l)\Indicator(l) dl, 
\end{equation}}where $\Indicator$ is the parameterized phase indicator function. \textcolor{Reviewer1}{The accuracy of the stream scheme is comparable to \ac{RKA} \citep{Rider1998} for the reversed single vortex test case. A second-order reconstruction algorithm \ac{ELVIRA} increases the accuracy of the advection. First-order convergent Youngs' reconstruction algorithm causes errors in handling thin filaments, due to the instabilities in the interface orientation. These instabilities are amplified and cause the interface to break up artificially. The accuracy and computational cost of the stream scheme is highly dependent on $N_s$. Using $N_s=10$ makes the computational efforts comparable to other dimensionally un-split algorithms. Results also show that the scheme suffers from \emph{wisps}. Wisps are very small errors in $\PlicFraction$ in cells that should be either completely empty or completely full. Wisps have orders of magnitude lower values than jetsam and flotsam first described by \citet{Noh1976} and later by \citet{Rider1998}. \citet{Harvie2000} handle wisps by applying a conservative wisp redistribution algorithm that takes into account the direction of the interface-normal vector. The conservative wisp redistribution algorithm redistributes wisps within the structured $27$ cell stencil of a multi-material cell, making it the only point in the stream scheme dependent on the \ac{CFL} condition.}

\begin{figure}[htb] 
  \centering
    \def\svgwidth{0.45\textwidth}
       {\footnotesize
         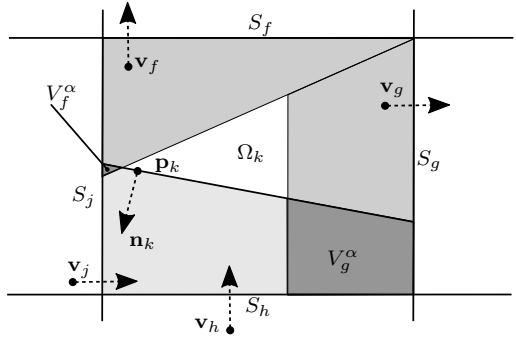
       }
       \caption{A schematic representation of the two-dimensional DDR scheme \citep{Harvie2001}.}
  \label{fig:ddrscheme}
\end{figure}


\textcolor{Reviewer1}{\citet{Harvie2001} have proposed the two-dimensional \ac{DDR} scheme on Cartesian meshes, that improves volume conservation compared to \citep{Rider1998,Harvie2000} using unique slopes for the donating regions at cell corners. The \ac{DDR} scheme constructs \emph{defined donating regions} shown schematically in \cref{fig:ddrscheme}, for all faces of a cell $c$ whose velocities are directed outward from the cell.\ The donating regions are intersected with the \ac{PLIC} interface to compute the fluxed phase-specific volume for each face.  Faces $f,g$ and $h$ are labeled in \cref{fig:ddrscheme} to distinguish their respective velocities. The volume of the donating region is a result of the \emph{total} volume conservation for the cell $c$. The conservation of the total volume for the cell $c$ indirectly introduces the \ac{CFL} criterion into the scheme. Preventing the characteristic overlap of the flux volumes in the \ac{RKA} (\cref{fig:riderkothe}) improves the volume conservation of the \ac{DDR} scheme together with the correction of the donating region for volume conservation. However, limiting the donating regions to a cell prevents the scheme from \emph{fluxing around the corner}.}

\citet{Cerne2002} have analyzed the numerical errors of the \GVOF{}. They have quantified the reconstruction errors for two-dimensional simulations on structured meshes in the form of \emph{reconstruction correctness}. They have also described the artificial distortion of the circular interface during translation on coarse meshes. Strongly under-resolved interfaces exhibit artificially high advection velocities. \citet{Cerne2002} propose the local dynamic \ac{AMR} \textcolor{Reviewer1R1}{for increasing overall accuracy based on a reconstruction-correctness criterion.}

\citet*{Scardovelli2003} have formalized their $2D$ \ac{EI-LE} method, proposed originally by \citet*{Aulisa2003}. The formalization models the steps of the \ac{EI-LE} scheme using linear maps. \citet{Scardovelli2003} extend the original \ac{EI-LE} scheme as a dimensionally un-split scheme that relies on a single advection and reconstruction step per time step. The scheme still requires the face-normal vectors to be aligned with the coordinate axes, because the method is structured. The \ac{EI-LE} is the first scheme that reports the absence of wisps and conserves the total area exactly. A result that deserves special attention is the reversed vortex test case \citep[figure 10]{Aulisa2003} with no visible wisps in the solution and a reduced geometrical advection error compared to previous methods, including the \ac{DDR} scheme. The extension of the \ac{EI-LE} method to unstructured meshes has been suggested by \citet*{Aulisa2003}, which requires a triangulation of the domain and a continuous area-preserving linear mapping. 

\begin{figure}[htb] 
  \centering
    \def\svgwidth{0.6\textwidth}
       {\footnotesize
         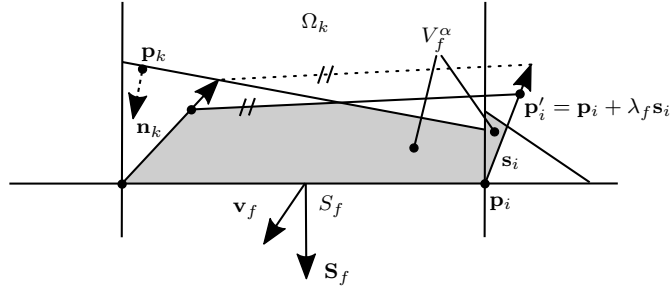
       }
       \caption{\textcolor{Reviewer1R1}{A schematic representation of the EMFPA scheme.}}
  \label{fig:emfpa}
\end{figure}

\citet{Lopez2004} have combined the volume conservation and numerical boundedness of the \ac{DDR} method by \citet{Harvie2000} and the around-the-corner flux calculation of the \ac{RKA} method by \citet{Rider1998} into a new \ac{EMFPA} scheme. Shown schematically in \cref{fig:emfpa}, the flux volume $\FluxVolume$ is defined by the positions of the swept face points $\Point_i'$ computed as 
\begin{equation}
  \Point_i' = \Point_i + \lambda_f\Slope_i,
\end{equation}
where $\lambda_f$ is a face-constant scalar coefficient and $\Slope_i$ is the displacement along the discrete Lagrange trajectory, computed using the \ac{IDW} interpolation 
\textcolor{Reviewer1R1}{\begin{equation}
    \Slope_i = \dfrac{1}{\sum_{f \in C_f(\Point_i)} w_{f}} \sum_{f \in C_f(\Point_i)} w_{f} \U_f^{n + \frac{1}{2}},
  \label{eqn:idw-slope}
\end{equation}
where 
\begin{equation}
    C_f(\Point_i) = \{ f \in F : \Point_i \in S_f \}
\end{equation}
is the point-face stencil, the set of indices of those faces $S_f$ of cells $\Omega_k$, which contain $\Point_i$. The weight $w_f$ in \cref{eqn:idw-slope} is then defined as
\begin{equation}
  w_{f} = \dfrac {1}{\| \Point_i  - \x_f\|},
\end{equation}
where $\x_f$ is the centroid of the face $S_f$.} The face velocity $\U_f$ that contributes to the slope is evaluated in an \emph{intermediate time step} $n+\frac{1}{2} \equiv t + 0.5\delta t$: the velocity field at this time is obtained from a linear interpolation between the current and the next time step, $n$ and $n+1$, respectively. Using the \ac{IDW} interpolation to compute the slope $\Slope_i$ of the discrete Lagrange trajectories introduces an interpolation error. On Cartesian equidistant meshes used by \citet{Lopez2004}, \cref{eqn:idw-slope} represents an arithmetic average, since all the distances from corner points to face centers are the same. The slope of the discrete Lagrange trajectory influences the overall accuracy of the scheme. The slope $\Slope_i$ and the face-constant parameter $\lambda_f$ determine the magnitude of the geometrical flux volume $\FluxVolume$ shown as a lightly shaded polygon in \cref{fig:emfpa}. \textcolor{Reviewer1R1}{Temporal second-order accuracy of $\FluxVolume$ is achieved by integrating \cref{eq:fluxvolmag} with a trapezoidal quadrature, namely}
\begin{equation}
  \FluxVolume = \int_t^{\tau + \delta \tau} \int_{S_f} \U(t) \cdot \vec{n} \, dS \, dt \approx 0.5 \delta t \left(\U_f^n + \U_f^{n+1} \right) \cdot \S_f.
  \label{eqn:fmfpa-fluxvol}
\end{equation}
\textcolor{Reviewer1R1}{Equation (\ref{eq:fluxvolmag}) must also be exactly satisfied in the discrete sense to ensure volume conservation.} This is achieved by introducing a face-constant parameter $\lambda_f$ that scales the flux volume. Calculation of the fluxed phase-specific volumes $\PhaseFluxVolumeContrib$ is done geometrically by approximating the solution of \cref{eq:phase-fluxvol}, 

Volume conservation is improved by the \ac{EMFPA} algorithm proposed by \citet{Lopez2004}, compared to the original \ac{RKA} algorithm proposed by \citet{Rider1998}, because of the reduction of the overlap between neighboring flux volumes. However, the face-constant volume conservation adjustment coefficient $\lambda_f$ may cause overshoots and undershoots. 
\citet{Lopez2004} show that their \ac{EMFPA} algorithm coupled with the \ac{SIR} algorithm results in an overall second-order convergence. They also emphasize the need for a second-order convergent reconstruction algorithm, because of its strong influence on the overall error. They do not quantify errors in volume conservation and numerical stability. The authors mention that some overshoots appear, but only in \textcolor{Reviewer1R1}{cells} where the slopes $\Slope_i$ are almost orthogonal to $\S_f$, in which case a local conservative redistribution algorithm is applied. Additionally, they state that wisps appear, but they do not have an effect on the computational efficiency of the algorithm. The effect of wisps on numerical stability is not addressed. 

A dimensionally un-split algorithm proposed by \citet{Pilliod2004} is based on the work of \citet*{Bell1988} and relies on the method of characteristics to integrate the flux volumes in time with either first or second-order accuracy. The scheme is two-dimensional and uses a face-constant velocity for the calculation of the characteristic lines. Unlike the \ac{DDR} scheme, and similar to the \ac{RKA} method, this method allows the calculation of the around-the-corner fluxes. \citet{Pilliod2004} reported volume conservation errors near machine epsilon for a translation of a circle and the rotation slotted disc by \citet{Zalesak1979}. \citet{Pilliod2004} emphasized the importance of using a second-order accurate reconstruction algorithm. Verification cases involving both spatially and temporally varying velocity fields are not presented, as well as the extension to three dimensions. 

\citet[section 4.1]{Dyadechko2005} have relied on the \acf{LE} \GVOF{} for advecting the volume fraction field in their \acf{MoF} method. Cell corner points are traced backward in time to compute $\VolFracskend$, using a $4$th-order \ac{RK} scheme. The method is two-dimensional, and the authors propose the use of \emph{bins} (\acf{AABB}) that simplify the geometry of polygonal cells to increase efficiency when computing phase-specific volumes that contribute to $\VolFracskend$. Because the Lagrangian backward tracing does not conserve the volume of the traced cell, \citet{Dyadechko2005} rely on a local conservative redistribution for correcting overshoots, undershoots and wisps. 

\begin{figure}[htb] 
  \centering
    \def\svgwidth{0.5\textwidth}
       {\footnotesize
         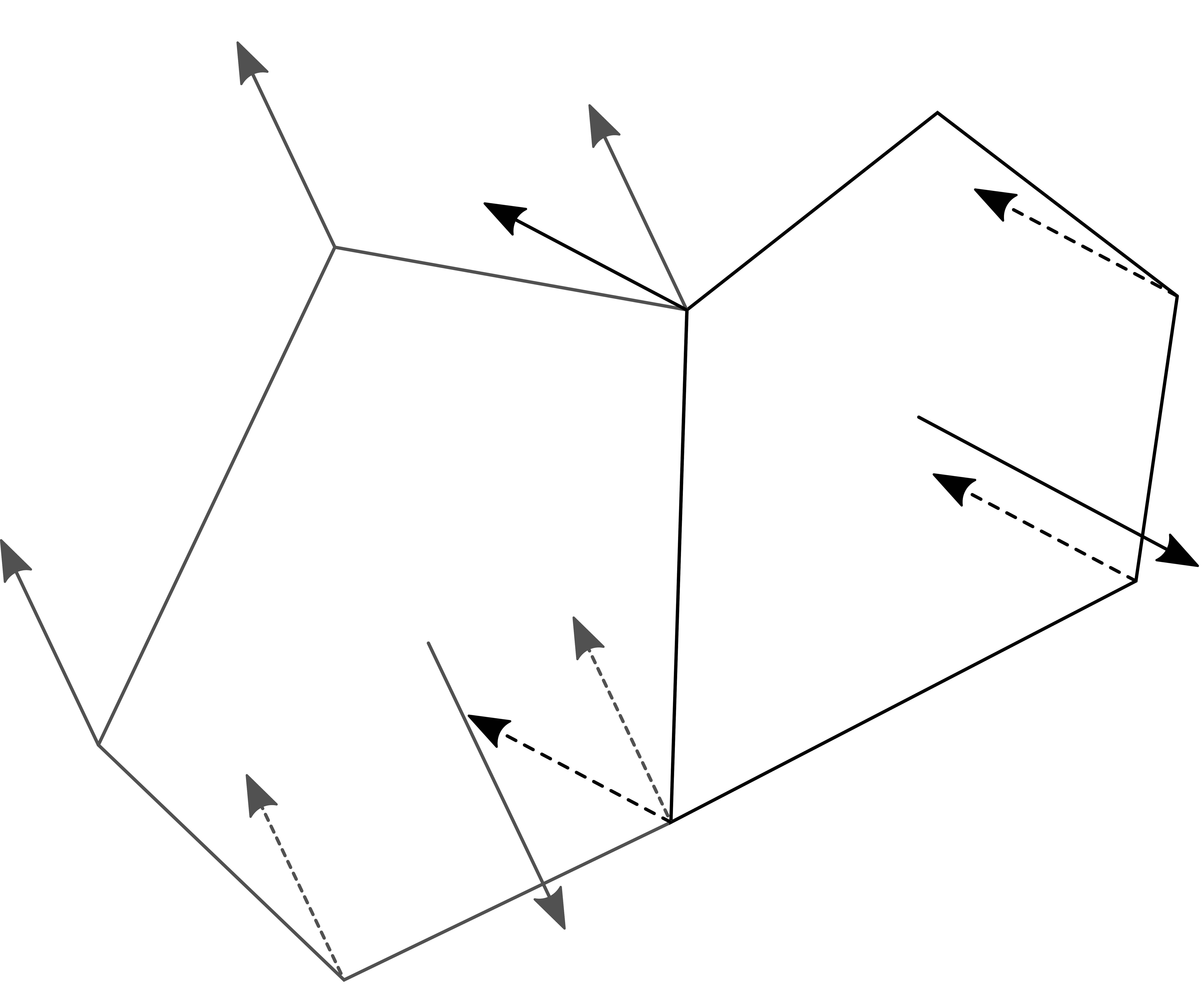
       }
       \caption{A schematic representation of the PCFSC advection scheme.} 
  \label{fig:pcfsc}
\end{figure}

\textcolor{Reviewer1}{\ac{PCFSC}, a three-dimensional extension of the \ac{RKA} of \citet{Rider1998} has been proposed by \citet{Liovic2006}. The \ac{PCFSC} algorithm has been developed on structured meshes, but it directly generalizes to unstructured meshes with convex cells. \Cref{fig:pcfsc} illustrates the \ac{PCFSC} algorithm. The dimensionally un-split advection is achieved by the flux-based un-split approach, with an important simplification of the flux volume: a single face-centered velocity vector $\U_f$ is used to construct the flux volume. A direct consequence of this is a \emph{flux volume bounded by planar polygons}. Consequentially, triangulation of the flux volume is not necessary, and, if $\U_f$ is divergence-free in the discrete sense, the flux volume does not have to be corrected to ensure volume conservation. Flux volumes bounded by planar polygons are additionally more easily intersected. However, as shown in \cref{fig:pcfsc}, using face-centered velocities $\U_f$ and $\U_g$ to sweep points of two edge-adjacent faces $f$ and $g$ results in non-unique Lagrange trajectories at cell-corner points $\Point_i$, which cause either overlaps or holes between flux volumes \emph{along the whole length of an edge}, shaded gray in \cref{fig:pcfsc}. Overlaps and holes between flux volumes cause overshoots, undershoots and wisps in $\VolFracskend$. To suppress these errors, \citet{Liovic2006} scale the fluxed phase volume with a scalar coefficient.}


\citet{Aulisa2007} have extended their \ac{EI-LE} scheme \citep{Aulisa2003,Scardovelli2003} to support three-dimensional computations on Cartesian meshes. The dimensionally split \ac{EILE-3D} and \ac{EILE-3DS} schemes conserve mass exactly for sphere translation and rotation test cases. \ac{EILE-3DS} delivers second-order convergence of the advection errors for the single vortex test case with a decreasing \ac{CFL} number. \citet{Aulisa2007} quantified volume conservation errors for spatially and temporally varying velocity fields. They show that the volume conservation errors of the \ac{EILE-3DS} method are converging from approximately $1e-03$ to $1e-06$ for their single vortex test case with increased mesh resolution and from $1e-04$ to $1e-09$ for the same test case with $32^3$ volumes and a decreasing \ac{CFL} number. 

\begin{figure}[t!] 
  \centering
    \def\svgwidth{0.45\textwidth}
       {\footnotesize
         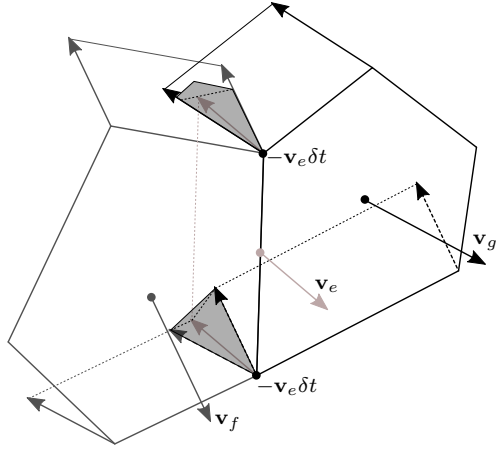
       }
       \caption{A schematic representation of the FMFPA-3D scheme.} 
  \label{fig:fmfpa}
\end{figure}

\textcolor{Reviewer1}{\Cref{fig:fmfpa} illustrates the flux volume calculation of the \ac{FMFPA-3D} advection scheme. The \ac{FMFPA-3D} scheme constructs a flux volume bounded by planar polygons, like the \ac{PCFSC} scheme. However, the \ac{FMFPA-3D} scheme goes one step further than the \ac{PCFSC} scheme, using a unique velocity at each \emph{edge center}. Each edge-centered velocity is interpolated from face-centered velocities using faces that share the edge. The faces of the flux volume created by sweeping the edges are planar, because the edge-centered velocity is constant for an edge. However, this does not solve the problem of overlaps or holes between flux volumes, because velocities are not unique at cell corner-points. For example, the edge $e$ with the edge-centered velocity $\U_e$ in \cref{fig:fmfpa} illustrates this issue: the end-points of this edge are shared with other edges, that have different velocities, so the velocities at cell corners are non-unique. The non-unique velocities cause overlaps or holes for flux volumes at cell corner points, shaded gray in \cref{fig:fmfpa} for the edge $e$. Compared to \ac{PCFSC} (\cref{fig:pcfsc}), overlaps of the \ac{FMFPA-3D} are much smaller. The flux volume shown in \cref{fig:fmfpa} still needs to be closed by connecting the swept edges of each swept face. The \ac{FMFPA-3D} scheme closes the flux volume with a plane in order to simplify its geometry by keeping the flux volume boundary planar. \citet{Hernandez2008} have show improved results compared to their $3D$ implementation of the \ac{RKA}.}

\textcolor{Reviewer1}{\citet{Zhang2008} have proposed the $2D$ \acf{PAM}, a $2D$ \ac{LE} \GVOF{} based on Lagrangian tracking and Eulerian remapping, similar to the \ac{MoF} method (\citep{Dyadechko2005,Ahn2009}). The \ac{PAM} scheme traces phase-specific volumes forward in time using the flow map (i.e. $\FlowMap{t^n}{t^{n+1}} (\Omega^+(t^n) \cap \Omega_k$)) and intersects the traced phase-specific volumes with the mesh $\Omega$ to compute $\VolFracskend$. Tracing points using an \ac{RK} scheme leads to errors in mass conservation, as velocities evaluated at the material polygon points are not divergence-free. \citet{Zhang2008} do not mention interpolation, so the points seem to be traced using the exact velocity $\U(x,t)$, in which case the errors resulting from velocity interpolation are avoided. After the tracing the material polygons (phase-specific volumes $\Omega^+ \cap \Cell_k$) forward, additional geometrical operations are required to correct material polygons to ensure volume conservation (\citep[section 3.2]{Zhang2008}), because the Lagrange advection is not inherently conserving volume. Topological changes are handled using merging algorithms for polygons (\citep[section 3.3]{Zhang2008}). Verification using temporally and spatially varying velocity fields shows that the \ac{PAM} method is more accurate than \ac{EMFPA} on structured equidistant meshes. The volume conservation within $Ev \in [1e-08,1e-05]$ depends on the mesh resolution (e.g. \citep[table 4]{Zhang2008}). The authors state that the method is overall approximately $20$\% slower than \ac{EMFPA}.}

\citet{Ahn2009} have extended the \ac{MoF} method proposed by \citet{Dyadechko2005} with local dynamic \acf{AMR} and show an overall second-order convergent solution for a set of standard verification cases, with different levels of local dynamic \ac{AMR}. Adaptive \ac{MoF} is implemented in $2D$ and the same problems in overshoots, undershoots and wisps are dealt with both local and global conservative error redistribution. \citet{Ahn2009ns} have coupled the \ac{MoF} method with the single-field two-phase \ac{NS} equation system.  

\textcolor{Reviewer1}{\citet{Zhang2013} proposes  the \acf{DRACS} scheme: a fourth-order accurate representation of the \acf{DR}. The increase in accuracy is due to flux-volumes having non-linear boundaries approximated with cubic splines. The \ac{DRACS} scheme uses (E)\ac{LVIRA} for piecewise linear interface reconstruction, so the fourth-order convergence is only shown for cases where no significant interface deformation occurs (solid body rotation, $2D$ shear with $T=0.5$s). In order to obtain an overall fourth-order accuracy for larger deformations, a higher-order volume conservative interface reconstruction should be used. \citet{Zhang2013} motivates the development of the higher-order \ac{DRACS} method by the need for an accurate curvature calculation required for two-phase flows. Only small interface deformation is presented, so it is not clear if topological changes and strongly deforming interfaces are handled robustly and accurately.}

\textcolor{Reviewer1}{\Citet{Chenadec2013} have proposed an alternative \ac{LE} \GVOF{} that is directly applicable to unstructured meshes with general polyhedral cells. \textcolor{Reviewer1R1}{\Citet{Chenadec2013} name their method \ac{HyLEM} and rely on \cref{eq:chitranspmat}: $\Omega_k$ is considered as a material volume.} This idea is interesting because it relies on mesh motion and PLIC reconstruction in order to compute  $\VolFracskend$, which simplifies somewhat the required intersection in $3D$. However, there are at least three difficulties in maintaining volume conservation. First, the condition given by \cref{eq:alphamat} is not upheld without correcting $\Phi_{t^n}^{t^n+1}(\Omega_k)$ for volume conservation, because discrete interpolated velocities are used to trace the cell $\Omega_k$ forward in time. Second, the one-to-many relationship required by \cref{eq:eulerremap}, together with the discontinuities of the \ac{PLIC} interface at cell faces, may cause volume conservation errors when computing necessary intersections in \cref{eq:eulerremap}. Third, a $3D$ calculation will create $\Phi_{t^n}^{t^n+1}(\Omega_k)$ as polyhedrons with non-planar faces, so the reconstruction algorithm needs to accommodate this. \citet{Chenadec2013} have left the generalization to $3D$ as future work and have shown for $2D$ Cartesian meshes that a second-order mass conservation error is achieved using higher-order \ac{RK} schemes. From numerical experiments, the authors have come to the conclusion that volume conservation errors systematically cancel out when the solutions of forward and backward Lagrangian backtracing are averaged, resulting in a combined forward/backward integration scheme for $\VolFracskend$. The effect of systematic cancellation is not surprising because harmonic velocity components are integrated \citep{Wiedeman2002}. \citet{Chenadec2013} reported third-order accurate volume conservation errors within $[1e-14,1e-05]$ for the trapezoidal rule for cases with strong interface deformation.}

\textcolor{Reviewer1}{\citet{Zhang2014} develop the \acf{iPAM}, a fourth-order accurate $2D$ method that replaces the cubic spline approximation for the nonlinear donating region boundary in the \ac{DRACS} scheme with multiple piecewise-linear segments. The \ac{iPAM} has the highest absolute accuracy and convergence-order among all other $2D$ \GVOF{}s. However, already the $2D$ \ac{iPAM} method is significantly more complex than other \GVOF{}s. The $2D$ interface approximation is extended with additional marker points, while coalescence and breakup are handled by directly changing the geometry of the interface, which makes it very similar to the Front Tracking method \citep{Tryggvason2001}. Nef polyhedrons are proposed for the extension of \ac{iPAM} to $3D$, for intersections of general polyhedrons \citep{Hachenberger2007}. However, the very large number of necessary polyhedron intersections prohibit the use of Nef polyhedrons because of the prohibitive computational costs. \ac{iPAM} ensures mass conservation by adjusting material polygons using edge manipulation and adding/removing points. The authors state that local volume conservation cannot be fulfilled for some cells (\citet[footnote, page 2370]{Zhang2014}), however they disregard this issue in favor of the fourth-order convergence of the volume fraction in the $L_1$ norm. The volume (mass) conservation error is not reported for the verification tests. \citet{Zhang2014} assume a Lipschitz continuous velocity field in space in their derivation and do not mention interpolation of the velocity field at cell-corner points, a source of an additional error. Still, \ac{iPAM} delivers a stable higher-order convergence, which makes it an attractive candidate for a possible extension to $3D$, especially if the statements regarding its efficiency are confirmed by \ac{HPC} measurements, and a higher-order numerical method is used for the two-phase Navier-Stokes system.} 

\textcolor{Reviewer1}{\citet{Maric2013} have used unique velocities at cell-corner points on unstructured meshes using \ac{IDW} interpolation, together with a Youngs' reconstruction algorithm and an iterative correction of the volumetric flux for volume conservation, similar to \ac{EMFPA} method \citep{Hernandez2008}.\ Harvie-Fletcher error redistribution algorithm is used to conservatively correct for overshoots, undershoots, and wisps. Local dynamic \acf{AMR} increases the accuracy of the advection significantly. Despite the first-order accuracy, the dynamic \ac{AMR} reduces the error substantially at a small fraction of the overall computational cost, compared to uniform mesh resolution.}

\citet{Owkes2014} and \citet{Jofre2014} have proposed Eulerian flux-based \GVOF{}s that also rely on unique discrete Lagrange trajectories at cell corner-points. The method proposed by \citet{Jofre2014} is directly applicable to unstructured meshes, whereas \citet{Owkes2014} state that the extension for unstructured meshes is straightforward. Those two methods share the same correction of the flux volume $\FluxVolume$ for volume conservation and the methods of \textcolor{Reviewer1}{\citet{Owkes2014} and \citet{Jofre2014}} rely on the \ac{LVIRA} algorithm for the second-order accurate \ac{PLIC} interface reconstruction. They both show second-order convergent geometrical advection errors, are numerically bounded and volume conservative.  

\textcolor{Reviewer1}{\citet{Comminal2015} propose a $2D$ \acf{CCU} \ac{LE} scheme. Contrary to \citep{Zhang2008, Zhang2013, Zhang2014}, where an exact velocity is used at cell corner points, \citet{Comminal2015} address the problem of interpolation errors. The \ac{CCU} method builds upon the ideas proposed in the \ac{GPCA} method by \citet{Cervone2009}. Higher-order accuracy is used for the vertex displacement integration, and vertex interpolation is second-order accurate (bilinear in \ac{GPCA}).  \citet{Comminal2015} show that the overall advection accuracy is influenced mostly by velocity interpolation at cell-corner points and temporal integration of the respective displacements. Compared to \ac{PAM} and \ac{iPAM} schemes, \ac{CCU} is much simpler because it replaces complex explicit manipulation of material polygons with a simple correction of the control volume pre-image for volume conservation similar to the one applied by \citet{Owkes2014} and \citet{Jofre2014} in $3D$. The topological changes of the interface are handled automatically by the interface reconstruction. Of course, the absolute accuracy and convergence of the \ac{iPAM} and \ac{PAM} is much higher compared with \ac{CCU}. However, extending the \ac{CCU} scheme to $3D$ on unstructured meshes would involve significantly simpler geometrical operations.}

\citet{Owkes2017} propose an extension of their earlier method \citep{Owkes2014}, using a staggered solution approach and sub-grid mesh resolution for enhancing mass and momentum conservation. The sub-grid resolution increases the effective resolution of the volume fraction transport of their original scheme in the same way as the dynamic \ac{AMR} does it in \citep{Ahn2009} and \citep{Maric2013}. 

\citet{Ivey2017} propose an un-split geometrical \ac{VOF} method that constructs the flux volume iteratively.  The iterative flux volume calculation removes self-intersections, as well as intersections with the neighboring flux volumes. The calculation of the flux volumes can be applied to non-convex star-shaped polyhedral cells. Their \ac{NIFPA}-1 scheme has results comparable to those of the \ac{OD} method \citep{Owkes2014}. Average CPU times are reported for the construction of the fluxed phase-specific volumes. 

\begin{figure}[thb]
  \centering
  \begin{subfigure}[]{0.49\textwidth}
    \centering
    \includegraphics[width=0.5\columnwidth]{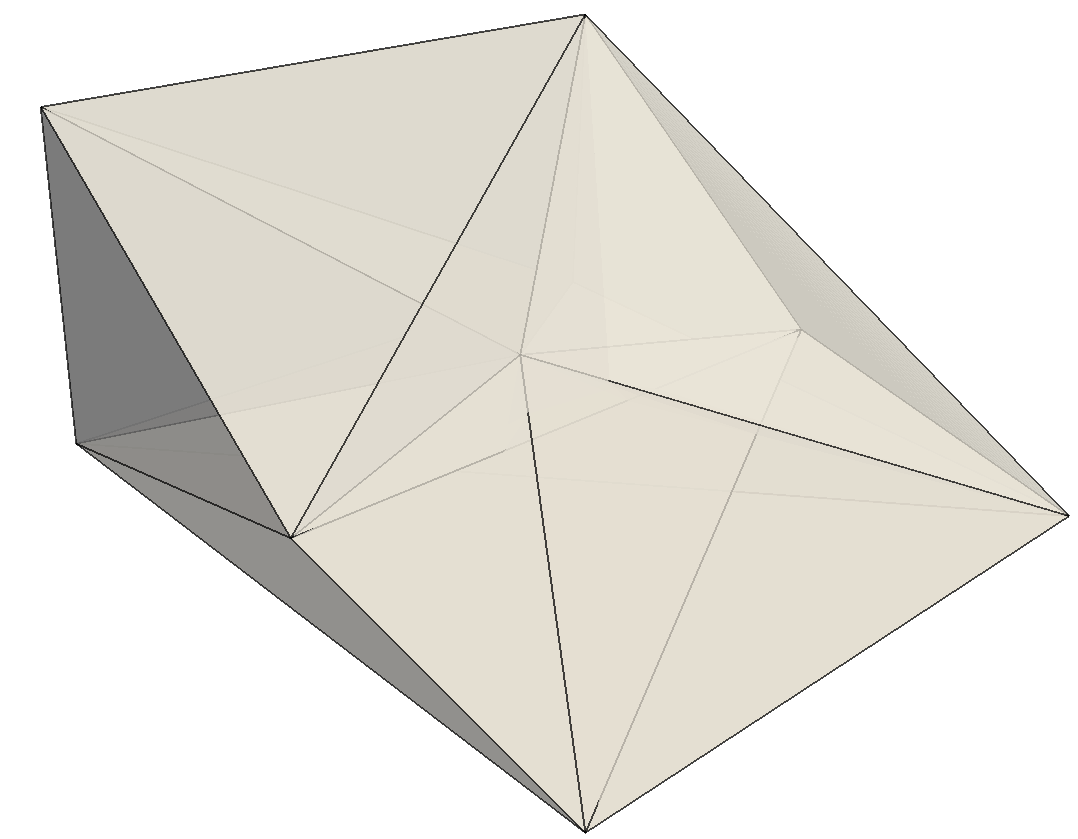}
    \caption{Oriented triangulation of a non-convex flux volume $\FluxVolume$.}
    \label{fig:fluxtris:orie}
  \end{subfigure}
  \begin{subfigure}[]{0.49\textwidth}
    \centering
    \includegraphics[width=0.5\columnwidth]{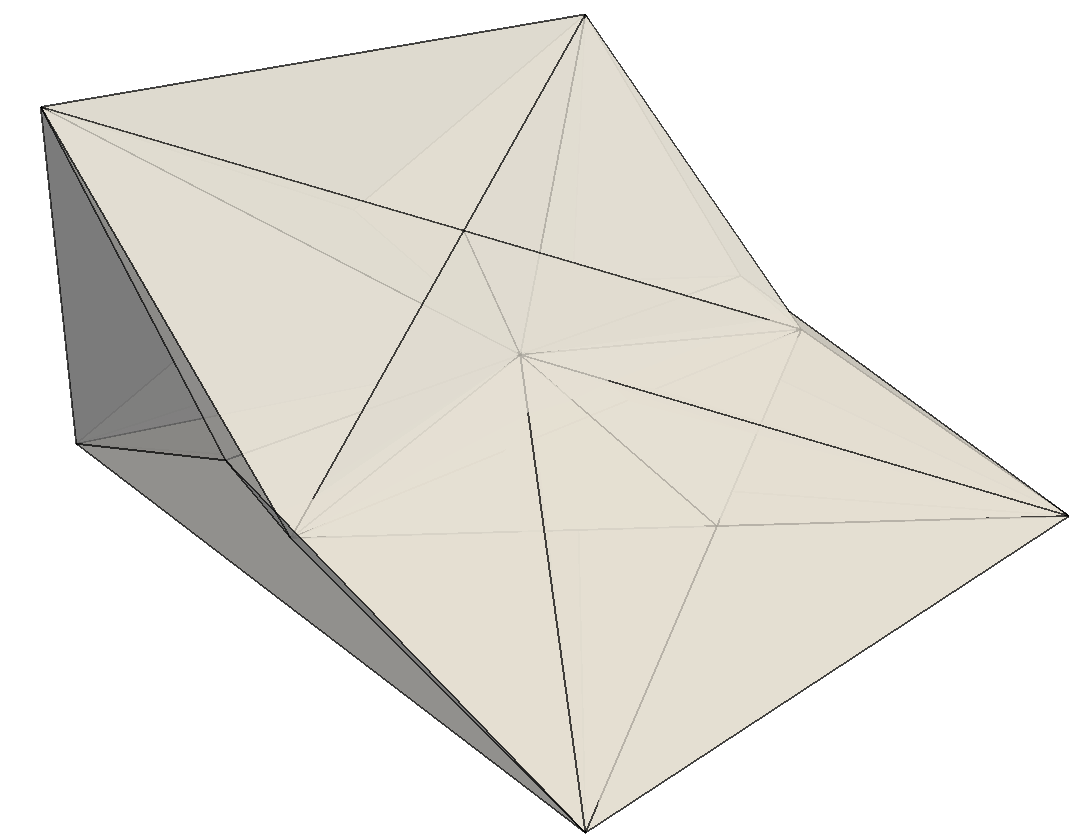}
    \caption{Barycentric triangulation of a non-convex flux volume $\FluxVolume$.}
    \label{fig:fluxtris:bary}
  \end{subfigure}
    \begin{subfigure}[b]{0.49\textwidth}
    \centering
    \def\svgwidth{0.45\textwidth}
       {\footnotesize
         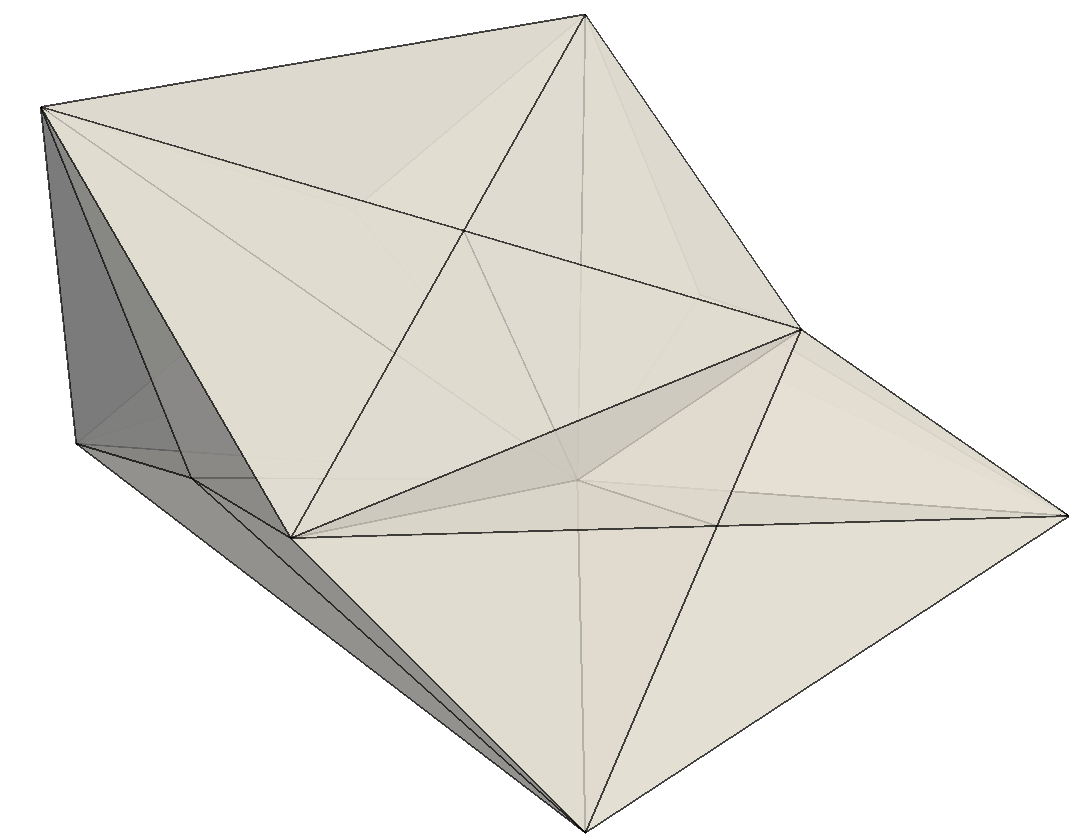
       }
    \caption{Flux-aware triangulation of a non-convex flux volume $\FluxVolume$.}
    \label{fig:fluxtris:flux}
  \end{subfigure}
    \begin{subfigure}[b]{0.49\textwidth}
    \centering
    \def\svgwidth{0.45\textwidth}
       {\footnotesize
         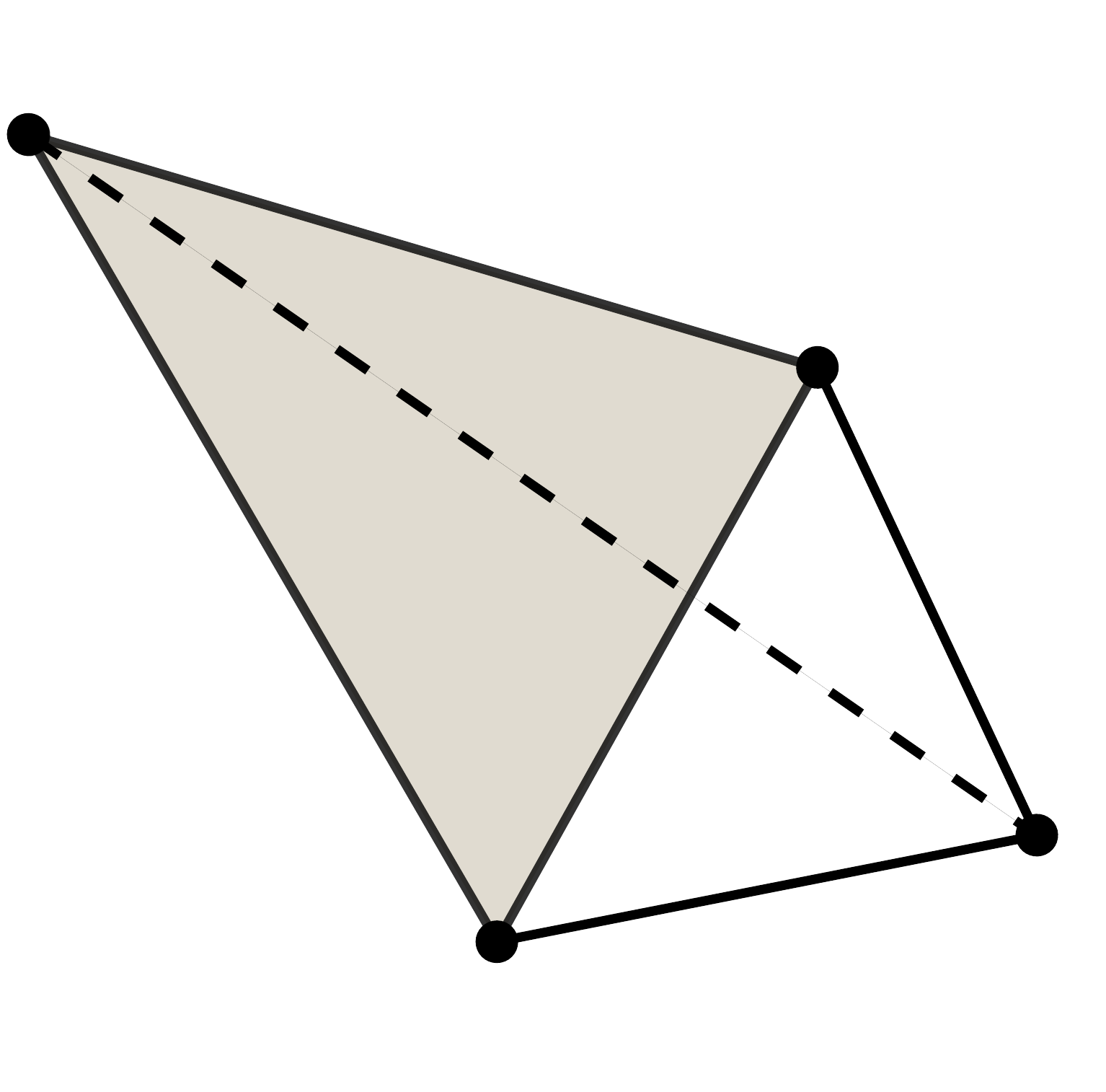
       }
        \caption{A tetrahedron given by the star-point $\mathbf{x}^*_f$ of $\FluxVolume$.}
    \label{fig:fluxtris:star-point}
  \end{subfigure}
  \caption{Different triangulations of a non-convex flux volume.}
  \label{fig:fluxtris}
\end{figure}
\citet{Maric2018} propose a flux-aware triangulation for non-convex flux volumes with non-planar faces (\cref{fig:fluxtris}). Either an \emph{oriented triangulation} (\cref{fig:fluxtris:orie}), or a \emph{centroid triangulation} (\cref{fig:fluxtris:bary}), are usually used to approximate flux volumes in the un-split \GVOF{}. The centroid triangulation uses centroids of (generally) non-planar ruled surfaces that bound the flux volume, to decompose the flux volume into tetrahedrons. On the other hand, the oriented triangulation decomposes the ruled surfaces of the flux volume into triangles, either clockwise or counter-clockwise with respect to the normal vector of each ruled surface. Both triangulations generate self-intersections of tetrahedrons, even for relatively simple star-shaped non-convex flux volumes, as shown for example in \cref{fig:fluxtris:orie,fig:fluxtris:bary}. \textcolor{Reviewer1R1}{The flux-aware triangulation (\cref{fig:fluxtris:flux}) correctly decomposes the flux volume $\FluxVolume$ into tetrahedrons by calculating a so-called \emph{star-point} of $\FluxVolume$. A star-point of a set is a point with the property that the linear segment between this point and any other point of the set lies inside the set. A star-point of a triangulated $\FluxVolume$ is a point that constructs a positive-valued mixed product of the vectors of a tetrahedron constructed from the star-point and any triangle in the triangulation of $\partial \FluxVolume$. An example of such a tetrahedron is emphasized in \cref{fig:fluxtris:flux} and it is shown in detail in \cref{fig:fluxtris:star-point}. This tetrahedron is constructed from the star-point that lies on the face $S_f$ ($\x^*_f$ in \cref{fig:fluxtris:star-point}), and the triangle given by the edge $(\x'_{f,i},\x'_{f,i+1})$ and the centroid $x'_f$ of the mapped face $\FlowMap{t_n}{t^{n+1}} S_f$. If the mixed product $(\x_f' - \x_f^*) \cdot ((\x'_{f,i} - \x^*_f)\times(\x'_{f,i} - \x^*_f))$ is positive \emph{for each triangle} in the triangulation of $\partial \FluxVolume$, a correct decomposition of the flux volume into tetrahedrons is constructed by the star-point $\x_f^*$. The tetrahedral decomposition of $\FluxVolume$ is then intersected to compute the fluxed phase-specific volume $\PhaseFluxVolumeContrib$ in \cref{eqn:phase-specific}. The possibility of both negative and positive contributions to $\PhaseFluxVolumeContrib$ caused by $\FluxVolume$ having a non-empty intersection with halfspaces $(\x_f,\Sf)$,$(\x_f,-\Sf)$ is addressed using local adaptive mesh refinement. Note that the separation of positive and negative contributions substantially decreases computational efficiency.} In addition to the flux-aware triangulation, a modification of the Swartz reconstruction algorithm is developed. An extrapolation based on Taylor series from cell centers to cell-corner points with second-order accuracy is proposed for velocities at cell-corner points. A global error redistribution algorithm stabilizes the solution. Compared to contemporary methods, the overall accuracy of the solution is increased by the flux-aware triangulation for verification cases on unstructured hexahedral meshes. Average computational time per time step of the UFVFC-Swartz scheme is similar to the \ac{OD} method \citep{Owkes2014}.

\subsection{Geometric/algebraic Volume-of-Fluid methods}

\textcolor{Reviewer1R1}{Recently, Volume-of-Fluid methods that reduce the dimensionality in the computation of $\PhaseFluxVolumeContrib$ from three to two dimensions have been introduced. We categorize these methods here as geometric/algebraic methods, to distinguish them from the fully three-dimensional geometric VOF methods. Geometric/algebraic VOF methods approximate the interface either geometrically or implicitly, using an approximation of the phase-indicator function. The reconstructed interface is then used for an algebraic calculation of the fluxed phase-specific volume $\PhaseFluxVolumeContrib$ for solving the reformulated volume fraction equation} 
\textcolor{Reviewer1R1}{\begin{equation}
  \int_{t^n}^{t^{n+1}} \int_{\Omega_k} \partial_t \VolFrac \, dV \, dt + \sum_f \int_{t^n}^{t^{n+1}} \int_{S_f} \Indicator(\x,t) \U \cdot \n \, dS \, dt = \int_{t^n}^{t^{n+1}} \int_{\Omega_k} \Indicator(\x,t) \div{\U} \, dV \, dt. 
  \label{eqn:geo-alg}
\end{equation}}
The term on the r.h.s. is a numerical correction given by \cref{eqn:rk:corr}, proposed originally by \citet{Rider1998}, that can be omitted when the velocity field $\U$ satisfies the discrete divergence-free condition.

\citet{Roenby2016} have proposed the isoAdvector method that calculates $\PhaseFluxVolumeContrib$ as  
\textcolor{Reviewer1R1}{\begin{equation}
    \PhaseFluxVolumeContrib = \int_{t^n}^{t^{n+1}} \int_{S_f} \Indicator(\x,t) (\U \cdot \n) \, dS \, dt = \, \U_{f}(t^n) \cdot \hat{\vec{n}_f} \int_{t^n}^{t^{n+1}}\int_{S_f} \Indicator(\x,t) \, dS \, dt + O(\delta t) + O(h^2).
  \label{eqn:iso:af}
\end{equation}}
In \cref{eqn:iso:af}, the velocity $\U_{f}(t)$ with which the interface moves across the face $f$ is formally first-order accurate in time. Velocity $\U_{f}$ is the second-order accurate average associated with the centroid of the face $S_f$ and the final order of accuracy in space depends on the choice of interpolation used for $\U_f(t)$. \textcolor{Reviewer1R1}{Alternatively, \citet{Roenby2016} use the approximation $\U_f \approx \vec{n}_f \frac{F_f}{\SfMag}$, where $F_f$ is the volumetric flux over $S_f$.} For polyhedral cells, the unit normal vector of $S_f$, $\vec{n}_f$, is constant over $S_f$ and it is \textcolor{Reviewer1}{computed from a centroid triangulation of the face $S_f$}. \textcolor{Reviewer1R1}{Complex three-dimensional calculations are reduced by isoAdvector to two-dimensional calculations, which significantly simplifies geometrical calculations and increases computational efficiency.} The core of the isoAdvector scheme is the evaluation of the integral
\textcolor{Reviewer1R1}{\begin{equation}
    \int_{t^n}^{t^{n+1}} \int_{S_f} \Indicator(\x,t) \, dS \, dt=  \int_{t^n}^{t^{n+1}} A_f(t) dt,
\end{equation}}\textcolor{Reviewer1}{where $A_f(t)$ is the area of the face $S_f$ that is inside the phase $1$ (\textcolor{Reviewer1R1}{"wetted" by phase $1$}) at time $t$}. The isoAdvector schemes relies on a geometric approximation of the interface and polynomial algebraic extrapolation to express $A_f(t)$. The interface that intersects the face $S_f$ is approximated with a line, and it defines $A_f$ at some point in time $t$. The evolution of $A_f(t)$ is approximated by computing multiple positions of the interface line by evolving the interface from the downwind cell in space using the cell-centered velocity $\U_k$, \textcolor{Reviewer1R1}{over sub-intervals in the time step $[t^n,  t^{n+1}]$ associated with partition points $t_s$.}. Using $A_f(t_s)$, $A_f(t)$ is extrapolated as a quadratic polynomial, whose exact integral over $t$ is then used in \cref{eqn:iso:af}. The reduction of dimensionality and the algebraic extrapolation of $A_f(t)$ result in a simplification of the scheme, compared to the fully geometrical LE schemes. The correction on the r.h.s. in \cref{eqn:iso:af} is not used by the isoAdvector scheme. \textcolor{Reviewer1}{Recently, \citet{Scheufler2019} have improved the isoAdvector scheme by introducing a classical PLIC interface whose normal is improved using the Reconstructed Distance Function (RDF) from \citet{Cummins2005}: a signed distance is computed at cell centers from the PLIC interface, and the gradient of the distance field is then used to improve the PLIC normal, iteratively. Results shown by \citet{Scheufler2019} are comparable to contemporary un-split \GVOF{} at significantly reduced computational costs.}

\citet{Xie2017} propose the \ac{THINC/QQ} method. It is based on a multidimensional quadratic polynomial approximation of a diffuse interface, modeled as a hyberbolic tangent function with a user-defined steepness parameter. The \ac{THINC/QQ} method also approximates the solution of \cref{eqn:geo-alg} with the r.h.s. term included. The \ac{THINC/QQ} method relies on a piecewise-quadratic approximation of the interface, while the geometrical \ac{LE} methods and the isoAdvector scheme rely on a piecewise-linear interface approximation. Like the isoAdvector scheme, the flux across the face $S_f$ is computed in $2D$, relying on the Gauss quadrature of the reconstructed quadratic approximation of the tangent hyperbolic function. \ac{THINC/QQ} uses a third-order accurate explicit Runge-Kutta scheme for the temporal integration. 

\textcolor{Reviewer1R1}{The reduction of computational complexity provided by geometric/algebraic methods does come at the cost of a lower overall accuracy and stronger restrictions regarding numerical stability, compared to the fully $3D$ geometric methods, as discussed in the next section. The question remaining for future research is, whether the stability restrictions of the geometric/algebraic schemes can be made less strict and their accuracy increased further in comparison to fully geometrical methods, or if the three-dimensional calculations used by the geometric methods can be made more efficient.}

\subsection{Advection comparison}

Because of the number of reviewed methods, two verification cases are chosen for method comparison: $2D$ shear (single vortex) and $3D$ deformation case. Results available in the literature are mostly directly comparable with each other. The verification of the advection is based on a set of commonly used error measures:\newline 
the volume conservation error
\begin{equation}
  E_v = \dfrac{|\sum_k V_k \PlicFraction(t)  - \sum_k V_k \PlicFraction(t_0)|}{\sum_k V_k \PlicFraction(t_0)},
  \label{eqn:ev}
\end{equation}
the geometrical (shape) error 
\begin{equation}
  E_g = \sum_k V_k | \PlicFraction(t)  - \PlicFraction^e |,
  \label{eqn:eg}
\end{equation}
the normalized variant of the shape error
\begin{equation}
  E_n = \dfrac{\sum_k V_k | \PlicFraction(t)  - \PlicFraction^e |}{\sum_k V_k \PlicFraction^e},
  \label{eqn:en}
\end{equation}
and the numerical boundedness (stability) error 
\begin{equation}
    E_b = \max(\max_k((\PlicFraction - 1), 0),\max_k((0 - \PlicFraction),0)). 
\end{equation}
For those publications that have defined the used CPU architecture and measured \emph{absolute CPU times}, an average CPU time used per time-step by the VOF method ($T_e$) is used for comparison. \textcolor{Reviewer1R1}{One should note that a direct comparison of algorithms on different CPU architectures is difficult, as a change in the architecture may cause a substantial change in the computational efficiency of most algorithms.} \textcolor{Reviewer1}{Additionally, performance measurements should be reported in absolute units\footnote{If relative units are reported, absolute values should be reported for the denominator used in the normalization.}, with sufficient statistical information (an average value does not suffice), performed on a dedicated computing node (to avoid the influence of the operating system), with disabled CPU scaling, and using compiler optimizations that are otherwise used for productive computations \citep{Hoefler2015}. Although using different CPU architectures may strongly impact performance measurements, differences in orders of magnitude cannot be justified by the difference in CPU architecture, especially when relatively similar CPU architectures are used.} 

Verification tests for the interface advection must be carefully selected because the velocity functions used in the literature are all compositions of harmonic functions. Temporal integration of such functions with the first-order accurate Euler quadrature is prone to artificial error cancellation \citep{Wiedeman2002}, resulting in an artificial second-order convergence. 

\subsubsection{$2D$ shear (single vortex)}

The $2D$ shear verification case was introduced originally by \citet{Leveque1996}. The test consists of a circular interface of radius $r=0.15$, with the center $c=(0.5, 0.75, 0)$. An explicitly prescribed velocity $\U(\x,t) = \U(u_x(t),u_y(t),0)$ is given as  
\begin{align}
  u_x(t) & = sin(2\pi y)sin^2(\pi x) cos\left(\dfrac{\pi t}{T}\right),  \\
  u_y(t) & = -sin(2 \pi x)sin^2(\pi y) cos\left(\dfrac{\pi t}{T}\right),
  \label{eqn:2Dshear}
\end{align}
in a \textcolor{Reviewer1R1}{unit square solution domain} with $CFL=1$. Results of the $2D$ shear case have recently been summarized for contemporary methods by \citet{Comminal2015} and their table is extended in \cref{tab:adv:shear2D} by additional results that were obtained for the same mesh resolutions.

The $2D$ shear results of the isoAdvector were computed with resolutions alternative to those reported in the literature, using $[100^2,200^2,400^2]$ cubical volumes, so they cannot be directly summarized in \cref{tab:adv:shear2D}. Additionally $CFL=0.5$ instead of $1$ was used. Volume conservation error $E_v$ is reported near machine tolerance, and numerical boundedness error $E_b$ is within $[~1e-08,1e-07]$. The $E_n$ advection errors of the isoAdvector scheme are $E_n = [4.7e-02, 1.2e-02, 2.3e-03]$, scaled with the area of the circular interface to $E_g = [3.32e-03, 8.48e-04, 1.63e-04]$, with respective convergence orders $(1.96, 2.38)$. Therefore, the isoAdvector shows second order convergence for $CFL=0.5$ for the $2D$ shear case with shape and numerical boundedness errors that are somewhat worse than those reported for the most recent fully geometrical \ac{LE} method. Reported total CPU time used for the verification cases of $[13, 60, 314]$ seconds confirms the computational efficiency of this geometric/algebraic VOF method. The isoAdvector-plicRDF by \citet{Scheufler2019} diverges with $CFL=1$, but recovers second-order convergence at $CFL=0.5$ \citep[table 3]{Scheufler2019}, cf. \cref{tab:adv:shear2D}. The isoAdvector-plicRDF method delivers better results than THINC/QQ using a higher $CFL=0.5$ condition. Reported execution times of the isoAdvector-plicRDF are significantly smaller than for other \ac{LE} un-split geometrical VoF methods. 

The $E_g$ errors reported for the \ac{THINC/QQ} method are approximately two times higher than $E_g$ errors computed by the most recent geometrical \ac{LE} methods. \citet{Xie2017} report the numerical boundedness error $E_b$ only for the circle translation verification case. $E_b$ is reported within $[\approx 1e-8, \approx 1e-7]$ for $CFL=0.2$ and $[\approx 1e-7, \approx 1e-2]$ for $CFL=0.8$ for different values of the interface thickness parameter $\beta$. 
\begin{table}[H]
  \centering
{\scriptsize
  \input{tables/Comminal2015-extended.tex}
}
    \caption{$E_g$ errors of the $2D$ shear (single vortex case) with $CFL=1$, with $N^2$ square volumes, \textcolor{Reviewer1R1}{with $(N)$ given in the first row of the table}.}
  \label{tab:adv:shear2D}
\end{table}
Geometrical \ac{LE} \ac{VOF} methods are stable for $CFL \le 1$, and show volume conservation and numerical stability (boundedness) errors near machine tolerance, while $CFL=0.15$ was used by \citet{Xie2017} for the $2D$ shear case. As the $CFL$ \textcolor{Reviewer1R1}{condition} determines the temporal accuracy, this makes it difficult to perform a direct comparison with other \ac{LE} methods. One might argue that the $CFL$ magnitudes are much smaller when the advection scheme is coupled to the \ac{NS} system, as the time steps become restricted by physical stability parameters. Then, $E_b\approx 1e-07$ for $CFL=0.2$ would allow stable simulations of multiphase problems with large density ratios with \ac{THINC/QQ}. Still, as volume conservation and numerical stability errors, as well as CPU times, are not reported for $2D$ shear and $3D$ deformation cases, it is difficult to fully compare the \ac{THINC/QQ} to standard geometrical \ac{LE} VOF methods.

Absolute computational times expressed either per time step, or as total execution time for the verification case, are not reported for many methods, making them difficult to compare in terms of computational efficiency. CPU time is often omitted with a claim that the interface advection \emph{usually} takes up only a fraction of the computational time compared to the pressure-velocity coupling algorithm. With the geometrical component added to any \ac{LE} scheme, this might not be true, depending on the cost of $3D$ geometrical operations. Average CPU time per time step required for the interface advection has been reported in \citep{Maric2013, Owkes2014, Owkes2017, Maric2018} and total computational times were reported in \citep{Ahn2009, Zhang2014, Roenby2016}. The CPU times reported by \citet{Zhang2014} for the \ac{iPAM} scheme with $h_L = 0.1h$ were $[984,4172,16709]$ for meshes with $[32^2, 64^2, 128^2]$ cells, with a reported \textcolor{Reviewer1R1}{third-order convergence} as listed in \cref{tab:adv:shear2D}. 

Overall, for the $2D$ shear case, the \ac{iPAM}, \ac{AMR}-\ac{MoF}, \ac{CCU} and \ac{GPCA} methods result in by far the best absolute error values compared to other methods. An interesting fact is that two of those methods are cell-based \ac{LE} methods, that avoid the construction and correction of flux polyhedrons and approach the problem of Eulerian re-mapping on the basis of cell pre-images. \textcolor{Reviewer1R1}{Cell-based methods decompose images of polyhedral (polygonal) cells into tetrahedrons (triangles) more easily, compared to the flux-based methods that require complex tetrahedral decomposition of flux volumes $\FluxVolume$.} Higher-order interpolation of the vertex displacements proposed first for the \ac{GPCA} and then for \ac{CCU} method is an additional factor that improves accuracy. Apart from \ac{iPAM} with its sub-grid interface resolution, \ac{AMR}-\ac{MoF} and \ac{UFVFC}-Swartz, other geometrical \ac{LE} methods rely on one or the other variant of the \ac{ELVIRA} reconstruction algorithm. \textcolor{Reviewer1R1}{Total errors in simulations with many interface cells are influenced substantially by the reconstruction error and for those cases, the \ac{MoF} reconstruction still delivers the most accurate results for the $2D$ shear case.} 

\subsubsection{$3D$ deformation}

The $3D$ deformation case was used by \citet{Enright2002} for the Particle Level Set (PLS) method and it was originally proposed by \citet{Smolarkiewicz1982}. It has subsequently been used by many authors to verify the geometrical \ac{VOF} method as well. The $3D$ shear verification consists of a sphere with radius $r=0.15$, centered at $(0.35,0.35,0.35)$, and the velocity given by 
\begin{align}
    \U(\x,\,t) = 
    \left( 
			\begin{array}{l}
			  2 \sin(2\pi y) \sin(\pi x)^2 \sin(2 \pi z) \cos \left(\frac{\pi t}{T}\right)\\
			  -\sin(2\pi x) \sin(\pi y)^2 \sin (2\pi z) \cos \left(\frac{\pi t}{T}\right)\\
			  -\sin(2\pi x) \sin(2 \pi y) \sin (\pi z)^2 \cos \left(\frac{\pi t}{T}\right)
			\end{array} 
    \right),
    \label{eqn:3Ddeformation}
\end{align}
where $T=3$ and $CFL=0.5$. 

\begin{table}[H]
  \centering
  \footnotesize
  \input{tables/voFoam-hex-deformation3D-Trapezoid-Taylor-Trapezoid-Swartz-2-1-30-30-cellInt-trapezoid-pointCellInt-Taylor-fluxInt-trapezoid.tex}
    \caption{Advection error $E_g$ comparison for the $3D$ deformation case with $T=3$ and $\max_{\Omega}(CFL)=0.5$ and $N^3$ cubical volumes. Bracketed $N$ values represent the double sub-grid scale resolution used for the advection by \citet{Owkes2017}. For cases with $CFL\ne0.5$, used $CFL$ value is added, and $CFL_\Sigma$ is the $CFL$ condition applied for interface cells $0 < \VolFrac_k < 1$.} 
  \label{tab:adv:3deform}
\end{table}

The difficulty in developing dimensionally un-split \ac{LE} \ac{VOF} methods in three-dimensions is reflected in the small number of methods with actual three-dimensional results (cf. \cref{tab:adv:3deform,tab:adv:shear2D}). Volume conservation and numerical boundedness errors are often not reported in the literature, as well as performance measurements in absolute CPU time, which complicates a direct method comparison.

The most accurate is the \acf{OD-S} because its volume fraction advection uses local dynamic mesh refinement. The differences between \ac{OD-S} and \ac{UFVFC}-Swartz are based on different interface reconstruction algorithms (\ac{ELVIRA} and simplified Swartz, respectively), different interpolation of cell-corner velocities, and different temporal integration. Double mesh resolution for the \ac{OD-S} given by local mesh refinement doubles the average CPU time per time-step, confirming the linear complexity in terms of the number of mixed cells. 

The benefits of adding the \acf{CBIR} reconstruction step to the \acf{FMFPA-3D} scheme by \citet{Hernandez2008,Lopez2008} can be seen in \cref{tab:adv:3deform}. The absolute error values are decreased somewhat by the \ac{CBIR} step, leading to a slightly increased convergence order. However, as CPU times are not reported, it is difficult to know whether this improvement increases computational costs significantly. Additionally, the errors of the \ac{FMFPA-3D}-(\ac{CLCIR},\ac{CBIR}) combinations are sorted with respect to their magnitudes in \cref{tab:adv:3deform}, ignoring the fact that $CFL=1$ was used by \citet{Lopez2008,Hernandez2008}. 

A second-order accurate temporal integration method will already resolve the temporal derivative of the velocity functions given by \cref{eqn:3Ddeformation,eqn:2Dshear} very accurately, because of the low frequency in the cosine temporal term.  Because of this, the reconstruction error has a stronger influence in the $3D$ deformation case, so using higher $CFL$ numbers reduces the absolute error magnitude. A comparison between \ac{FMFPA-3D} and other methods is based on the same $CFL$ numbers by \citet{Lopez2008,Hernandez2008}. 

$CFL=0.25$ was used for the \ac{THINC/QQ} method by \citet{Xie2017}, and no $E_v,E_b,T_e$ is reported for the $3D$ deformation case, which somewhat complicates direct comparison. Considering the aforementioned temporal integration resolution, the third-order Runge-Kutta scheme used in \ac{THINC/QQ} certainly resolves the temporal derivative with sufficient accuracy, and the use of the smaller $CFL$ number might be related to the stability limits of the higher-order flux reconstruction. Absolute error magnitudes of the \ac{THINC/QQ} are approximately two times larger than the errors reported by the most recent geometrical \ac{LE} \ac{VOF} methods. 

\textcolor{Reviewer1}{\citet{Roenby2016} report $E_v,E_b$ errors as well as the total CPU time. The results of the isoAdvector scheme are worse than those of \ac{THINC/QQ} scheme on coarser mesh resolutions, and better than \ac{THINC/QQ} on finer mesh resolutions, \textcolor{Reviewer1R1}{but still twice as large than the error magnitudes reported for other contemporary geometrical \ac{LE} schemes}. However, CPU times reported for the isoAdvector in \citet{Roenby2016} show significant improvement in computational efficiency compared to geometrical \ac{LE} schemes. The isoAdvector-plicRDF method by \citet{Scheufler2019} delivers similar results to \ac{LE} geometrical VoF methods on coarser mesh resolutions, but is on average two times less accurate on finer mesh resolutions. This is compensated with higher computational efficiency compared to other methods. Note that the results of isoAdvector-plicRDF are computed for the 3D deformation case in \cref{tab:adv:3deform} for $\max(CFL_\Sigma)$, the maximal $CFL$ condition at the interface and not for the whole domain, which somewhat complicates a direct comparison.}

Data and tables summarized in this section are available at \url{http://dx.doi.org/10.25534/tudatalib-162}.

%% file: figures/rider-kothe.pdf_tex
\begingroup%
  \makeatletter%
  \providecommand\color[2][]{%
    \errmessage{(Inkscape) Color is used for the text in Inkscape, but the package 'color.sty' is not loaded}%
    \renewcommand\color[2][]{}%
  }%
  \providecommand\transparent[1]{%
    \errmessage{(Inkscape) Transparency is used (non-zero) for the text in Inkscape, but the package 'transparent.sty' is not loaded}%
    \renewcommand\transparent[1]{}%
  }%
  \providecommand\rotatebox[2]{#2}%
  \newcommand*\fsize{\dimexpr\f@size pt\relax}%
  \newcommand*\lineheight[1]{\fontsize{\fsize}{#1\fsize}\selectfont}%
  \ifx\svgwidth\undefined%
    \setlength{\unitlength}{1462.86135864bp}%
    \ifx\svgscale\undefined%
      \relax%
    \else%
      \setlength{\unitlength}{\unitlength * \real{\svgscale}}%
    \fi%
  \else%
    \setlength{\unitlength}{\svgwidth}%
  \fi%
  \global\let\svgwidth\undefined%
  \global\let\svgscale\undefined%
  \makeatother%
  \begin{picture}(1,0.4661883)%
    \lineheight{1}%
    \setlength\tabcolsep{0pt}%
    \put(0.08654035,0.46106137){\color[rgb]{0,0,0}\makebox(0,0)[lt]{\lineheight{0}\smash{\begin{tabular}[t]{l} \end{tabular}}}}%
    \put(0,0){\includegraphics[width=\unitlength,page=1]{rider-kothe.pdf}}%
    \put(0.71062908,0.35465984){\color[rgb]{0,0,0}\makebox(0,0)[lt]{\lineheight{0}\smash{\begin{tabular}[t]{l}$-\delta t \U_g(t)$\end{tabular}}}}%
    \put(0.2305113,0.0914824){\color[rgb]{0,0,0}\makebox(0,0)[lt]{\lineheight{0}\smash{\begin{tabular}[t]{l}$S_f$\end{tabular}}}}%
    \put(0.37407443,0.0399158){\color[rgb]{0,0,0}\makebox(0,0)[lt]{\lineheight{0}\smash{\begin{tabular}[t]{l}$\S_f$\end{tabular}}}}%
    \put(0.13313927,0.1982384){\color[rgb]{0,0,0}\makebox(0,0)[lt]{\lineheight{0}\smash{\begin{tabular}[t]{l}$-\delta t \U_f(t)$\end{tabular}}}}%
    \put(0.78748765,0.09246147){\color[rgb]{0,0,0}\makebox(0,0)[lt]{\lineheight{0}\smash{\begin{tabular}[t]{l}$S_g$\end{tabular}}}}%
    \put(0,0){\includegraphics[width=\unitlength,page=2]{rider-kothe.pdf}}%
    \put(0.06675468,0.2508625){\color[rgb]{0,0,0}\rotatebox{-9.8378404}{\makebox(0,0)[lt]{\lineheight{0}\smash{\begin{tabular}[t]{l}$\PlicNormal$\end{tabular}}}}}%
    \put(0.0941737,0.38912876){\color[rgb]{0,0,0}\rotatebox{-10.237152}{\makebox(0,0)[lt]{\lineheight{0}\smash{\begin{tabular}[t]{l}$\PlicPosition$\end{tabular}}}}}%
    \put(0,0){\includegraphics[width=\unitlength,page=3]{rider-kothe.pdf}}%
    \put(0.70695085,0.06799821){\color[rgb]{0,0,0}\rotatebox{-34.27903}{\makebox(0,0)[lt]{\lineheight{0}\smash{\begin{tabular}[t]{l}$\n_l$\end{tabular}}}}}%
    \put(0.76971714,0.18343042){\color[rgb]{0,0,0}\rotatebox{-33.898376}{\makebox(0,0)[lt]{\lineheight{0}\smash{\begin{tabular}[t]{l}$\Point_l$\end{tabular}}}}}%
    \put(0.41519345,0.40289423){\color[rgb]{0,0,0}\makebox(0,0)[lt]{\lineheight{0}\smash{\begin{tabular}[t]{l}$\Omega_k$\end{tabular}}}}%
    \put(0.83009688,0.40289423){\color[rgb]{0,0,0}\makebox(0,0)[lt]{\lineheight{0}\smash{\begin{tabular}[t]{l}$\Omega_l$\end{tabular}}}}%
    \put(0.54417882,0.01600982){\color[rgb]{0,0,0}\makebox(0,0)[lt]{\lineheight{0}\smash{\begin{tabular}[t]{l}$\PhaseFluxVolumeContrib$\end{tabular}}}}%
    \put(0,0){\includegraphics[width=\unitlength,page=4]{rider-kothe.pdf}}%
    \put(0.79488291,0.2496326){\color[rgb]{0,0,0}\makebox(0,0)[lt]{\lineheight{0}\smash{\begin{tabular}[t]{l}$-\delta t \U_f(t)$\end{tabular}}}}%
    \put(0,0){\includegraphics[width=\unitlength,page=5]{rider-kothe.pdf}}%
    \put(0.46700385,0.35429434){\color[rgb]{0,0,0}\makebox(0,0)[lt]{\lineheight{0}\smash{\begin{tabular}[t]{l}$\partial V_f$\end{tabular}}}}%
  \end{picture}%
\endgroup%

%% file: figures/harvie-stream.pdf_tex
\begingroup%
  \makeatletter%
  \providecommand\color[2][]{%
    \errmessage{(Inkscape) Color is used for the text in Inkscape, but the package 'color.sty' is not loaded}%
    \renewcommand\color[2][]{}%
  }%
  \providecommand\transparent[1]{%
    \errmessage{(Inkscape) Transparency is used (non-zero) for the text in Inkscape, but the package 'transparent.sty' is not loaded}%
    \renewcommand\transparent[1]{}%
  }%
  \providecommand\rotatebox[2]{#2}%
  \newcommand*\fsize{\dimexpr\f@size pt\relax}%
  \newcommand*\lineheight[1]{\fontsize{\fsize}{#1\fsize}\selectfont}%
  \ifx\svgwidth\undefined%
    \setlength{\unitlength}{1031.95074463bp}%
    \ifx\svgscale\undefined%
      \relax%
    \else%
      \setlength{\unitlength}{\unitlength * \real{\svgscale}}%
    \fi%
  \else%
    \setlength{\unitlength}{\svgwidth}%
  \fi%
  \global\let\svgwidth\undefined%
  \global\let\svgscale\undefined%
  \makeatother%
  \begin{picture}(1,0.63432842)%
    \lineheight{1}%
    \setlength\tabcolsep{0pt}%
    \put(-1.2732676,1.04429937){\color[rgb]{0,0,0}\makebox(0,0)[lt]{\begin{minipage}{2.94814514\unitlength}\raggedright  \end{minipage}}}%
    \put(0,0){\includegraphics[width=\unitlength,page=1]{harvie-stream.pdf}}%
    \put(0.43783055,0.33117655){\color[rgb]{0,0,0}\makebox(0,0)[lt]{\lineheight{0}\smash{\begin{tabular}[t]{l}$\PhaseFluxVolumeContrib(i)$\end{tabular}}}}%
    \put(0.28885727,0.62706063){\color[rgb]{0,0,0}\makebox(0,0)[lt]{\lineheight{0}\smash{\begin{tabular}[t]{l} \end{tabular}}}}%
    \put(0,0){\includegraphics[width=\unitlength,page=2]{harvie-stream.pdf}}%
    \put(0.14536561,0.49840622){\color[rgb]{0,0,0}\makebox(0,0)[lt]{\lineheight{0}\smash{\begin{tabular}[t]{l}$\PlicPosition$\end{tabular}}}}%
    \put(0.07731379,0.3949626){\color[rgb]{0,0,0}\makebox(0,0)[lt]{\lineheight{0}\smash{\begin{tabular}[t]{l}$\PlicNormal$\end{tabular}}}}%
    \put(0.23667895,0.04497357){\color[rgb]{0,0,0}\makebox(0,0)[lt]{\lineheight{0}\smash{\begin{tabular}[t]{l}$S_f$\end{tabular}}}}%
    \put(0.52653302,0.02201265){\color[rgb]{0,0,0}\makebox(0,0)[lt]{\lineheight{0}\smash{\begin{tabular}[t]{l}$\S_f$\end{tabular}}}}%
    \put(0,0){\includegraphics[width=\unitlength,page=3]{harvie-stream.pdf}}%
    \put(0.81124567,0.50087224){\color[rgb]{0,0,0}\rotatebox{11.205384}{\makebox(0,0)[lt]{\lineheight{0}\smash{\begin{tabular}[t]{l}$w_i$\end{tabular}}}}}%
    \put(0,0){\includegraphics[width=\unitlength,page=4]{harvie-stream.pdf}}%
    \put(0.74179792,0.28670147){\color[rgb]{0,0,0}\makebox(0,0)[lt]{\lineheight{0}\smash{\begin{tabular}[t]{l}$l$\end{tabular}}}}%
    \put(0,0){\includegraphics[width=\unitlength,page=5]{harvie-stream.pdf}}%
    \put(0.67571897,0.49582056){\color[rgb]{0,0,0}\rotatebox{27.158038}{\makebox(0,0)[lt]{\lineheight{0}\smash{\begin{tabular}[t]{l}$L_i$\end{tabular}}}}}%
    \put(0.71295175,0.07262004){\color[rgb]{0,0,0}\makebox(0,0)[lt]{\lineheight{0}\smash{\begin{tabular}[t]{l}$i$\end{tabular}}}}%
    \put(0.81363418,0.07455765){\color[rgb]{0.14117647,0.10980392,0.10980392}\makebox(0,0)[lt]{\lineheight{0}\smash{\begin{tabular}[t]{l}$N_s$\end{tabular}}}}%
    \put(0,0){\includegraphics[width=\unitlength,page=6]{harvie-stream.pdf}}%
    \put(0.09869143,0.19837661){\color[rgb]{0,0,0}\makebox(0,0)[lt]{\lineheight{0}\smash{\begin{tabular}[t]{l}$\Omega^+_k$\end{tabular}}}}%
    \put(0,0){\includegraphics[width=\unitlength,page=7]{harvie-stream.pdf}}%
    \put(1.83597314,-0.15224602){\color[rgb]{0,0,0}\makebox(0,0)[lt]{\lineheight{0}\smash{\begin{tabular}[t]{l} \end{tabular}}}}%
    \put(0.57274647,0.07262004){\color[rgb]{0.14117647,0.10980392,0.10980392}\makebox(0,0)[lt]{\lineheight{0}\smash{\begin{tabular}[t]{l}$i-1$\end{tabular}}}}%
  \end{picture}%
\endgroup%

%% file: figures/harvie-ddr.pdf_tex
\begingroup%
  \makeatletter%
  \providecommand\color[2][]{%
    \errmessage{(Inkscape) Color is used for the text in Inkscape, but the package 'color.sty' is not loaded}%
    \renewcommand\color[2][]{}%
  }%
  \providecommand\transparent[1]{%
    \errmessage{(Inkscape) Transparency is used (non-zero) for the text in Inkscape, but the package 'transparent.sty' is not loaded}%
    \renewcommand\transparent[1]{}%
  }%
  \providecommand\rotatebox[2]{#2}%
  \newcommand*\fsize{\dimexpr\f@size pt\relax}%
  \newcommand*\lineheight[1]{\fontsize{\fsize}{#1\fsize}\selectfont}%
  \ifx\svgwidth\undefined%
    \setlength{\unitlength}{1432.17819214bp}%
    \ifx\svgscale\undefined%
      \relax%
    \else%
      \setlength{\unitlength}{\unitlength * \real{\svgscale}}%
    \fi%
  \else%
    \setlength{\unitlength}{\svgwidth}%
  \fi%
  \global\let\svgwidth\undefined%
  \global\let\svgscale\undefined%
  \makeatother%
  \begin{picture}(1,0.65748566)%
    \lineheight{1}%
    \setlength\tabcolsep{0pt}%
    \put(0.35091214,0.68889541){\color[rgb]{0,0,0}\makebox(0,0)[lt]{\lineheight{0}\smash{\begin{tabular}[t]{l} \end{tabular}}}}%
    \put(1.57616269,0.01389242){\color[rgb]{0,0,0}\makebox(0,0)[lt]{\lineheight{0}\smash{\begin{tabular}[t]{l} \end{tabular}}}}%
    \put(0,0){\includegraphics[width=\unitlength,page=1]{harvie-ddr.pdf}}%
    \put(0.45465693,0.34533748){\color[rgb]{0,0,0}\makebox(0,0)[lt]{\lineheight{0}\smash{\begin{tabular}[t]{l}$\Cell_k$\end{tabular}}}}%
    \put(0,0){\includegraphics[width=\unitlength,page=2]{harvie-ddr.pdf}}%
    \put(0.29399464,0.32620176){\color[rgb]{0,0,0}\makebox(0,0)[lt]{\lineheight{0}\smash{\begin{tabular}[t]{l}$\PlicPosition$\end{tabular}}}}%
    \put(0.24400312,0.17548271){\color[rgb]{0,0,0}\makebox(0,0)[lt]{\lineheight{0}\smash{\begin{tabular}[t]{l}$\PlicNormal$\end{tabular}}}}%
    \put(0.80559463,0.33246294){\color[rgb]{0,0,0}\makebox(0,0)[lt]{\lineheight{0}\smash{\begin{tabular}[t]{l}$S_g$\end{tabular}}}}%
    \put(0.08065641,0.46025778){\color[rgb]{0,0,0}\makebox(0,0)[lt]{\lineheight{0}\smash{\begin{tabular}[t]{l}$\PhaseFluxVolumeContrib$\end{tabular}}}}%
    \put(-0.66418933,0.87605878){\color[rgb]{0,0,0}\makebox(0,0)[lt]{\begin{minipage}{2.12427517\unitlength}\raggedright  \end{minipage}}}%
    \put(0,0){\includegraphics[width=\unitlength,page=3]{harvie-ddr.pdf}}%
    \put(0.73378384,0.47627497){\color[rgb]{0,0,0}\makebox(0,0)[lt]{\lineheight{0}\smash{\begin{tabular}[t]{l}$\U_g$\end{tabular}}}}%
    \put(0.25409986,0.52483784){\color[rgb]{0,0,0}\makebox(0,0)[lt]{\lineheight{0}\smash{\begin{tabular}[t]{l}$\U_f$\end{tabular}}}}%
    \put(0,0){\includegraphics[width=\unitlength,page=4]{harvie-ddr.pdf}}%
    \put(0.47545298,0.59918863){\color[rgb]{0,0,0}\makebox(0,0)[lt]{\lineheight{0}\smash{\begin{tabular}[t]{l}$S_f$\end{tabular}}}}%
    \put(0,0){\includegraphics[width=\unitlength,page=5]{harvie-ddr.pdf}}%
    \put(0.37165958,0.01448885){\color[rgb]{0,0,0}\makebox(0,0)[lt]{\lineheight{0}\smash{\begin{tabular}[t]{l}$\U_h$\end{tabular}}}}%
    \put(0.47058026,0.04500433){\color[rgb]{0,0,0}\makebox(0,0)[lt]{\lineheight{0}\smash{\begin{tabular}[t]{l}$S_h$\end{tabular}}}}%
    \put(0.13169229,0.26174444){\color[rgb]{0,0,0}\makebox(0,0)[lt]{\lineheight{0}\smash{\begin{tabular}[t]{l}$S_j$\end{tabular}}}}%
    \put(0,0){\includegraphics[width=\unitlength,page=6]{harvie-ddr.pdf}}%
    \put(0.12305002,0.12479332){\color[rgb]{0,0,0}\makebox(0,0)[lt]{\lineheight{0}\smash{\begin{tabular}[t]{l}$\U_j$\end{tabular}}}}%
    \put(0.63066763,0.13792736){\color[rgb]{0,0,0}\makebox(0,0)[lt]{\lineheight{0}\smash{\begin{tabular}[t]{l}$V^\alpha_{g}$\end{tabular}}}}%
  \end{picture}%
\endgroup%

%% file: figures/hernandez2004.pdf_tex
\begingroup%
  \makeatletter%
  \providecommand\color[2][]{%
    \errmessage{(Inkscape) Color is used for the text in Inkscape, but the package 'color.sty' is not loaded}%
    \renewcommand\color[2][]{}%
  }%
  \providecommand\transparent[1]{%
    \errmessage{(Inkscape) Transparency is used (non-zero) for the text in Inkscape, but the package 'transparent.sty' is not loaded}%
    \renewcommand\transparent[1]{}%
  }%
  \providecommand\rotatebox[2]{#2}%
  \newcommand*\fsize{\dimexpr\f@size pt\relax}%
  \newcommand*\lineheight[1]{\fontsize{\fsize}{#1\fsize}\selectfont}%
  \ifx\svgwidth\undefined%
    \setlength{\unitlength}{1637.17419434bp}%
    \ifx\svgscale\undefined%
      \relax%
    \else%
      \setlength{\unitlength}{\unitlength * \real{\svgscale}}%
    \fi%
  \else%
    \setlength{\unitlength}{\svgwidth}%
  \fi%
  \global\let\svgwidth\undefined%
  \global\let\svgscale\undefined%
  \makeatother%
  \begin{picture}(1,0.45495835)%
    \lineheight{1}%
    \setlength\tabcolsep{0pt}%
    \put(0,0){\includegraphics[width=\unitlength,page=1]{hernandez2004.pdf}}%
    \put(0.20664774,0.44885026){\color[rgb]{0,0,0}\makebox(0,0)[lt]{\lineheight{0}\smash{\begin{tabular}[t]{l} \end{tabular}}}}%
    \put(1.27848061,-0.14163343){\color[rgb]{0,0,0}\makebox(0,0)[lt]{\lineheight{0}\smash{\begin{tabular}[t]{l} \end{tabular}}}}%
    \put(0,0){\includegraphics[width=\unitlength,page=2]{hernandez2004.pdf}}%
    \put(0.76074844,0.25465463){\color[rgb]{0,0,0}\makebox(0,0)[lt]{\lineheight{0}\smash{\begin{tabular}[t]{l}$\Point_{i}' = \Point_i + \lambda_f \Slope_i$\end{tabular}}}}%
    \put(0,0){\includegraphics[width=\unitlength,page=3]{hernandez2004.pdf}}%
    \put(0.45946208,0.11238778){\color[rgb]{0,0,0}\makebox(0,0)[lt]{\lineheight{0}\smash{\begin{tabular}[t]{l}$S_f$\end{tabular}}}}%
    \put(0.46775394,0.01259048){\color[rgb]{0,0,0}\makebox(0,0)[lt]{\lineheight{0}\smash{\begin{tabular}[t]{l}$\S_f$\end{tabular}}}}%
    \put(0.70993957,0.11867337){\color[rgb]{0,0,0}\makebox(0,0)[lt]{\lineheight{0}\smash{\begin{tabular}[t]{l}$\Point_i$\end{tabular}}}}%
    \put(0.60991462,0.35921109){\color[rgb]{0,0,0}\makebox(0,0)[lt]{\lineheight{0}\smash{\begin{tabular}[t]{l}$\PhaseFluxVolumeContrib$\end{tabular}}}}%
    \put(0.79684155,0.36867622){\color[rgb]{0,0,0}\makebox(0,0)[lt]{\lineheight{0}\smash{\begin{tabular}[t]{l} \end{tabular}}}}%
    \put(0.77378876,0.11404791){\color[rgb]{0,0,0}\makebox(0,0)[lt]{\lineheight{0}\smash{\begin{tabular}[t]{l} \end{tabular}}}}%
    \put(0.72787333,0.17738766){\color[rgb]{0,0,0}\makebox(0,0)[lt]{\lineheight{0}\smash{\begin{tabular}[t]{l}$\Slope_i$\end{tabular}}}}%
    \put(0.33366661,0.10904978){\color[rgb]{0,0,0}\makebox(0,0)[lt]{\lineheight{0}\smash{\begin{tabular}[t]{l}$\U_f$\end{tabular}}}}%
    \put(0,0){\includegraphics[width=\unitlength,page=4]{hernandez2004.pdf}}%
    \put(0.19220538,0.22506019){\color[rgb]{0,0,0}\makebox(0,0)[lt]{\lineheight{0}\smash{\begin{tabular}[t]{l}$\PlicNormal$\end{tabular}}}}%
    \put(0.19942642,0.33646349){\color[rgb]{0,0,0}\makebox(0,0)[lt]{\lineheight{0}\smash{\begin{tabular}[t]{l}$\PlicPosition$\end{tabular}}}}%
    \put(0,0){\includegraphics[width=\unitlength,page=5]{hernandez2004.pdf}}%
    \put(0.43436004,0.37983171){\color[rgb]{0,0,0}\makebox(0,0)[lt]{\lineheight{0}\smash{\begin{tabular}[t]{l}$\Omega_k$\end{tabular}}}}%
    \put(0.72616432,0.11896064){\color[rgb]{0,0,0}\makebox(0,0)[lt]{\lineheight{0}\smash{\begin{tabular}[t]{l} \end{tabular}}}}%
    \put(0,0){\includegraphics[width=\unitlength,page=6]{hernandez2004.pdf}}%
  \end{picture}%
\endgroup%

%% file: figures/Liovic2006.pdf_tex
\begingroup%
  \makeatletter%
  \providecommand\color[2][]{%
    \errmessage{(Inkscape) Color is used for the text in Inkscape, but the package 'color.sty' is not loaded}%
    \renewcommand\color[2][]{}%
  }%
  \providecommand\transparent[1]{%
    \errmessage{(Inkscape) Transparency is used (non-zero) for the text in Inkscape, but the package 'transparent.sty' is not loaded}%
    \renewcommand\transparent[1]{}%
  }%
  \providecommand\rotatebox[2]{#2}%
  \newcommand*\fsize{\dimexpr\f@size pt\relax}%
  \newcommand*\lineheight[1]{\fontsize{\fsize}{#1\fsize}\selectfont}%
  \ifx\svgwidth\undefined%
    \setlength{\unitlength}{1818.35870361bp}%
    \ifx\svgscale\undefined%
      \relax%
    \else%
      \setlength{\unitlength}{\unitlength * \real{\svgscale}}%
    \fi%
  \else%
    \setlength{\unitlength}{\svgwidth}%
  \fi%
  \global\let\svgwidth\undefined%
  \global\let\svgscale\undefined%
  \makeatother%
  \begin{picture}(1,0.81982895)%
    \lineheight{1}%
    \setlength\tabcolsep{0pt}%
    \put(0,0){\includegraphics[width=\unitlength,page=1]{Liovic2006.pdf}}%
    \put(0.30046783,0.26798916){\color[rgb]{0,0,0}\makebox(0,0)[lt]{\lineheight{0}\smash{\begin{tabular}[t]{l}$\U_f$\end{tabular}}}}%
    \put(0.71156911,0.45123576){\color[rgb]{0,0,0}\makebox(0,0)[lt]{\lineheight{0}\smash{\begin{tabular}[t]{l}$\U_g$\end{tabular}}}}%
    \put(0,0){\includegraphics[width=\unitlength,page=2]{Liovic2006.pdf}}%
    \put(0.25707022,0.68841516){\color[rgb]{0,0,0}\makebox(0,0)[lt]{\lineheight{0}\smash{\begin{tabular}[t]{l}$-\U_f\delta t$\end{tabular}}}}%
    \put(0.72471194,0.77585666){\color[rgb]{0,0,0}\makebox(0,0)[lt]{\lineheight{0}\smash{\begin{tabular}[t]{l}$-\U_g\delta t$\end{tabular}}}}%
    \put(-1.77883885,2.16988362){\color[rgb]{0,0,0}\makebox(0,0)[lt]{\begin{minipage}{1.61925845\unitlength}\raggedright \end{minipage}}}%
  \end{picture}%
\endgroup%

%% file: figures/Hernandez2008.pdf_tex
\begingroup%
  \makeatletter%
  \providecommand\color[2][]{%
    \errmessage{(Inkscape) Color is used for the text in Inkscape, but the package 'color.sty' is not loaded}%
    \renewcommand\color[2][]{}%
  }%
  \providecommand\transparent[1]{%
    \errmessage{(Inkscape) Transparency is used (non-zero) for the text in Inkscape, but the package 'transparent.sty' is not loaded}%
    \renewcommand\transparent[1]{}%
  }%
  \providecommand\rotatebox[2]{#2}%
  \newcommand*\fsize{\dimexpr\f@size pt\relax}%
  \newcommand*\lineheight[1]{\fontsize{\fsize}{#1\fsize}\selectfont}%
  \ifx\svgwidth\undefined%
    \setlength{\unitlength}{1779.91680908bp}%
    \ifx\svgscale\undefined%
      \relax%
    \else%
      \setlength{\unitlength}{\unitlength * \real{\svgscale}}%
    \fi%
  \else%
    \setlength{\unitlength}{\svgwidth}%
  \fi%
  \global\let\svgwidth\undefined%
  \global\let\svgscale\undefined%
  \makeatother%
  \begin{picture}(1,0.8742808)%
    \lineheight{1}%
    \setlength\tabcolsep{0pt}%
    \put(-1.89934009,2.21674782){\color[rgb]{0,0,0}\makebox(0,0)[lt]{\begin{minipage}{1.65423052\unitlength}\raggedright \end{minipage}}}%
    \put(0,0){\includegraphics[width=\unitlength,page=1]{Hernandez2008.pdf}}%
    \put(0.40522008,0.04455283){\color[rgb]{0,0,0}\makebox(0,0)[lt]{\lineheight{0}\smash{\begin{tabular}[t]{l}$\U_f$\end{tabular}}}}%
    \put(0.91000254,0.39606885){\color[rgb]{0,0,0}\makebox(0,0)[lt]{\lineheight{0}\smash{\begin{tabular}[t]{l}$\U_g$\end{tabular}}}}%
    \put(0,0){\includegraphics[width=\unitlength,page=2]{Hernandez2008.pdf}}%
    \put(0.60071935,0.30893764){\color[rgb]{0,0,0}\makebox(0,0)[lt]{\lineheight{0}\smash{\begin{tabular}[t]{l}$\U_e$\end{tabular}}}}%
    \put(0,0){\includegraphics[width=\unitlength,page=3]{Hernandez2008.pdf}}%
    \put(0.50567124,0.55512149){\color[rgb]{0,0,0}\makebox(0,0)[lt]{\lineheight{0}\smash{\begin{tabular}[t]{l}$-\U_e\delta t$\end{tabular}}}}%
    \put(0,0){\includegraphics[width=\unitlength,page=4]{Hernandez2008.pdf}}%
    \put(0.48670464,0.10758998){\color[rgb]{0,0,0}\makebox(0,0)[lt]{\lineheight{0}\smash{\begin{tabular}[t]{l}$-\U_e\delta t$\end{tabular}}}}%
    \put(0,0){\includegraphics[width=\unitlength,page=5]{Hernandez2008.pdf}}%
  \end{picture}%
\endgroup%

%% file: figures/triangulation-nonconvex-flux.pdf_tex
\begingroup%
  \makeatletter%
  \providecommand\color[2][]{%
    \errmessage{(Inkscape) Color is used for the text in Inkscape, but the package 'color.sty' is not loaded}%
    \renewcommand\color[2][]{}%
  }%
  \providecommand\transparent[1]{%
    \errmessage{(Inkscape) Transparency is used (non-zero) for the text in Inkscape, but the package 'transparent.sty' is not loaded}%
    \renewcommand\transparent[1]{}%
  }%
  \providecommand\rotatebox[2]{#2}%
  \newcommand*\fsize{\dimexpr\f@size pt\relax}%
  \newcommand*\lineheight[1]{\fontsize{\fsize}{#1\fsize}\selectfont}%
  \ifx\svgwidth\undefined%
    \setlength{\unitlength}{808.5bp}%
    \ifx\svgscale\undefined%
      \relax%
    \else%
      \setlength{\unitlength}{\unitlength * \real{\svgscale}}%
    \fi%
  \else%
    \setlength{\unitlength}{\svgwidth}%
  \fi%
  \global\let\svgwidth\undefined%
  \global\let\svgscale\undefined%
  \makeatother%
  \begin{picture}(1,0.77829314)%
    \lineheight{1}%
    \setlength\tabcolsep{0pt}%
    \put(0,0){\includegraphics[width=\unitlength,page=1]{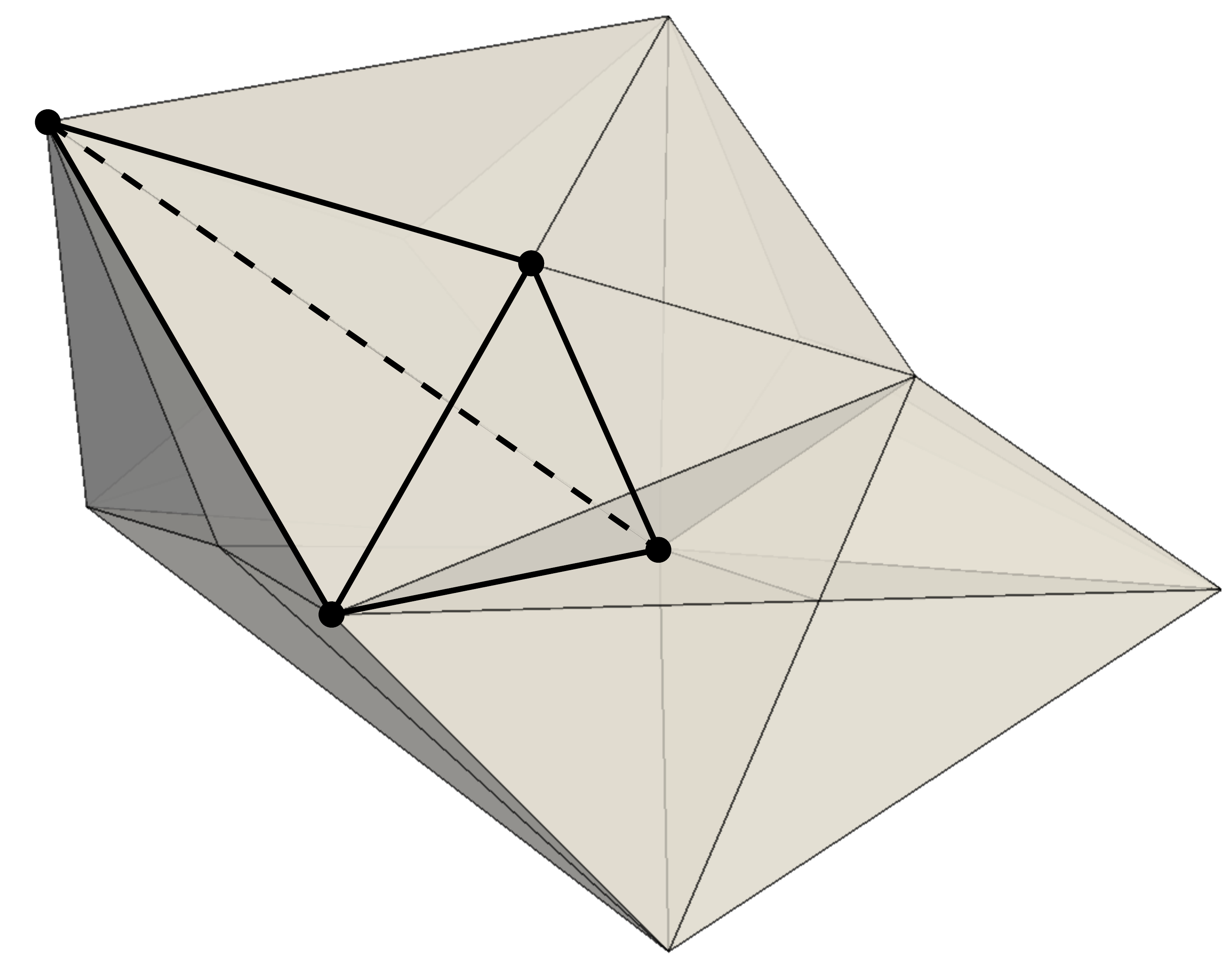}}%
  \end{picture}%
\endgroup%

%% file: figures/star-point.pdf_tex
\begingroup%
  \makeatletter%
  \providecommand\color[2][]{%
    \errmessage{(Inkscape) Color is used for the text in Inkscape, but the package 'color.sty' is not loaded}%
    \renewcommand\color[2][]{}%
  }%
  \providecommand\transparent[1]{%
    \errmessage{(Inkscape) Transparency is used (non-zero) for the text in Inkscape, but the package 'transparent.sty' is not loaded}%
    \renewcommand\transparent[1]{}%
  }%
  \providecommand\rotatebox[2]{#2}%
  \newcommand*\fsize{\dimexpr\f@size pt\relax}%
  \newcommand*\lineheight[1]{\fontsize{\fsize}{#1\fsize}\selectfont}%
  \ifx\svgwidth\undefined%
    \setlength{\unitlength}{441.66819763bp}%
    \ifx\svgscale\undefined%
      \relax%
    \else%
      \setlength{\unitlength}{\unitlength * \real{\svgscale}}%
    \fi%
  \else%
    \setlength{\unitlength}{\svgwidth}%
  \fi%
  \global\let\svgwidth\undefined%
  \global\let\svgscale\undefined%
  \makeatother%
  \begin{picture}(1,0.98673177)%
    \lineheight{1}%
    \setlength\tabcolsep{0pt}%
    \put(0.90696675,0.13449028){\makebox(0,0)[lt]{\lineheight{1.25}\smash{\begin{tabular}[t]{l}$\mathbf{x}^*_f$\end{tabular}}}}%
    \put(0,0){\includegraphics[width=\unitlength,page=1]{star-point.pdf}}%
    \put(0.00110172,0.91503938){\makebox(0,0)[lt]{\lineheight{1.25}\smash{\begin{tabular}[t]{l}$x'_{f,i}$\end{tabular}}}}%
    \put(0.41730418,0.03218895){\makebox(0,0)[lt]{\lineheight{1.25}\smash{\begin{tabular}[t]{l}$x'_{f,i+1}$\end{tabular}}}}%
    \put(0.7353083,0.69792356){\makebox(0,0)[lt]{\lineheight{1.25}\smash{\begin{tabular}[t]{l}$x'_f$\end{tabular}}}}%
  \end{picture}%
\endgroup%

%% file: tables/Comminal2015-extended.tex
\begin{tabular}{rllllllll}
\toprule
   T &  Method                                    &  $E_g$ (32)  &  $E_g$ (64) & O (32) &  $E_g$ (128) & O (64) & $E_g$ (256) & O (128) \\
\midrule
 0.5 & RK \citep{Rider1998}                  &  7.29e-4  &  1.42e-4 &   2.36 &  3.90e-5  &   1.86 &        - &       - \\
     & RK \citep{Rider1998}                  &  7.29e-4  &  1.42e-4 &   2.36 &  3.90e-5  &   1.86 &        - &       - \\
     & Stream \citep{Harvie2000}             &  5.51e-4  &  1.10e-4 &   2.32 &  3.38e-5  &   1.71 &        - &      -  \\
     & EMFPA-SIR \citep{Lopez2004}           &  4.45e-4  &  7.99e-5 &   2.48 &  2.04e-5  &   1.97 &        - &      -  \\
     & UFVFC-Swartz \citep{Maric2018}        &  6.56e-4  &  9.89e-5 &   2.73 &  2.29e-5  &   2.11 &  4.38e-6 &   2.39  \\ 
     & MZ \citep{Mencinger2011}              &  4.68e-4  &  6.91e-5 &   2.76 &  2.07e-5  &   1.74 &        - &      -  \\
     & GPCA \citep{Cervone2009}              &  4.12e-4  &  7.32e-5 &   2.41 &  1.93e-5  &   1.93 &        - &      -  \\
     & OD \citep{Owkes2014}                  &        -  &  -       &      - &         - &      - &        - &      -  \\
     & NIFPA-1 \citep{Ivey2017}              &        -  &  -       &      - &         - &      - &        - &      -  \\
     & CCU \citep{Comminal2015}              &  3.20e-4  &  7.68e-5 &   2.06 &  1.32e-5  &   2.54 &  2.45e-6 &   2.43  \\
     & iPAM ($h_L=0.1h$) \citep{Zhang2014}   &  4.07e-5  &  4.96e-6 &   3.04 &  5.86e-7  &   3.08 &        - &      -  \\

 2.0 & RK \citep{Rider1998}                  &  2.36e-3  &  5.85e-4 &   2.01 &  1.31e-4  &   2.16 &        - &      -  \\
     & Stream \citep{Harvie2000}             &  2.37e-3  &  5.65e-4 &   2.07 &  1.32e-4  &   2.10 &        - &      -  \\
     & EMFPA-SIR \citep{Lopez2004}           &  2.14e-3  &  5.39e-4 &   1.99 &  1.29e-4  &   2.06 &        - &      -  \\
     & MZ \citep{Mencinger2011}              &  2.11e-3  &  5.28e-4 &   2.00 &  1.28e-4  &   2.05 &        - &      -  \\
     & GPCA \citep{Cervone2009}              &  2.18e-3  &  5.32e-4 &   2.05 &  1.29e-4  &   2.03 &        - &      -  \\
     & CCU \citep{Comminal2015}              &  1.86e-3  &  4.18e-4 &   2.15 &  9.62e-5  &   2.12 &  1.97e-5 &   2.29  \\
     & iPAM ($h_L=0.1h$) \citep{Zhang2014}   &  8.34e-5  &  1.05e-5 &   2.99 &  1.37e-6  &   2.94 &        - &      -  \\

8.0 & THINC/QQ \textbf{(CFL=0.15)} \citep{Xie2017}      &  6.70e-2  &  1.52e-2 &   1.98 &  3.06e-3  &   2.27 &        - &      -  \\
    & isoAdvector-plicRDF \textbf{(CFL=0.5)} \citep{Scheufler2019}  &  -  &  1.26e-02 & 2.27 &  2.61e-03 &   2.19 &  5.71e-4 &   2.44  \\
    & NIFPA-1 \citep{Ivey2017}               &        -  &  1.14e-2 &      - &  2.68e-03 &   2.01 & 5.37e-04 &   2.32  \\
    & RK \citep{Rider1998}                   &  4.78e-2  &  6.96e-3 &   2.78 &  1.44e-03 &   2.27 &        - &      -  \\
    & Stream \citep{Harvie2000}              &  3.72e-2  &  6.79e-3 &   2.45 &  1.18e-03 &   2.52 &        - &      -  \\
    & EMFPA-SIR \citep{Lopez2004}            &  3.77e-2  &  6.58e-3 &   2.52 &  1.07e-03 &   2.62 &  2.35e-4 &   2.19  \\
    & MZ \citep{Mencinger2011}               &  5.42e-2  &  7.85e-3 &   2.79 &  1.05e-03 &   2.90 &        - &      -  \\
    & GPCA \citep{Cervone2009}               &        -  &        - &      - &  1.17e-03 &      - &        - &      -  \\
    & OD \citep{Owkes2014}                   &        -  &  7.58e-3 &      - &  1.88e-03 &   2.01 &  4.04e-4 &   2.22  \\
    & UFVFC-Swartz \citep{Maric2018}         &  3.78e-2  &  5.74e-3 &   2.72 &  1.45e-03 &   1.98 &  3.77e-4 &   1.95  \\ 
    & OD-S \citep{Owkes2014}                 &        -  &  1.04e-2 &   2.95 &  1.344e-03 &  1.94 &  3.50e-4 &   2.22  \\
    & CCU \citep{Comminal2015}               &  3.81e-2  &  4.58e-3 &   3.06 &  1.00e-03 &   2.20 &  1.78e-4 &   2.59  \\
    & \acs{AMR}-\ac{MoF}\citep{Ahn2009}       &  2.33e-2  &  3.15e-3 &   2.88 &  5.04e-04 &   2.64 &        - &      -  \\
    & iPAM ($h_L=0.1h$) \citep{Zhang2014}    &  6.21e-4  &  7.85e-5 &   2.99 &  9.89e-06  &   2.99 &       - &     -   \\
\bottomrule
\end{tabular}

%% file: tables/voFoam-hex-deformation3D-Trapezoid-Taylor-Trapezoid-Swartz-2-1-30-30-cellInt-trapezoid-pointCellInt-Taylor-fluxInt-trapezoid.tex
\begin{tabular}{llrrrrll}
\toprule
    &                                        &                      $E_v$       &  $E_b$ &      $E_g$  &   O($E_g$) & $T_e$ & $T_r$ \\   
CFL=0.5                                      & N &                  &     &         &        &       &      \\
\midrule
isoAdvector-plicRDF \citep{Scheufler2019}     & 32 \textbf{$CFL_\Sigma=0.5$} & 1.08e-16         & 2.12e-14         & 8.36e-3   & -  &  0.058       &  0.049 \\
THINC/QQ \citep{Xie2017}                     & 32 \textbf{$CFL=0.25$} & -         & -         & 7.96e-3  &  1.46  &  -      & - \\
Jofre et al. \citep{Jofre2014}                            & 32                   & -         & -         & 6.92e-3  &  -     &  -      & - \\
Owkes et al. \citep{Owkes2014}                            & 32                   & 2.79e-15  & 2.34e-17  & 6.98e-3  &  1.73  &  0.78   & - \\
\acs{PCFSC}-\acs{CVTNA} \citep{Liovic2006}     & 32                   & -         & -         & 7.41e-3  &  1.90  &  -      & - \\
\acs{FMFPA-3D}-\acs{CLCIR} \citep{Lopez2008}   & 32 \textbf{$CFL=1.0$}  & -         & -         & 6.85e-3  &  1.53  & -       & - \\
\acs{NIFPA}-1 \citep{Ivey2017}                & 33                   & -         & -         & 6.71e-3  &  1.58  &  -      & - \\
\acs{FMFPA-3D}-\acs{CBIR} \citep{Lopez2008}    & 32 \textbf{$CFL=1.0$}  & -         & -         & 6.64e-3  &  1.67  & -       & - \\
UFVFC-Swartz \citep{Maric2018}               & 32                   & 2.46e-15  & 0.0       & 5.86e-3  &  1.91  &  0.69   & 0.14 \\
Owkes et al. \citep{Owkes2017}                            & 32 (64)              & 3.07e-14  & 1.82e-17  & 2.31e-3  &  2.02  &  3.822  & - \\
\midrule
isoAdvector-plicRDF \citep{Scheufler2019}     & 64 \textbf{$CFL_\Sigma=0.5$} & 9.75e-16         & 6.41e-14         & 3.25e-3  &  1.36 &  0.15       &  0.13 \\
isoAdvector \citep{Roenby2016}                           & 64                   & 1.50e-13  & 2.6e-10   & 3.00e-3  &  2.31  & -       & - \\
THINC/QQ \citep{Xie2017}                     & 64 \textbf{$CFL=0.25$} & -         & -         & 2.89e-3  &  1.67  &  -      & - \\
Jofre et al. \citep{Jofre2014}                            & 64                   & -         & -         & 2.43e-3  &  1.51  & -       & - \\
\acs{FMFPA-3D}-\acs{CLCIR} \citep{Lopez2008}   & 64 \textbf{$CFL=1.0$}  & -         & -         & 2.38e-3  &  2.47  & -       & - \\
Owkes et al. \citep{Owkes2014}                            & 64                   & 1.675e-14 & 2.752e-17 & 2.10e-3  &  1.89  & 2.85    & - \\
\acs{NIFPA}-1 \citep{Ivey2017}                & 65                   & -         & -         & 2.24e-3  &  2.16  & -       & - \\
\acs{FMFPA-3D}-\acs{CBIR} \citep{Lopez2008}    & 64 \textbf{$CFL=1.0$}  & -         & -         & 2.09e-3  &  2.57  & -       & - \\
\acs{PCFSC}-\acs{CVTNA} \citep{Liovic2006}     & 64                   & -         & -         & 1.99e-3  &  2.69  & -       & - \\
UFVFC-Swartz \citep{Maric2018}               & 64                   & 6.01e-15  & 0.0       & 1.56e-3  &  2.34  & 2.81    & 0.51 \\
Owkes et al. \citep{Owkes2017}                            & 64 (128)             & 6.01e-14  & 1.11e-16  & 5.68e-4  &  2.05  & 9.799   & - \\
\midrule
THINC/QQ \citep{Xie2017}                     & 128 \textbf{$CFL=0.25$}& -         & -         & 9.05e-4  &  -     &  -      & - \\
isoAdvector-plicRDF \citep{Scheufler2019}     & 128 \textbf{$CFL_\Sigma=0.5$} & 3.705e-15         & 1.11e-15       & 6.57e-4   &  2.31 &  0.52       &  0.39 \\
isoAdvector \citep{Roenby2016}                           & 128                  & 1.20e-12  & 1.10e-7   & 6.04e-4  &  2.33  & -       & - \\
Jofre et al. \citep{Jofre2014}                            & 128                  & -         & -         & 6.37e-4  &  -     & -       & -  \\
Owkes et al. \citep{Owkes2014}                            & 128                  & 1.68e-14  & 2.75e-17  & 5.63e-4  &  -     & 12.2    & - \\
\acs{NIFPA}-1 \citep{Ivey2017}                & 129                  & -         & -         & 4.99e-4  &  1.89  & -       & - \\
\acs{FMFPA-3D}-\acs{CLCIR} \citep{Lopez2008}   & 128 \textbf{$CFL=1.0$} & -         & -         & 4.31e-4  &  2.47  & -       & - \\
\acs{PCFSC}-\acs{CVTNA} \citep{Liovic2006}     & 128                  & -         & -         & 3.09e-4  &  2.14  & -       & - \\
\acs{FMFPA-3D}-\acs{CBIR} \citep{Lopez2008}    & 128 \textbf{$CFL=1.0$} & -         & -         & 3.52e-4  &  2.59  & -       & - \\
UFVFC-Swartz \citep{Maric2018}               & 128                  & 1.56e-14  & 0.0       & 3.08e-4  &  -     & 12.00   & 2.37 \\
Owkes et al. \citep{Owkes2017}                            & 128 (256)            & 9.33e-14  & 3.33e-20  & 1.37e-4  &  1.51  & 26.636  & - \\
\midrule
isoAdvector \citep{Roenby2016}                           & 256                  & 9.40e-9    & 1.10e-10 & 1.20e-4  &  -  & -       & - \\
isoAdvector-plicRDF \citep{Scheufler2019}     & 256 \textbf{$CFL_\Sigma=0.5$} & 2.030e-14         & 2.366e-16       & 9.54e-5   &  2.78 &  2.63       &  1.67 \\
\bottomrule
\end{tabular}

%% file: sections/conclusions.tex
\section{Conclusions}

The presented review of numerous methods used for the reconstruction of the \ac{PLIC} interface and the advection of the volume fraction field outlined in \cref{sec:recon,sec:advect} shows that the \ac{LE} \GVOF{}s are actively being researched. Main obstacles in developing accurate, robust and efficient \GVOF{}s in $3D$ with support for unstructured meshes are the complex geometrical operations required for handling non-convex geometrical objects with non-planar faces, as well as the increase in computational complexity caused by the unstructured mesh topology. Although an extension of a \GVOF{} to support $3D$ computation on unstructured meshes is often claimed to be straightforward, the number of publications that actually do report results on $3D$ unstructured meshes shows that this is not the case. 

It seems that a direct comparison of methods based solely on the data available in the literature is often not possible. Interface reconstruction methods are very difficult to compare, as different error measures, initialization methods and validation cases are employed. The advection tests seem to be more standardized. However, different initialization algorithms that may play a role on fine mesh resolutions, different $CFL$ values, and different mesh resolutions are often used for the advection as well. \textcolor{Reviewer1R1}{Normalized time units are often used to report performance measurements, without providing the absolute CPU time used for the normalization, making it difficult to compare the methods directly.}

\textcolor{Reviewer1}{Reducing the dimensionality of geometrical calculations by the isoAvector and THINC/QQ schemes is an exciting approach because it significantly reduces the complexity and computational costs of the geometrical \ac{VOF} methods while keeping the errors comparable with other contemporary methods. Reduction of computational costs is especially relevant for unstructured meshes, where a significant computational overhead is introduced by $3D$ intersections when calculating fluxed phase-specific volumes. Convergence does seem to be better with lower CFL numbers for \textcolor{Reviewer1R1}{the isoAdvector and THINC/QQ methods}, as those methods avoid the calculation of the around-the-corner flux, and they have a reduced Lagrangean aspect of the calculation compared to fully un-split methods. However, hydrodynamic stability criteria imposed by surface-tension forces are often imposing much stricter time steps than the CFL criterion, at least in the context of two-phase DNS.}

Benefits of higher order of accuracy to the interpolation of velocity at cell-corner points and temporal integration is confirmed by $2D$ results of the cell-based \ac{LE} geometrical \ac{VOF} methods:  \ac{AMR}-\ac{MoF} \citep{Ahn2009}, \ac{GPCA} \citep{Cervone2009} and \ac{CCU}\citep{Comminal2015}. The highest accuracy shown in $2D$ by the \ac{iPAM} method \citep{Zhang2014} and \ac{AMR}-\ac{MoF} \citep{Ahn2009} and in $3D$ by the \ac{OD-S} method \citep{Owkes2017} shows the effect on the absolute error magnitude caused by adding sub-grid scale resolution and additional advection information to interface (\ac{iPAM}, and \ac{MoF}, respectively) or to the volume fraction advection (\ac{OD}-S).

\textcolor{Reviewer1}{Conversely to increasing the order of accuracy, \textcolor{Reviewer1R1}{reduced geometrical complexity of the geometric/algebraic VOF methods coupled} with dynamic \ac{AMR} might, in our opinion, present an effective way to achieve accurate, robust, and efficient solutions for an extensive range of two-phase DNS problems, \textcolor{Reviewer1R1}{as simpler algorithms are more straightforward to develop and implement in a scientific software, and are likely to have better parallel computational efficiency. Similarly, the high level of accuracy of three-dimensional geometric VOF methods \citep{Owkes2014,Jofre2014,Owkes2017,Ivey2017,Maric2018} motivates a further development of computationally efficient three-dimensional geometric calculations.}}